\documentclass[iop]{emulateapj}
\def\actaa{Acta Astronomica}

\begin{document}

\shorttitle{RR Lyrae Metallicity-Light Curve Relation}
\shortauthors{Ngeow et al.}

\title{Palomar Transient Factory and RR Lyrae: Metallicity-Light Curve Relation Based on ab-Type RR Lyrae in Kepler Field}

\author{Chow-Choong Ngeow\altaffilmark{1}, Po-Chieh Yu\altaffilmark{1}, Eric Bellm\altaffilmark{2}, Ting-Chang Yang\altaffilmark{1}, Chan-Kao Chang\altaffilmark{1}, Adam Miller\altaffilmark{3,5,6}, Russ Laher\altaffilmark{4}, Jason Surace\altaffilmark{4} and Wing-Huen Ip\altaffilmark{1}}

\altaffiltext{1}{Graduate Institute of Astronomy, National Central University, Jhongli 32001, Taiwan} \email{Corresponding/contact author: cngeow@astro.ncu.edu.tw}
\altaffiltext{2}{Cahill Center for Astronomy and Astrophysics, California Institute of Technology, Pasadena, CA 91125, USA}
\altaffiltext{3}{Division of Physics, Mathematics, and Astronomy, California Institute of Technology, Pasadena, CA 91125, USA}
\altaffiltext{4}{Spitzer Science Center, California Institute of Technology, Pasadena, CA 91125, USA}
\altaffiltext{5}{Jet Propulsion Laboratory, California Institute of Technology, Pasadena, CA 9110, USA}
\altaffiltext{6}{Hubble Fellow}

\begin{abstract}

The wide-field synoptic sky surveys, known as the Palomar Transient Factory (PTF) and the intermediate Palomar Transient Factory (iPTF), will accumulate a large number of known and new RR Lyrae. These RR Lyrae are good tracers to study the substructure of the Galactic halo if their distance, metallicity, and galactocentric velocity can be measured. Candidates of halo RR Lyrae can be identified from their distance and metallicity before requesting spectroscopic observations for confirmation. This is because both quantities can be obtained via their photometric light curves, because the absolute $V$-band magnitude for RR Lyrae is correlated with metallicity, and the metallicity can be estimated using a metallicity-light curve relation. To fully utilize the PTF and iPTF light-curve data in related future work, it is necessary to derive the metallicity-light curve relation in the native PTF/iPTF $R$-band photometric system. In this work, we derived such a relation using the known ab-type RR Lyrae located in the {\it Kepler} field, and it is found to be $[Fe/H]_{PTF}  =  -4.089 - 7.346 P + 1.280 \phi_{31}$ (where $P$ is pulsational period and $\phi_{31}$ is one of the Fourier parameters describing the shape of the light curve), with a dispersion of $0.118$~dex. We tested our metallicity-light curve relation with new spectroscopic observations of a few RR Lyrae in the {\it Kepler} field, as well as several data sets available in the literature. Our tests demonstrated that the derived metallicity-light curve relation could be used to estimate metallicities for the majority of the RR Lyrae, which are in agreement with the published values.

\end{abstract}

\keywords{stars: variables: RR Lyrae --- distance scale --- stars: abundances}

\section{Introduction}

The Palomar Transient Factory \citep[PTF, 2009-2012, see][]{law09,rau09} and its successor, the intermediate Palomar Transient Factory (iPTF, 2013-2016),\footnote{{\tt http://www.ptf.caltech.edu/iptf}} are dedicated wide-field synoptic sky survey projects with aims of detecting various types of transients in the Universe. Given the synoptic nature of PTF/iPTF surveys, a large number of known or new RR Lyrae with homogeneous light curve data is expected to be found in PTF/iPTF data. Since RR Lyrae are population II standard candles at which they have roughly a constant absolute magnitude in $V$ band ($M_V$), RR Lyrae have been used in various distance-scale studies such as tracing the Galactic halo structure \citep[for example, see][]{watkins09,sesar10}. Therefore, we have initiated a program to investigate the properties of RR Lyrae in the PTF/iPTF native $R$-band photometric system (hereafter $R_{PTF}$). 

It is well-known that $M_V$ for RR Lyrae is correlated with metallicity \citep[for example, see][]{mcnamara99,caputo00,demarque00,sandage06}, where the metallicity is mostly measured or expressed in terms of $[Fe/H]$, then the distance to an RR Lyrae can be deduced by knowing its metallicity and hence its $M_V$ value.\footnote{Note that the application of the $M_V$-$[Fe/H]$ relation to derive distance is still prone to several issues such as reddening to individual stars, the form of the $M_V$-$[Fe/H]$ relation (linear, quadratic, or two relations for metal-rich and metal-poor RR Lyraes), and evolutionary effects (such as empirical diagnostics to quantify the evolution away from the zero age horizontal branch). Recent discussions on these issues can be found, for example, in \citet{braga15} and \citet{marconi15}. Detailed investigations of these issues, however, are beyond the scope of this paper and will be addressed in subsequent papers. The current paper, which represents the first paper in a series, only deals with the $[Fe/H]$ part in the $M_V$-$[Fe/H]$ relation.} The best way to obtain $[Fe/H]$ is via a spectroscopic technique; however, this can be quite expensive in terms of telescope time. Fortunately, photometric $[Fe/H]$ for RR Lyrae can be estimated via the metallicity-light curve relation, at which the light curves for RR Lyrae can be fitted with a truncated sine-series of Fourier decomposition \citep[for example, see][]{simon81,deb09}\footnote{Note that their Fourier decomposition is expressed as a cosine-series, instead of a sine-series.}:

\begin{eqnarray}
m(t) & = & m_0 + \sum_{i=1}^{n} A_i \sin \left( \frac{2i\pi t}{P} + \phi_i \right), 
\end{eqnarray}

\noindent where $n$ is the order of fitting, $P$ is pulsation period in days, and $t$ is time of observation. The mean magnitude $m_0$, amplitude $A_i$, and phase $\phi_i$ values at given $i{\mathrm{th}}$-order can be obtained by fitting the observed light-curve data with Equation (1). The light-curve parameters, or Fourier parameters, can be expressed in terms of $A_i$ and $\phi_i$:

\begin{eqnarray}
R_{ij} & = & \frac{A_i}{A_j};\ \ \ \ \phi_{ij} \ = \ \phi_i - i \phi_j.
\end{eqnarray} 

The metallicity-light curve relation for fundamental mode $ab$-type RR Lyrae (hereafter RRab) was first quantitatively studied in \citet{simon88}. Later, such a relation was derived, or fitted, from field RRab in \citet{kovacs95}, and subsequently extended by \citet{jurcsik96}. In \citet{jurcsik96}, such a relation is given as $[Fe/H]_V = -5.038-5.394P+1.345\phi_{31}$, where $\phi_{31}=\phi_3-3\phi_1$ is calculated using Equation (1) and (2) based on the $V$-band light curves. A preliminary updated version of the \citet{jurcsik96} relation is presented in \citet{mv16}, which incorporates RR Lyrae in globular clusters. Other metallicity-light curve relations based on the $V$-band light curves can be found, for example, in \citet{sandage04}. Besides the $V$-band light curves, \citet{smolec05} gives the relation based on $I$-band light curves. \citet{wu06} and \citet{delee08} derived a similar relation as in \citet{jurcsik96} for unfiltered, or white light, CCD observations, and for the $g$-band SDSS (Sloan Digitized Sky Survey) data, respectively. \citet{watkins09}, \citet{sesar10}, and \citet{oluseyi12} further developed the metallicity-light curve relation in the SDSS photometric system with additional terms in the relation. Similarly, \citet{nemec11} and \citet{nemec13} derived such a relation in the {\it Kepler} magnitude ($K_p$) system. Alternatively, \citet{deb10} and \citet{skowron16} derived the Fourier interrelations to convert the $\phi_{31}$ parameters in $I$ band to $V$ band and then applied the metallicity-light curve relations mentioned. The validity of such a metallicity-light curve relation has been tested and verified, for example, in \citet{jurcsik03}, \citet{gratton04}, \citet{kocacs05}, \citet{wu06}, and \citet{kunder08}.

To fully utilize the RR Lyrae found in the PTF/iPTF data for future distance-scale work, it is necessary to derive the metallicity-light curve relation in the native $R_{PTF}$-band photometric system. Even though, in principle, it is possible to apply the transformation Equation provided in \citet{ofek12a} to convert the PTF/iPTF photometry in the $g_{PTF}$-band to the Johnson-Cousin $V$-band and hence to apply the \citet{jurcsik96} relation; in practice, this is difficult to achieve because (a) this transformation requires the $(V-R_c)$ color curves for the RR Lyrae found in PTF/iPTF to be available, but we generally do not have such data, and (b) the majority of the data taken in PTF/iPTF are in the $R_{PTF}$-band. Therefore, direct derivation of the metallicity-light curve relation in the native $R_{PTF}$-band is desirable. In this work, we present the derivation of such a relation by using the known RR Lyrae located in the {\it Kepler} field, because these RR Lyrae possess very precise and accurate period determination based on the {\it Kepler} light curves and spectroscopic measurements of $[Fe/H]$ \citep{nemec13}. A brief description of the PTF/iPTF project is presented in Section 2. The PTF/iPTF data for RR Lyrae in the {\it Kepler} field were mentioned in Section 3, followed by the construction of the light curves in Section 4. Based on the PTF/iPTF light curves, we derived the $\phi_{31}$ Fourier parameters in Section 5. The $R_{PTF}$-band metallicity-light curve relation will be derived and tested in Section 6. Finally, a discussion and our conclusions will be presented in Section 7. Throughout the paper, the Fourier parameter $\phi_{31}$ is based on the sine-series as shown in Equation (1), which can be converted to the cosine-based $\phi_{31}$ by subtracting $\pi$ (that is, $\phi_{31}^{\mathrm{cosine}} = \phi_{31}^{\mathrm{sine}} - \pi$). Also, as a reminder, throughout the paper, a $\pm0.25$~dex difference at $[Fe/H]=-1.5$~dex corresponds to roughly a factor of two difference in metal abundance by mass.

\section{Brief Description of the PTF/iPTF Project}

The PTF/iPTF project mainly utilizes the 48 inch Samuel Oschin Telescope located at the Palomar Observatory, known as the P48 Telescope, to search for transients. The P48 Telescope is equipped with a wide-field mosaic camera consists of eleven $2K\times 4K$ CCD\footnote{The original CFHT 12k mosaic camera, which consists of 12 CCD; however one of them is out of function.} \citep{rahmer08}, for the surveys carried out by both PTF and iPTF. The pixel scale of each CCD is $1.01$ arc-second per pixel, hence providing a total field of view (FOV) of $\sim7.26$-degree$^2$ for a single PTF/iPTF image. Observations with the P48 Telescope were mainly done in the Mould $R$-band filter (i.e. the $R_{PTF}$ filter), with occasional observations carried out in the $g_{PTF}$ or $H\alpha$ filters. The nominal exposure time for PTF/iPTF images is 60~s, which can reach to a depth of $20.5$~mag in the $R_{PTF}$ band (with a $3$sigma detection). 

The PTF/iPTF imaging data from the P48 Telescope was reduced and processed with two different pipelines \citep{law09}. One of the pipelines, based on the image subtraction technique, is tailored for quick discovery of transients; while another pipeline, hosted at the Infrared Processing and Analysis Center (IPAC), will fully reduce the raw images and provide catalogs of all detected objects in the images. Photometric calibration of the detected objects was also included in the IPAC pipeline. Further details of the IPAC pipeline and the photometric calibration procedure can be found in \citet{ofek12a,ofek12b}, \citet{laher14}, and \cite{surace15}, and will not be repeated here. Since the main goals for PTF and iPTF include the detections of transients (such as supernovae) in the local Universe and pursue for new discoveries with dedicated and well-designed experiments, hence the cadence carried out in PTF and iPTF varies from 90~s to a few days. Time-series data from PTF/iPTF has not only has been used for the search of transients, but also in the studies of other time-domain phenomena such as variable stars \citep[for an example, see][]{vanE11} and asteroids \citep[for an example, see][]{chang14}.

\section{RR Lyrae in the Kepler Field and the PTF/iPTF Light-Curve Data}

\begin{deluxetable*}{lcccccll}
\tabletypesize{\scriptsize}
\tablecaption{Basic Information for RR Lyrae in the {\it Kepler} Field\tablenotemark{a} \label{tab1}}
\tablewidth{0pt}
\tablehead{
\colhead{KIC\tablenotemark{b}} &
\colhead{R.A. (J2000)} &
\colhead{Decl. (J2000)} &
\colhead{$P$ [days]} &
\colhead{$t_o$} &
\colhead{$\langle Kp \rangle$} &
\colhead{Type\tablenotemark{c}} &
\colhead{Other Name}
}
\startdata
7198959 & 19:25:27.912 & +42:47:03.72 & 0.566788   & 2455278.2263 & 7.862  & RRab-B & RR Lyr \\
11125706& 19:00:58.774 & +48:44:42.30 & 0.6132200  & 2454981.0658 & 11.367 & RRab-B & KIC 11125706 \\
3733346 & 19:08:27.228 & +38:48:46.19 & 0.6820264  & 2454964.7403 & 12.684 & RRab-NB& NR Lyr \\
6936115 & 19:10:22.250 & +42:27:31.57 & 0.52739847 & 2454953.2656 & 12.876 & RRab-NB& FN Lyr \\
11802860& 19:00:48.000 & +50:05:31.27 & 0.6872160  & 2454954.2160 & 13.053 & RRab-NB& AW Dra \\
6763132 & 19:07:48.374 & +42:17:54.67 & 0.5877887  & 2454954.0702 & 13.075 & RRab-NB& NQ Lyr \\
9591503 & 19:33:00.912 & +46:14:22.85 & 0.5713866  & 2454953.5624 & 13.293 & RRab-NB& V894 Cyg \\
9947026 & 19:19:57.958 & +46:53:21.41 & 0.5485905  & 2454953.7832 & 13.300 & RRab-NB& V2470 Cyg \\
7030715 & 19:23:24.527 & +42:31:42.34 & 0.68361247 & 2454953.8427 & 13.452 & RRab-NB& KIC 7030715 \\
6100702 & 18:50:37.730 & +41:25:25.72 & 0.4881457  & 2454953.8399 & 13.458 & RRab-NB& KIC 6100702 \\
7021124 & 19:10:26.681 & +42:33:37.04 & 0.6224925  & 2454965.6471 & 13.550 & RRab-NB& KIC 7021124 \\
10789273& 19:14:03.905 & +48:11:58.60 & 0.48027971 & 2455807.9302 & 13.770 & RRab-B & V838 Cyg \\
10136603& 19:20:18.888 & +47:07:48.54 & 0.4337747  & 2455778.7060 & 14.066 & RRab-NB& V839 Cyg \\
7505345 & 18:53:25.903 & +43:09:16.45 & 0.4737027  & 2455124.7072 & 14.080 & RRab-B & V355 Lyr \\
7988343 & 19:59:50.669 & +43:42:15.52 & 0.5811436  & 2454964.6700 & 14.494 & RRab-NB& V1510 Cyg \\
5559631 & 19:52:52.740 & +40:47:35.45 & 0.62070001 & 2454975.5439 & 14.643 & RRab-B & V783 Cyg \\
12155928& 19:18:00.490 & +50:45:17.93 & 0.43638507 & 2455120.8363 & 15.033 & RRab-B & V1104 Cyg \\
4484128 & 19:45:39.024 & +39:30:53.42 & 0.5478642  & 2454970.2834 & 15.363 & RRab-B & V808 Cyg \\
6070714 & 19:56:22.906 & +41:20:23.53 & 0.5340941  & 2454964.8067 & 15.370 & RRab-NB& V784 Cyg \\
5299596 & 19:51:16.999 & +40:26:45.20 & 0.5236377  & 2454964.5059 & 15.392 & RRab-NB& V782 Cyg \\
3864443 & 19:40:06.963 & +38:58:20.35 & 0.4869538  & 2454976.3672 & 15.593 & RRab-B & V2178 Cyg \\
10136240& 19:19:45.279 & +47:06:04.44 & 0.5657781  & 2454964.7551 & 15.648 & RRab-NB& V1107 Cyg \\
9508655 & 18:49:08.369 & +46:11:54.96 & 0.5942369  & 2454964.7820 & 15.696 & RRab-NB& V350 Lyr \\
9658012 & 19:41:20.004 & +46:23:28.64 & 0.533206   & 2455779.9450 & 16.001 & RRab-NB& KIC 9658012 \\
9697825 & 19:01:58.634 & +46:26:45.74 & 0.5575765  & 2454988.9332 & 16.001 & RRab-B & V360 Lyr \\
7742534 & 19:10:53.403 & +43:24:54.94 & 0.4564851  & 2454964.7860 & 16.002 & RRab-NB& V368 Lyr \\
6183128 & 18:52:50.359 & +41:33:49.47 & 0.561691   & 2455245.1590 & 16.260 & RRab-B & V354 Lyr \\
3866709 & 19:42:07.997 & +38:54:42.30 & 0.47070609 & 2454964.6037 & 16.265 & RRab-NB& V715 Cyg \\
8344381 & 18:46:08.640 & +44:23:13.99 & 0.5768288  & 2454964.9231 & 16.421 & RRab-NB& V346 Lyr \\
9578833 & 19:09:40.637 & +46:17:18.17 & 0.5270283  & 2455326.1915 & 16.537 & RRab-B & V366 Lyr \\
7257008 & 18:47:27.408 & +42:49:52.68 & 0.51177516 & 2455758.5859 & 16.542 & RRab-B & KIC 7257008 \\
7671081 & 19:09:36.634 & +43:21:49.97 & 0.5046123  & 2454996.3226 & 16.653 & RRab-B & V450 Lyr \\
9001926 & 18:52:01.805 & +45:18:31.61 & 0.5568016  & 2455082.6820 & 16.914 & RRab-B & V353 Lyr \\
9973633 & 19:58:49.068 & +46:50:56.83 & 0.51075    & 2455780.3655 & 16.999 & RRab-B & KIC 9973633 \\
9717032 & 19:38:19.155 & +46:27:47.06 & 0.5569092  & 2455779.8956 & 17.194 & RRab-NB& KIC 9717032 \\
6186029 & 18:58:25.692 & +41:35:49.45 & 0.5131158  & 2455160.5957 & 17.401 & RRab-B & V445 Lyr \\
7176080 & 18:49:24.434 & +42:44:45.56 & 0.5070740  & 2454964.9588 & 17.433 & RRab-NB& V349 Lyr 
\enddata
\tablenotetext{a}{Information taken from \citet{nemec13}.}
\tablenotetext{b}{{\it Kepler} Input Catalog.}
\tablenotetext{c}{RRab-NB: Non-Blazhko RRab stars; RRab-B: Blazhko RRab stars.} 
\end{deluxetable*}

\begin{deluxetable}{llcc}
\tabletypesize{\scriptsize}
\tablecaption{Summary of PTF/iPTF $R_{PTF}$-band Observations \label{tab2}}
\tablewidth{0pt}
\tablehead{
\colhead{KIC} &
\colhead{Observing Window\tablenotemark{a}} &
\colhead{$N_{60s}$\tablenotemark{b}} &
\colhead{$N_{10s}$\tablenotemark{b}} 
}
\startdata
7198959 & 2011-03-16 to 2014-10-05             &  51 &  0 \\
11125706& 2012-08-04 to 2014-10-05, {\tt iPTF} &  44 & 54 \\
3733346 & 2012-08-04 to 2012-08-06             &   7 &  0 \\
6936115 & 2010-05-21 to 2014-10-05, {\tt iPTF} &  79 & 52 \\
11802860& 2012-08-04 to 2014-10-05, {\tt iPTF} &  75 & 52 \\
6763132 & 2010-05-19 to 2014-06-20, {\tt iPTF} &  60 & 54 \\
9591503 & 2010-05-27 to 2014-10-05, {\tt iPTF} & 203 & 51 \\
9947026 & 2012-07-11 to 2014-10-05, {\tt iPTF} &  52 & 53 \\
7030715 & 2010-05-21 to 2014-10-05, {\tt iPTF} &  79 & 52 \\
6100702 & 2010-05-19 to 2014-06-20, {\tt iPTF} &  60 & 54 \\
7021124 & 2010-05-21 to 2014-10-05, {\tt iPTF} &  79 & 52 \\ 
10789273& 2012-08-04 to 2014-10-05, {\tt iPTF} &  44 & 53 \\
10136603& 2013-07-11 to 2015-10-05, {\tt iPTF} &  52 & 53 \\
7505345 & 2011-03-16 to 2014-06-20, {\tt iPTF} &  50 & 53 \\
7988343 & 2011-03-16 to 2014-08-18, {\tt iPTF} &  81 & 49 \\
5559631 & 2011-08-06 to 2014-10-05             &  46 &  0 \\
12155928& 2012-08-04 to 2014-10-05, {\tt iPTF} &  75 & 52 \\
4484128 & $\cdots$                             &   0 &  0 \\
6070714 & 2011-08-06 to 2014-10-05             &  34 &  0 \\
5299596 & 2011-08-02 to 2014-10-05, {\tt iPTF} &  89 & 51 \\
3864443 & 2011-08-02 to 2014-10-05, {\tt iPTF} &  47 & 51 \\
10136240& 2012-07-11 to 2014-10-05, {\tt iPTF} &  52 & 53 \\
9508655 & 2011-07-11 to 2014-06-20             & 152 &  0 \\
9658012 & 2010-05-27 to 2014-10-05, {\tt iPTF} & 204 & 51 \\
9697825 & 2011-07-11 to 2014-06-20             & 152 &  0 \\
7742534 & 2011-03-16 to 2014-10-05             &  80 &  0 \\
6183128 & 2010-05-19 to 2014-06-20, {\tt iPTF} &  60 &  0 \\
3866709 & 2011-08-02 to 2014-10-05, {\tt iPTF} &  47 & 51 \\
8344381 & 2011-03-16 to 2014-06-20, {\tt iPTF} &  50 & 53 \\
9578833 & 2012-07-11 to 2014-10-05, {\tt iPTF} &  52 & 53 \\
7257008 & 2011-03-16 to 2014-06-20, {\tt iPTF} &  50 & 53 \\
7671081 & 2011-03-16 to 2014-10-05             &  80 &  0 \\
9001926 & 2011-07-11 to 2014-06-20             & 152 &  0 \\
9973633 & 2011-07-26 to 2014-10-05, {\tt iPTF} &  46 & 52 \\
9717032 & 2010-05-27 to 2014-10-05, {\tt iPTF} & 204 & 51 \\
6186029 & 2010-05-19 to 2014-06-20, {\tt iPTF} &  47 & 54 \\
7176080 & 2011-03-16 to 2014-06-20, {\tt iPTF} &  50 & 53 
\enddata
\tablenotetext{a}{First and last day of PTF data with nominal 60~s exposure, including dedicated iPTF experiment (2015-05-29 to 2015-05-31 and 2015-06-05) labeled as {\tt iPTF} if applicable.}
\tablenotetext{b}{$N_{60s}$: number of $R_{PTF}$-band PTF/iPTF {\tt SExtractor} catalogs with the nominal 60~s exposure; $N_{10s}$: number of $R_{PTF}$-band PTF/iPTF {\tt SExtractor} catalogs with 10~s exposure taken during the dedicated iPTF experiment.} 
\end{deluxetable}

In total, there have been $41$ RR Lyrae found in {\it Kepler} field, including 21 non-Blazhko RRab stars, 16 Blazhko RRab stars, and 4 first overtone $c$-type RR Lyrae (hereafter RRc). In this work, we excluded the 4 RRc stars, because the small number of them in the sample did not permit a meaningful statistical analysis of their metallicity-light curve relation. For the RRab stars, Table \ref{tab1} and \ref{tab2} listed out their basic information and summarized the available PTF/iPTF data in the $R_{PTF}$ band, respectively. Note that a portion of the PTF data on the {\it Kepler} field is publicly available, which belongs to the first data release\footnote{{\tt http://www.ptf.caltech.edu/page/first\_data\_release}} of the PTF data. The difference in number of catalogs between the full and publicly available data ranges from $4$ (for KIC6186029) to $79$ (for KIC7988343); for the majority of them, the difference is less than $10$.  

Using the publicly available catalog data for 16 non-Blazhko RRab stars in {\it Kepler} field, \citet{ngeow15} attempted to derive a preliminary metallicity-light curve relation in the $R_{PTF}$ band. However, it was found that the PTF light curves for 8 of them displayed a larger scatter in their light curves. These RR Lyrae have a mean $R_{PTF}$ magnitude of $\sim 14$~mag or brighter, which is close to the saturation limit of PTF data \citep{vanE11,ofek12a}. A provisional metallicity-light curve relation based on these 8 bright RRab stars exhibited a dispersion of $0.76$~dex. In contrast, the remaining 8 RRab stars, with mean $R_{PTF}$ magnitudes fainter than $\sim14$~mag, showed a much tighter PTF light curves, and the dispersion of the derived metallicity-light curve relation is $0.13$~dex. One finding of \citet{ngeow15} is that the bright RRab stars should not be used in the derivation of the metallicity-light curve relation because their photometry will be affected by saturation limits in PTF/iPTF surveys. Excluding the prototype RR Lyr (KIC7198959) itself and V808 Cyg (KIC4484128, because it does not have data in PTF/iPTF), there are $\sim15$ RRab stars (mixed of both Blazhko and non-Blazhko stars) that have mean $R_{PTF}$ magnitudes brighter than $\sim 14$~mag, which is almost half of the total 29 RRab stars that have spectroscopic $[Fe/H]$ measurements \citep{nemec13}. To maximize usable RRab stars in deriving the metallicity-light curve relation, we launched a dedicated iPTF experiment to re-observe these bright RRab stars with a 10~s exposure time (so their light curves will not suffer from the saturation limit), in opposition to the nominal 60~s exposure time set in the regular PTF/iPTF surveys.

\subsection{The Dedicated iPTF Experiment}

A dedicated iPTF experiment was carried out in four nights from 2015 May 29-31 and June 05. PTF fields that covered most of the bright RRab stars (except the prototype RR Lyr itself because it is too bright to observe with P48 Telescope) were selected to be observed repeatedly in these nights. Each of the PTF fields were observed $\sim9$ to $\sim15$ times per night, with a cadence of $\sim18$ to $\sim20$ minutes. In contrast, these RRab stars were observed one to seven times per night during the regular PTF/iPTF surveys, with nightly cadence varying for 1 night to $\sim15$ nights. Except the exposure time, which was reduced to 10~s, observations and data reduction for these PTF fields within our dedicated iPTF experiment were done with the same P48 Telescope, CCD camera, $R_{PTF}$-band filter, and IPAC reduction pipeline as the regular PTF/iPTF surveys. Since some of the faint RRab stars fell within the footprint of the selected PTF fields, we included them in our analysis as mentioned in the next section. The number of catalogs for all RRab stars in the {\it Kepler} field from our dedicated iPTF experiment was listed in the last column of Table \ref{tab2}. Similarly, the number of catalogs from regular PTF/iPTF surveys was given in the third column of Table \ref{tab2}.

\section{Light-Curves Construction}

Catalog data from the IPAC pipeline for RRab stars in our sample were downloaded from the PTF/IPAC data archive hosted at the NASA/IPAC Infrared Science Archive (IRSA).\footnote{{\tt http://irsa.ipac.caltech.edu/applications/ptf/}} These $R_{PTF}$-band PTF/iPTF {\tt SExtractor} \citep{bertin96} catalog data, including both of the catalogs from regular PTF/iPTF surveys (with a nominal 60~s exposure time) and the dedicated iPTF experiment as mentioned previously, were stored in FITS binary table format. A {\tt python} script\footnote{An example of such a script can be found in {\tt http://phares.caltech.edu/iptf/\\ iptf\_SummerSchool\_2014/Miller2\_problems.html}} was used to extract the $R_{PTF}$-band light curves for our RRab stars. This was done by matching the detected sources in the catalogs to the input RRab stars with a match radius of two~arc-second. The heliocentric Julian date (HJD), photometric magnitude, and magnitude error of the matched sources were saved into {\tt python} arrays. The PTF~$R_{PTF}$-band magnitudes were constructed by adding the {\tt MAG\_AUTO} and {\tt ZEROPOINT} in PTF catalogs \citep[for more details, see][]{ofek12a}. 

\begin{figure*}
  $\begin{array}{ccc}
    \includegraphics[angle=0,scale=0.18]{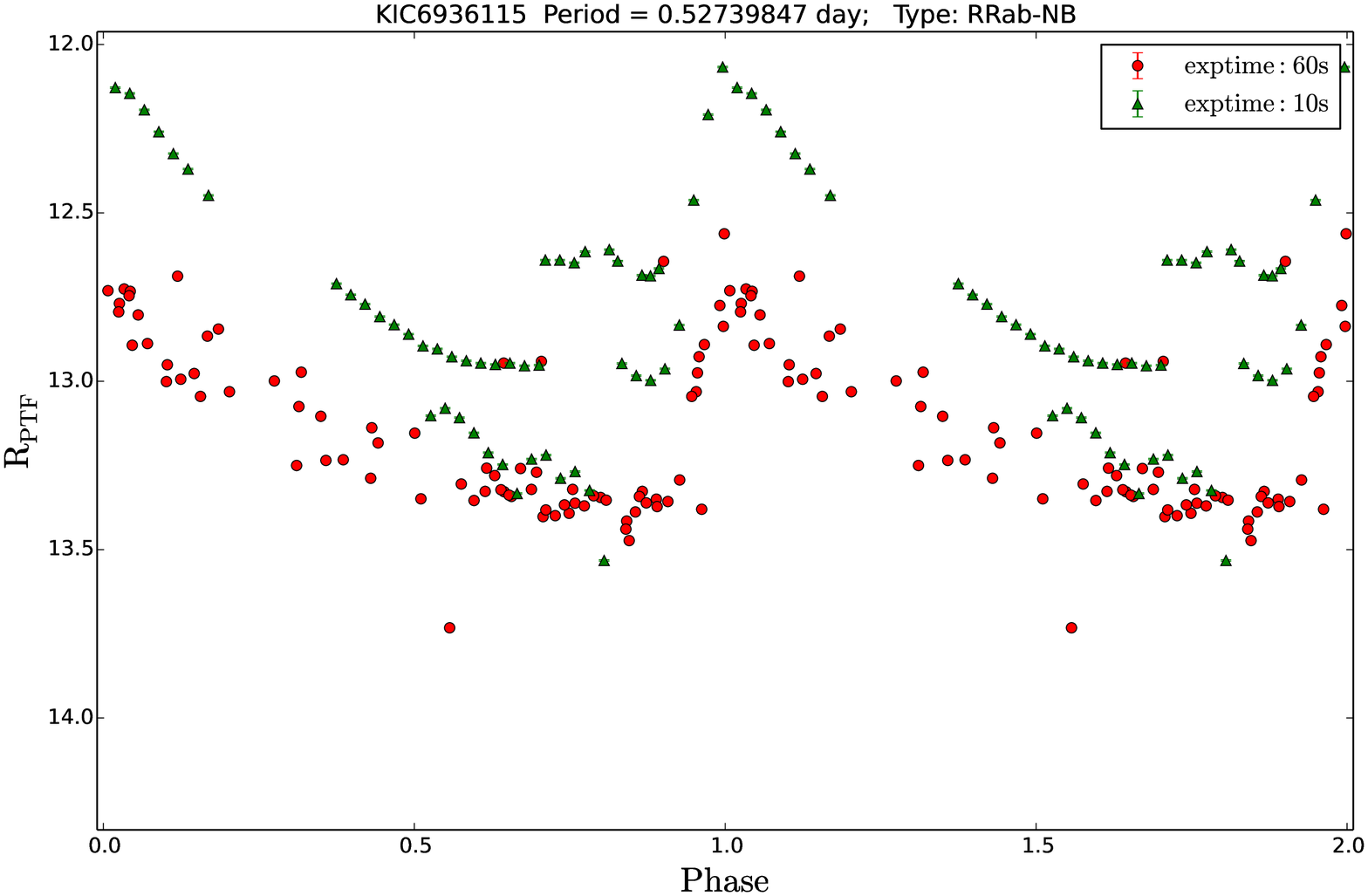} & 
    \includegraphics[angle=0,scale=0.18]{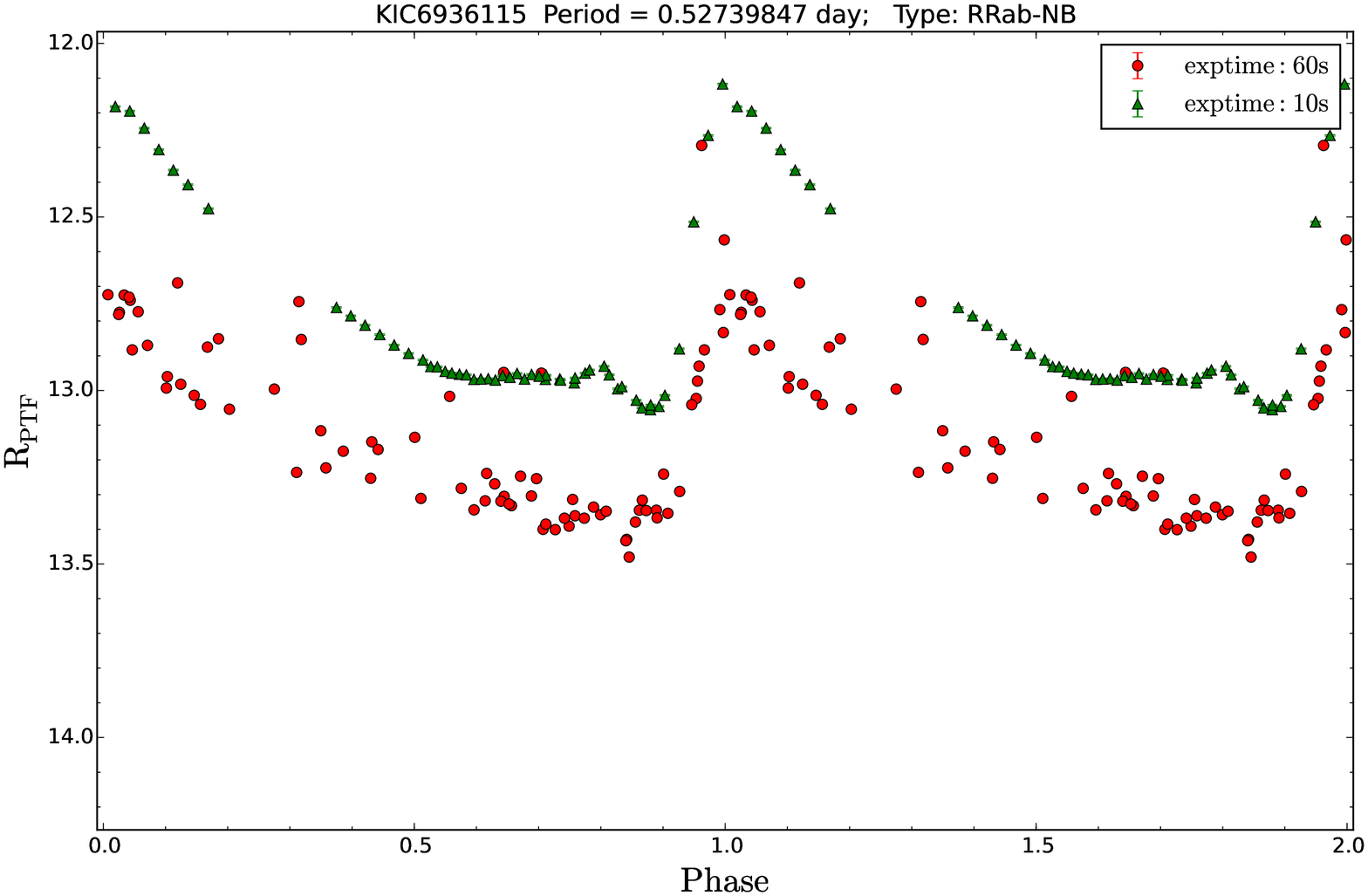} &
    \includegraphics[angle=0,scale=0.18]{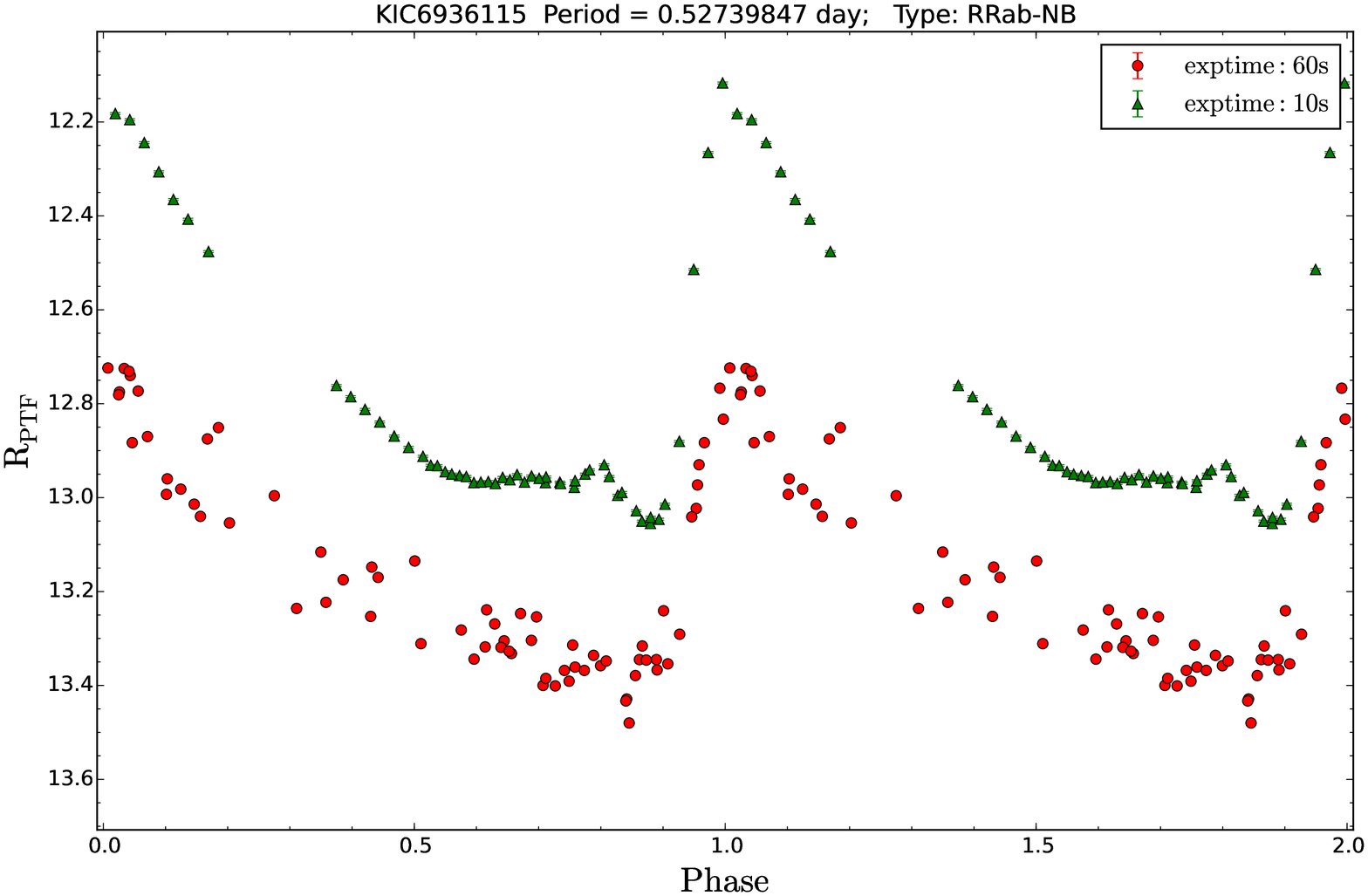} \\
    \includegraphics[angle=0,scale=0.18]{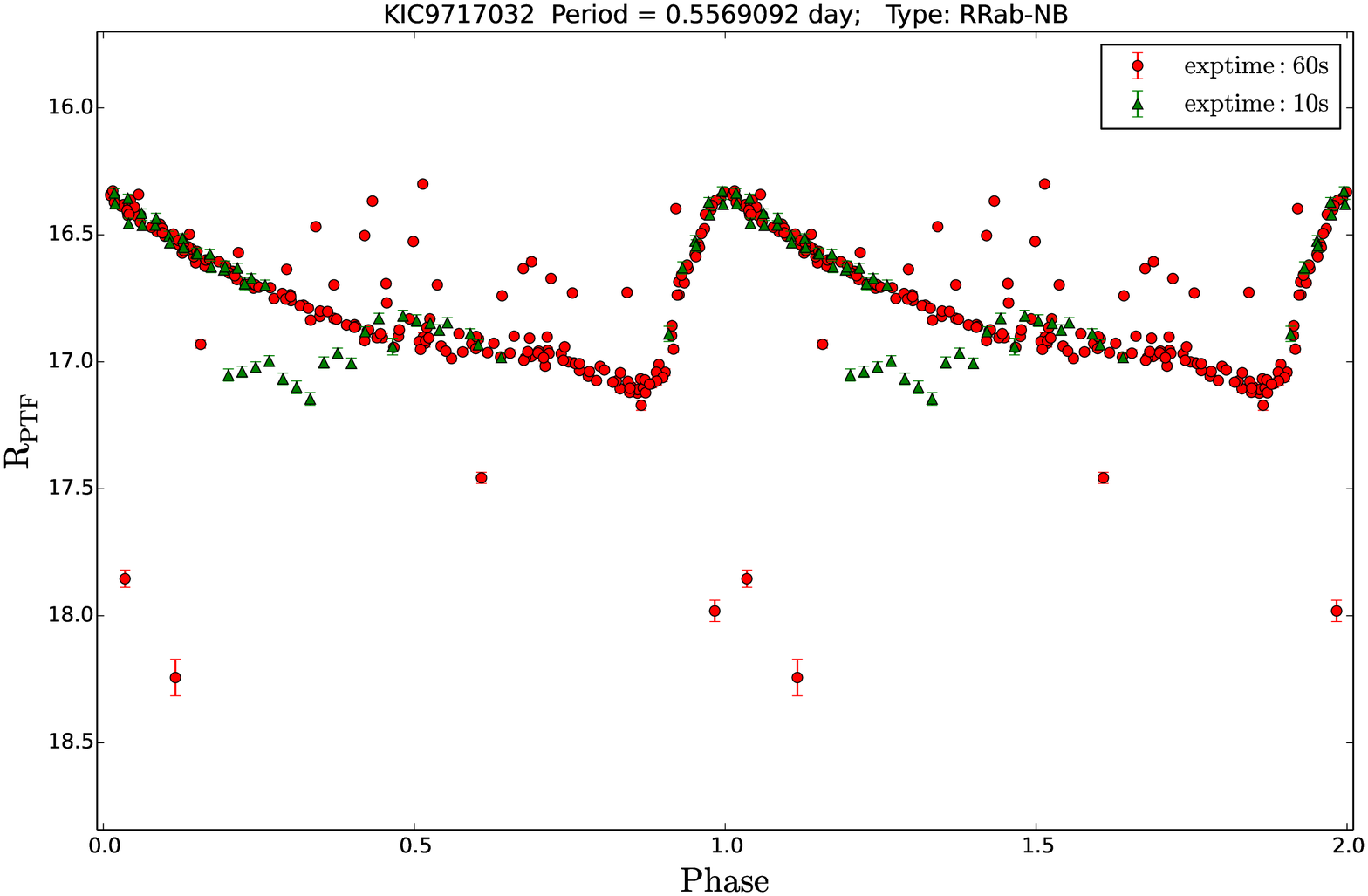} & 
    \includegraphics[angle=0,scale=0.18]{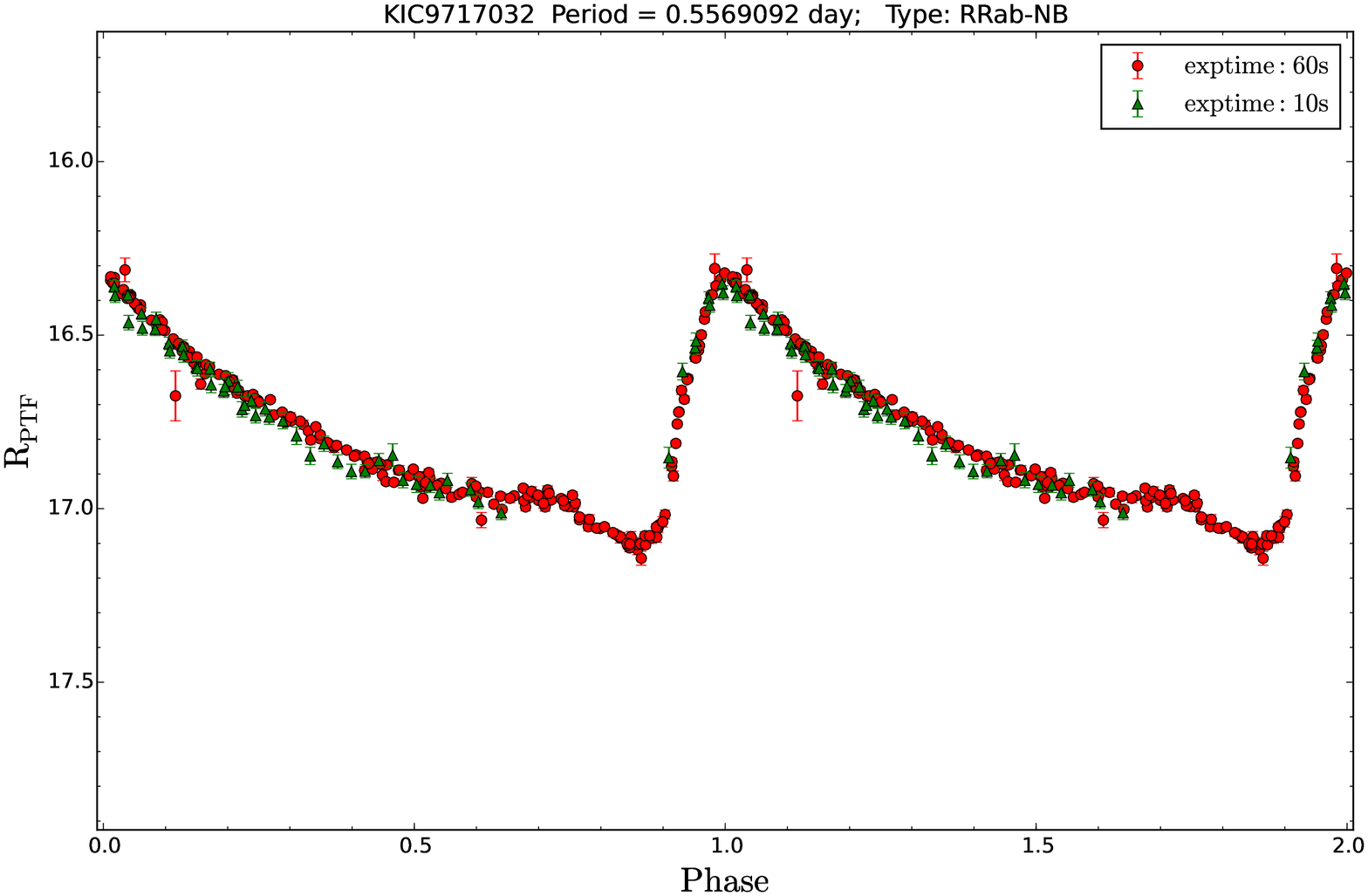} &
    \includegraphics[angle=0,scale=0.18]{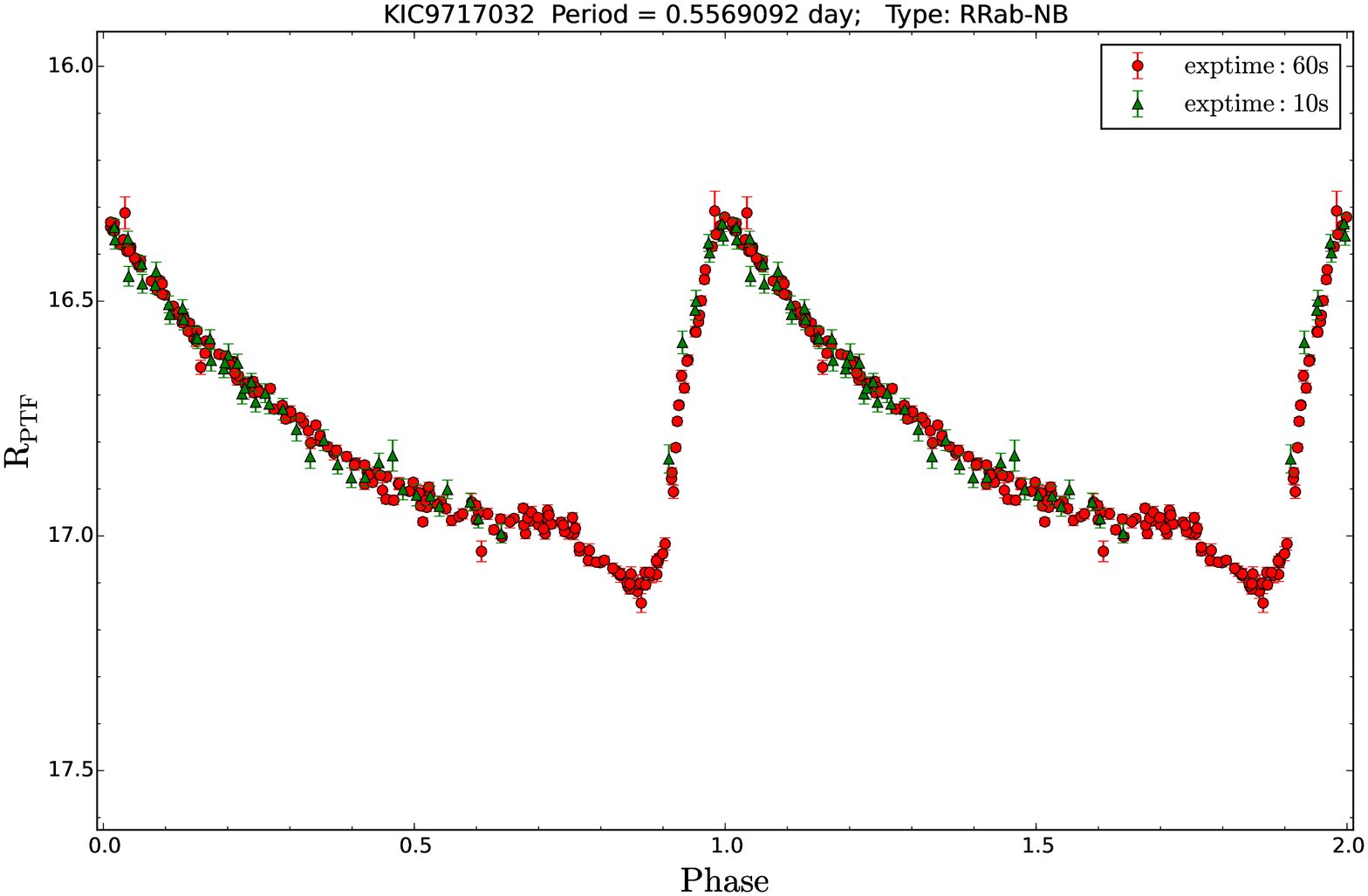} \\
  \end{array}$ 
  \caption{{\bf Left Panel:} example of extracted light curves for a bright RRab star (top panel) and a faint RRab star (bottom panel) from PTF/iPTF {\tt SExtractor} catalogs without any further photometric refinement. {\bf Middle Panel:} refined light curves after applying the differential photometry technique as described in the text. {\bf Right Panel:} further refinement of light curves after removing obvious outlier(s) from the 60~s light curves and adding a small magnitude shift to the 10~s light curves. Red filled circles and green triangles are for data with 60~s exposure time (from regular PTF/iPTF surveys) and 10~s exposure time (from the dedicated iPTF experiment).}
  \label{fig_v0}
\end{figure*}

Since none of the images involved in this work were taken under photometric conditions as defined in \citet[][i.e. the corresponding flag is $PHTCALFL=0$]{ofek12a}, the extracted ``raw'' light curves displayed numerous outliers. The left panel of Figure \ref{fig_v0} presents examples of the ``raw'' light curve for a bright and a faint non-Blazhko RR Lyrae. For the 10~s light curves, plots in the left panels of Figure \ref{fig_v0} show that one (or two) night from the dedicated iPTF experiment might be affected by weather. Furthermore, the upper-left panel of Figure \ref{fig_v0} displayed a vertical shift between the light curves taken with the 10~s and 60~s exposure time for the bright RR Lyrae. As discussed in \citet{ngeow15}, the 60~s data was affected by saturation, hence some fluxes were lost when using the aperture photometry. In contrast, the faint RR Lyrae shown in lower left panel of Figure \ref{fig_v0} does not have this problem.

To remedy the problem of large scatter shown in the ``raw'' light curves taken under the non-photometric condition, we employed a differential photometric technique \citep[e.g., see][]{honeycutt1992} to construct differential light curves for both of the 10~s and 60~s data. In addition to the reduction of nightly data, the IPAC pipeline also created stacked reference images and the associated {\tt SExtractor} reference catalogs \citep[see][for more details on how the reference images were created]{laher14}. We selected a subset of reference stars given in the {\tt SExtractor} reference catalogs for the RR Lyrae listed in Table \ref{tab1}\footnote{The only exception is for RR Lyrae KIC 3733346, because there are only 7 $R_{PTF}$-band images taken with the PTF/iPTF observations. For this RR Lyrae, we selected the best seeing image and catalog as reference.} to determine the mean magnitude differences, or relative zero-points $\Delta m$, between the reference stars and the nightly reduced catalogs. These reference stars have to meet with the following selection criteria: (a) exclude the targeted RR Lyrae stars in the reference catalogs; (b) $FLAGS=0$ in the {\tt SExtractor} catalog; (c) {\tt SExtractor} parameter $CLASS\_STAR>0.95$ (for stars-galaxies separation); (d) $15 < MAG\_AUTO < 17$ in the reference catalogs for the 60~s data (or $13 < MAG\_AUTO < 15$ for the 10~s data) such that stars with good enough signal-to-noise ratios were retained; and (e) more than 20 detections in nightly single-epoch catalogs (only for a few RR Lyrae stars, do the number of detections need to be tuned to a smaller value). For each of RR Lyrae stars, we then loop over the nightly single-epoch catalogs and cross-matched to the corresponding reference stars using a 2~arc-second search radius. Furthermore, we removed reference stars that might be variables using the following procedure: (1) calculate the mean variance of the photometric errors based on the ``raw'' light curves, $\langle \sigma^2_m\rangle$; (2) calculate the variance of the ``raw'' light curves using the median absolute deviation (MAD) algorithm, $\sigma^2_{LC}$; and (3) remove stars with $|\sigma^2_{LC}-\langle \sigma^2_m\rangle|>0.1$. For the remaining reference stars (for each of the RR Lyrae), the $\Delta m$ values are taken to be the median difference between the magnitudes from reference stars and the magnitudes in each single-epoch catalogs. The final adopted $\Delta m$ were then applied to the ``raw'' light curves to construct the differential light curves. The middle panels of Figure \ref{fig_v0} present the improvement of the light curves based on our procedures for the two example RR Lyrae. 

Several RR Lyrae in our sample still displayed few obvious outliers in the refined differential light curves and/or small offsets between the 60~s and 10~s light curves (see the middle panels of Figure \ref{fig_v0}). Those outliers were manually removed, and a small magnitude offset (which is smaller than 0.05~mag) is added to the 10~s light curves if needed. The right panels of Figure \ref{fig_v0} present the final refined light curves for the two exampled RR Lyrae. The final adopted differential light curves for the Blazhko and non-Blazhko RRab stars in our sample will be displayed in the next section. Note that the magnitudes in these light curves are not in absolute scale, as the a common zeropoint of $R_{PTF}=27.0$ was adopted when constructing the reference catalogs \citep{laher14}. Nevertheless, this would not affect the determination of Fourier parameters from these differential light curves, because Fourier parameters are independent of the global photometric calibration. 

\subsection{Excluded RRab Stars}

\begin{figure}
  \epsscale{1.0}
  \plotone{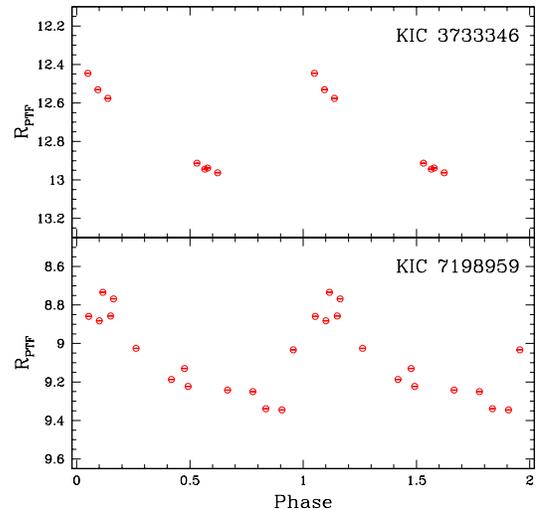}
  \caption{PTF 60~s $R_{PTF}$-band light curves for two bright RR Lyrae: KIC 3733346 (top panel) and KIC 7198959 (the prototype RR Lyr itself, bottom panel). Both light curves were excluded in our analysis.} 
  \label{fig_bad2}
\end{figure}

\begin{figure*}
  \epsscale{1.0}
  \plottwo{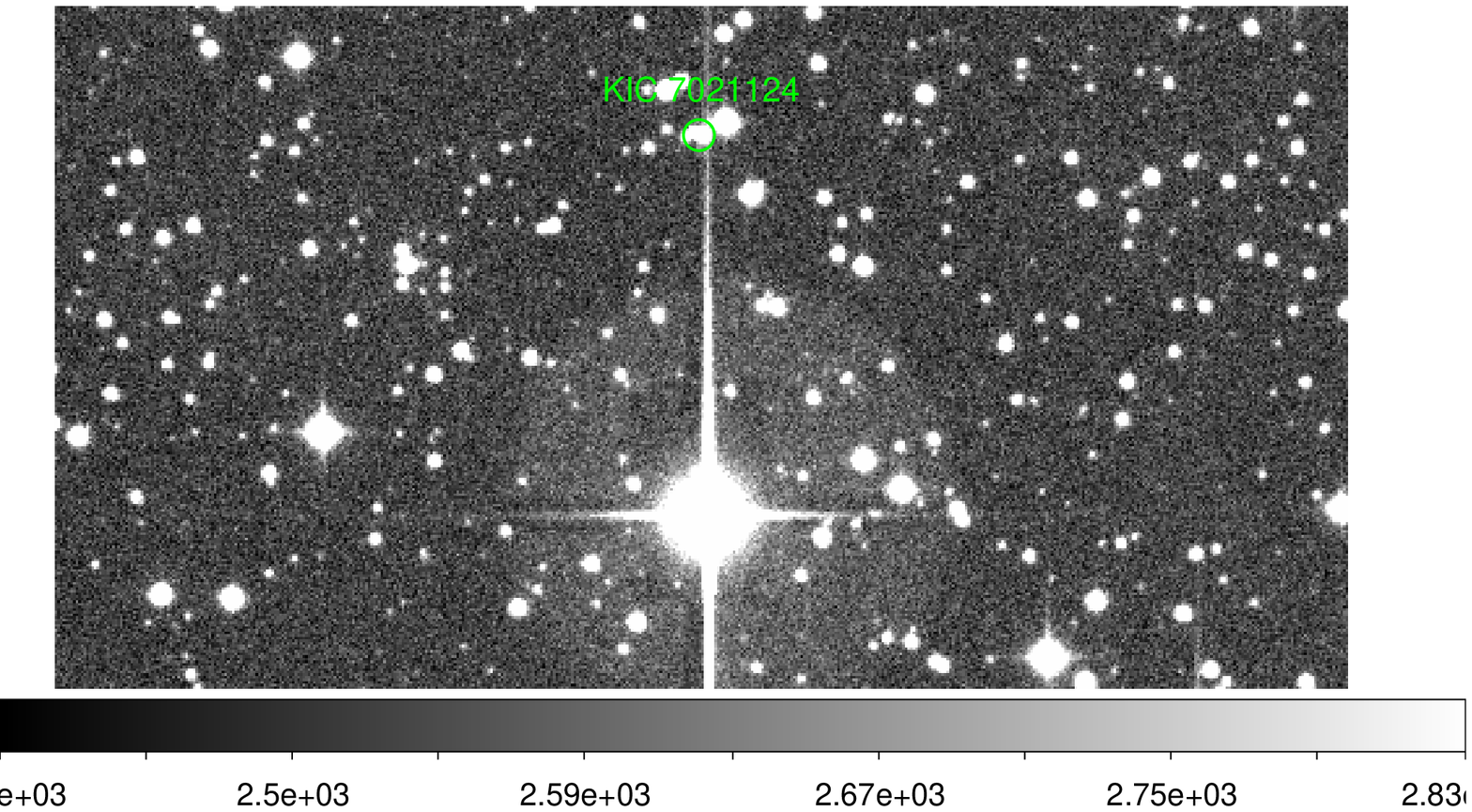}{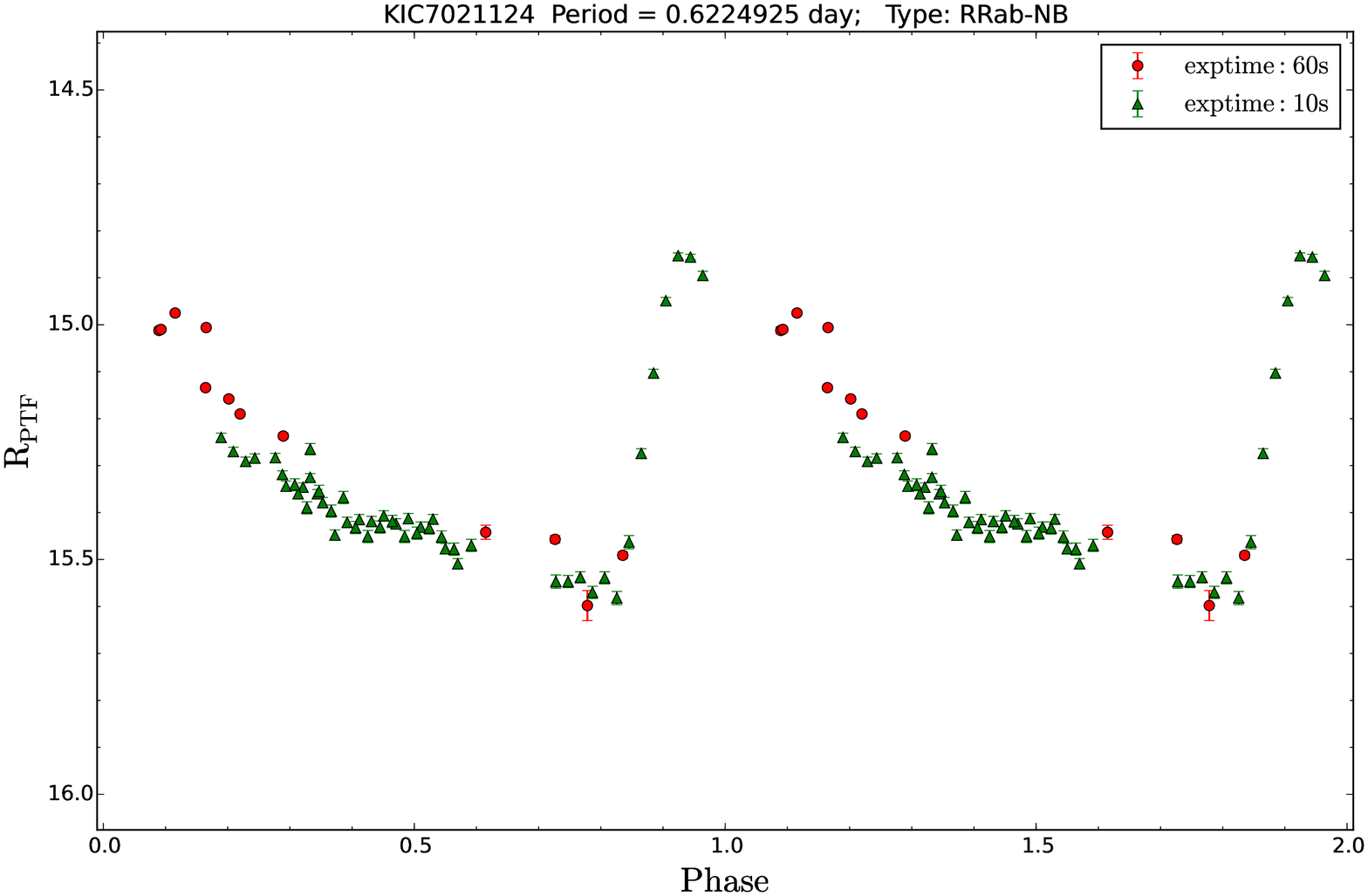}
  \caption{{\bf Left Panel:} a portion of the $R_{PTF}$-band image showing the influence of a diffraction spike from a nearby star on RR Lyrae KIC 7021124. {\bf Right Panel:} the 60~s (red circles) and 10~s (green triangles) differential light curves for this RR Lyrae.} 
  \label{fig_7021124}
\end{figure*}

\begin{figure*}
  \epsscale{1.0}
  \plottwo{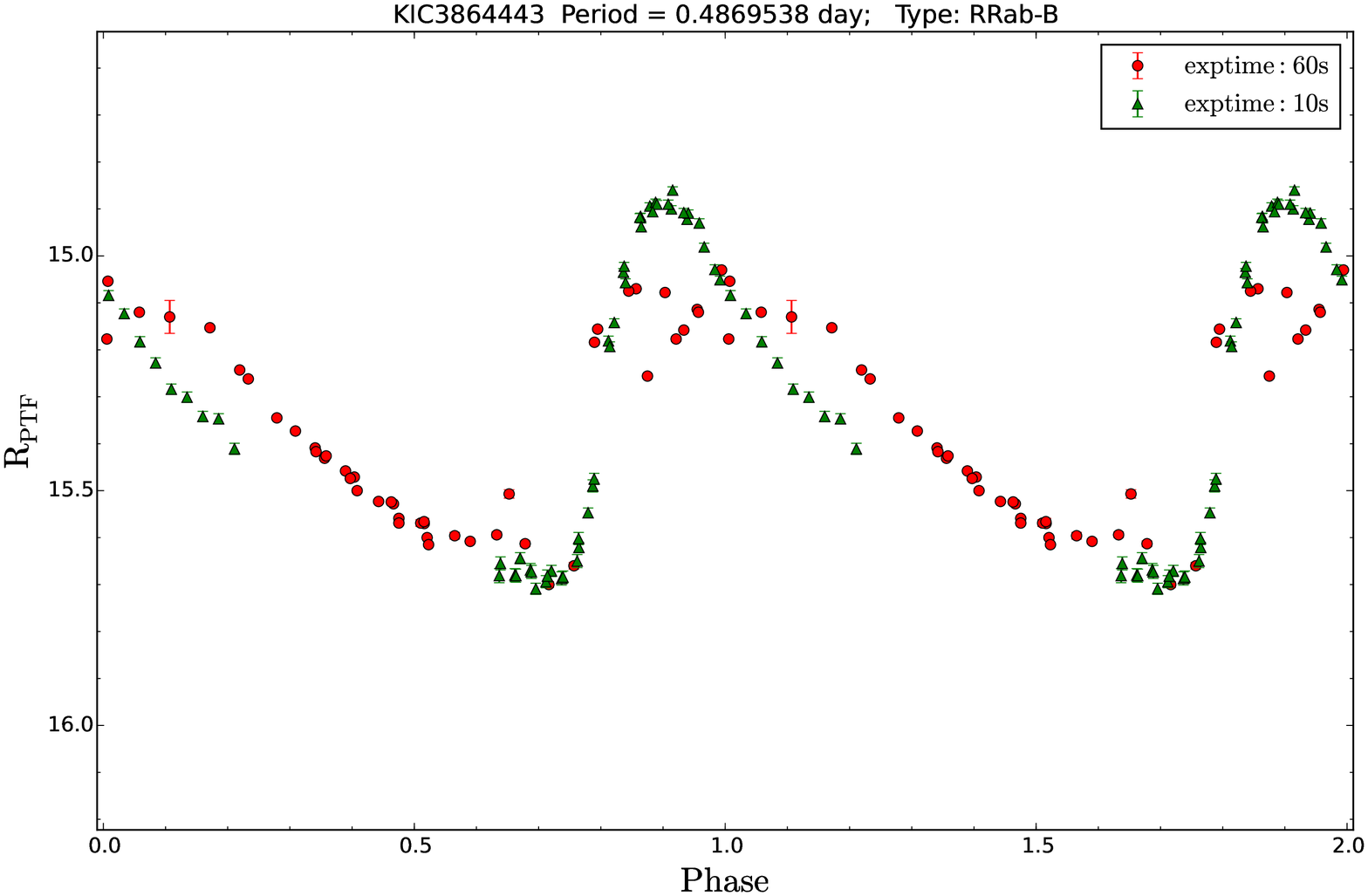}{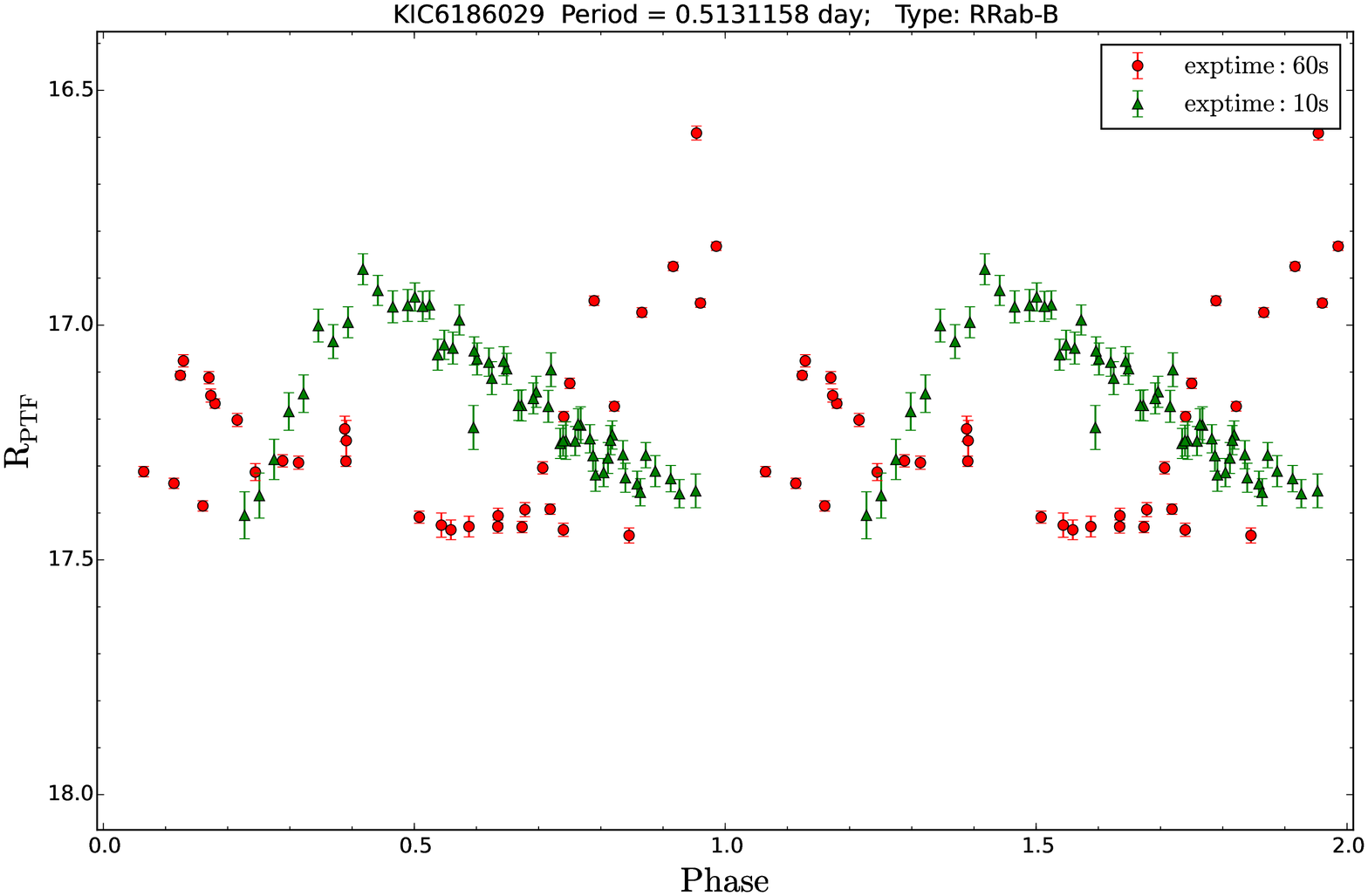}
  \caption{PTF $R_{PTF}$-band light curves for the two extreme Blazhko RR Lyrae identified in \citet{nemec13}. The red circles and green triangles are for the 60~s and 10~s, respectively, differential light curves.} 
  \label{fig_extreme}
\end{figure*}

We excluded the following RRab stars from our sample due to various reasons described below.

KIC 3733346: this bright RR Lyrae star only has seven data points in the $R_{PTF}$-band light curve (see top panel of Figure \ref{fig_bad2}) and hence does not permit a meaningful fitting of the Fourier parameter.

KIC 7198959: the prototype RR Lyr is too bright to be included in the 10~s observation, and the photometry (even after applying the differential light-curve technique) is severely affected by saturation. The light curve of this RR Lyrae is shown in the bottom panel of Figure \ref{fig_bad2}.

KIC 4484128: this RR Lyrae is located within the footprint of CCD 03 --- the only CCD chip that is out of function at the beginning of PTF/iPTF surveys, and hence no data collected from the PTF and iPTF observations.

KIC 7021124: photometry of this RR Lyrae was affected by the diffraction spike from a nearby bright star and a very close star with similar brightness, as shown in the left panel of Figure \ref{fig_7021124}. Therefore, we excluded this RR Lyrae in our sample. Differential light curves for this RR Lyrae were displayed in the right panel of Figure \ref{fig_7021124}.

KIC 3864443 and KIC 6186029: based on the long-term and almost continuous observations from {\it Kepler}, \citet{nemec13} identified these two RR Lyrae as extreme Blazhko stars because they exhibit large amplitudes and phase modulations when compared to other Blazhko RR Lyrae in the {\it Kepler} field. Figure \ref{fig_extreme} shows their $R_{PTF}$-band light curves, which also displayed obvious Blazhko modulations. Note that \citet{nemec13} excluded them in their analysis; therefore, we also excluded them in our sample. Further analysis and discussion on KIC 6186029 (V445 Lyr) can be found in \citet{guggenberger12}.

\section{Deriving the $\phi_{31}$ Fourier Parameters}

\subsection{For Non-Blazhko RRab Stars}

\begin{figure}
  %\epsscale{0.8}
  \plotone{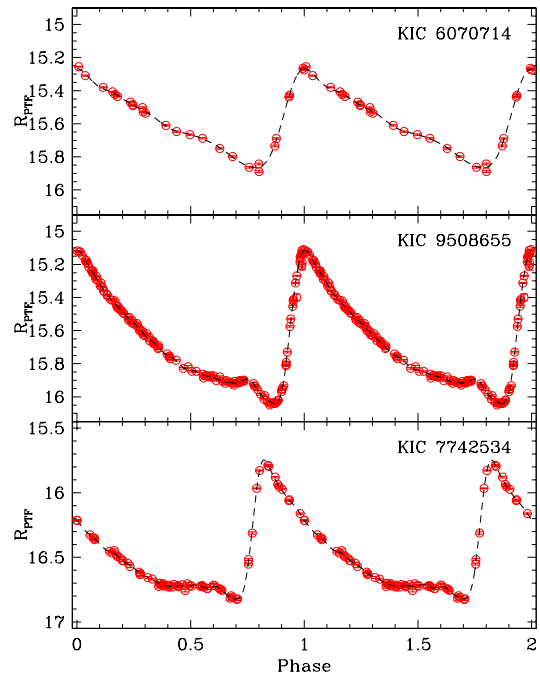}
  \caption{Differential light curves for three non-Blazhko RR Lyrae stars in our sample after applying the differential photometry technique as described in the text. These three RR Lyrae only have the 60~s observations. The dashed curves are the fitted light curves using Fourier expansion as given in Equation (1).} 
  \label{fig_3nonBL}
\end{figure}

\begin{figure*}
  $\begin{array}{cccc}
    \includegraphics[angle=0,scale=0.21]{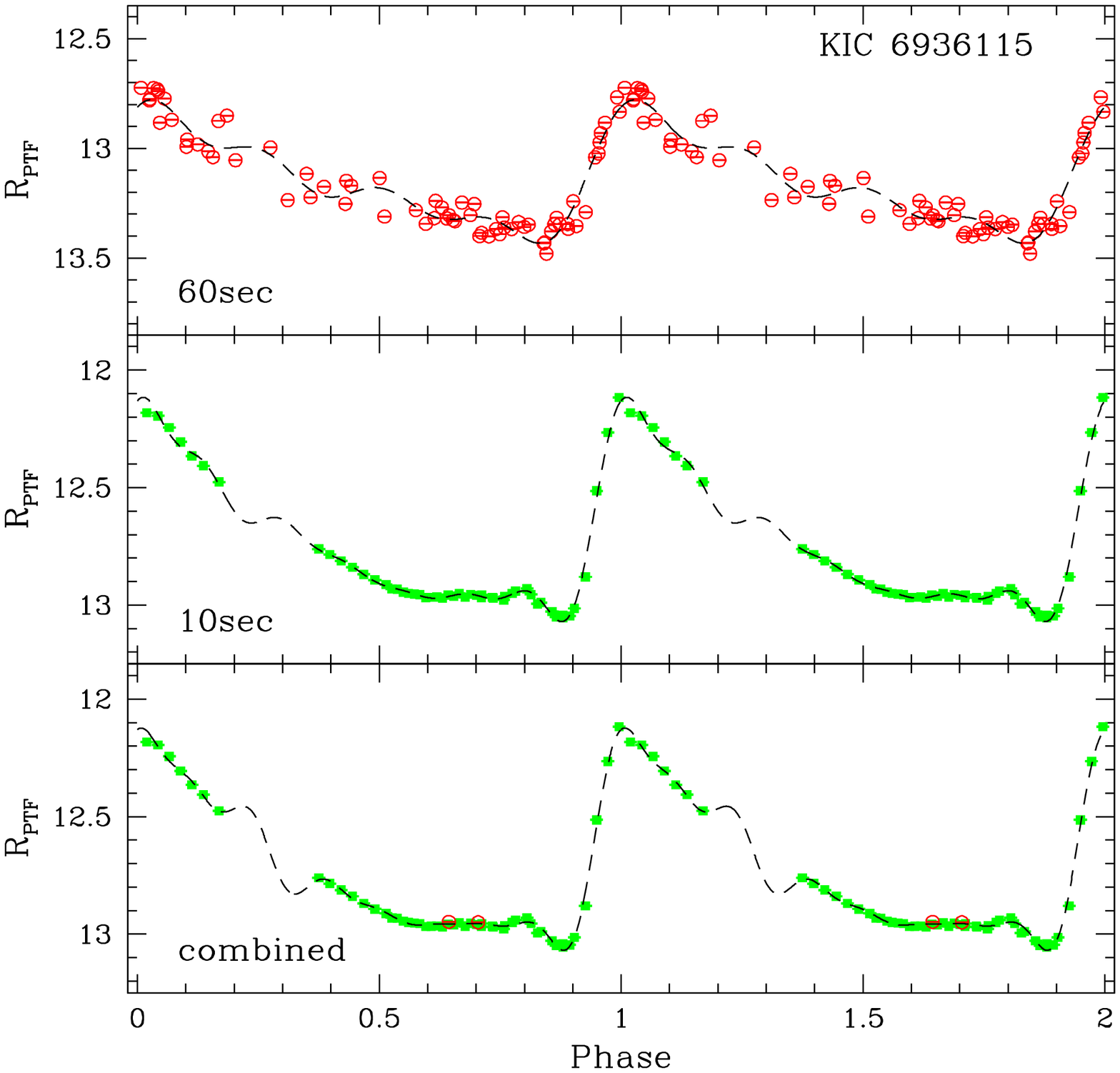} &
    \includegraphics[angle=0,scale=0.21]{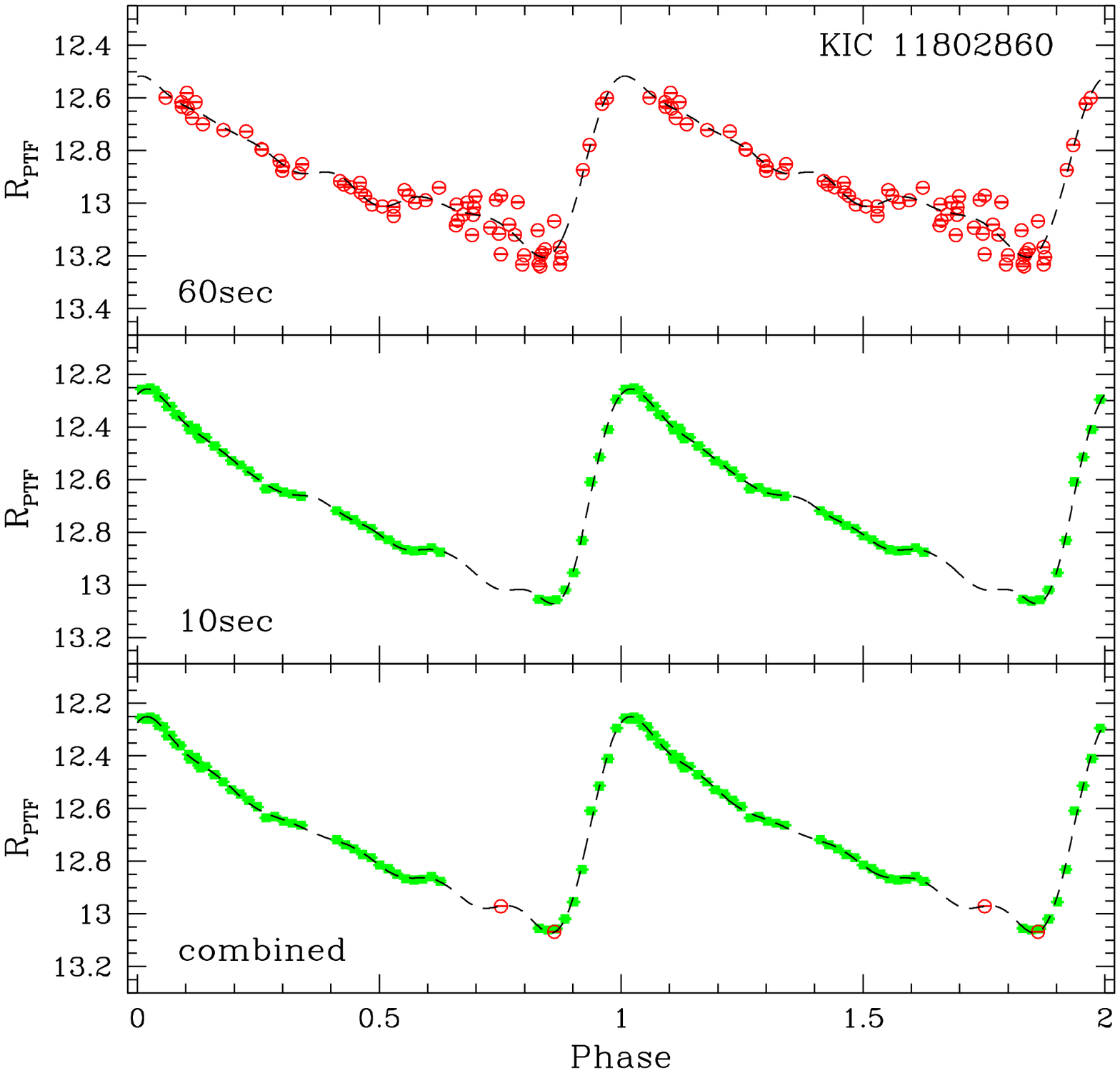} &
    \includegraphics[angle=0,scale=0.21]{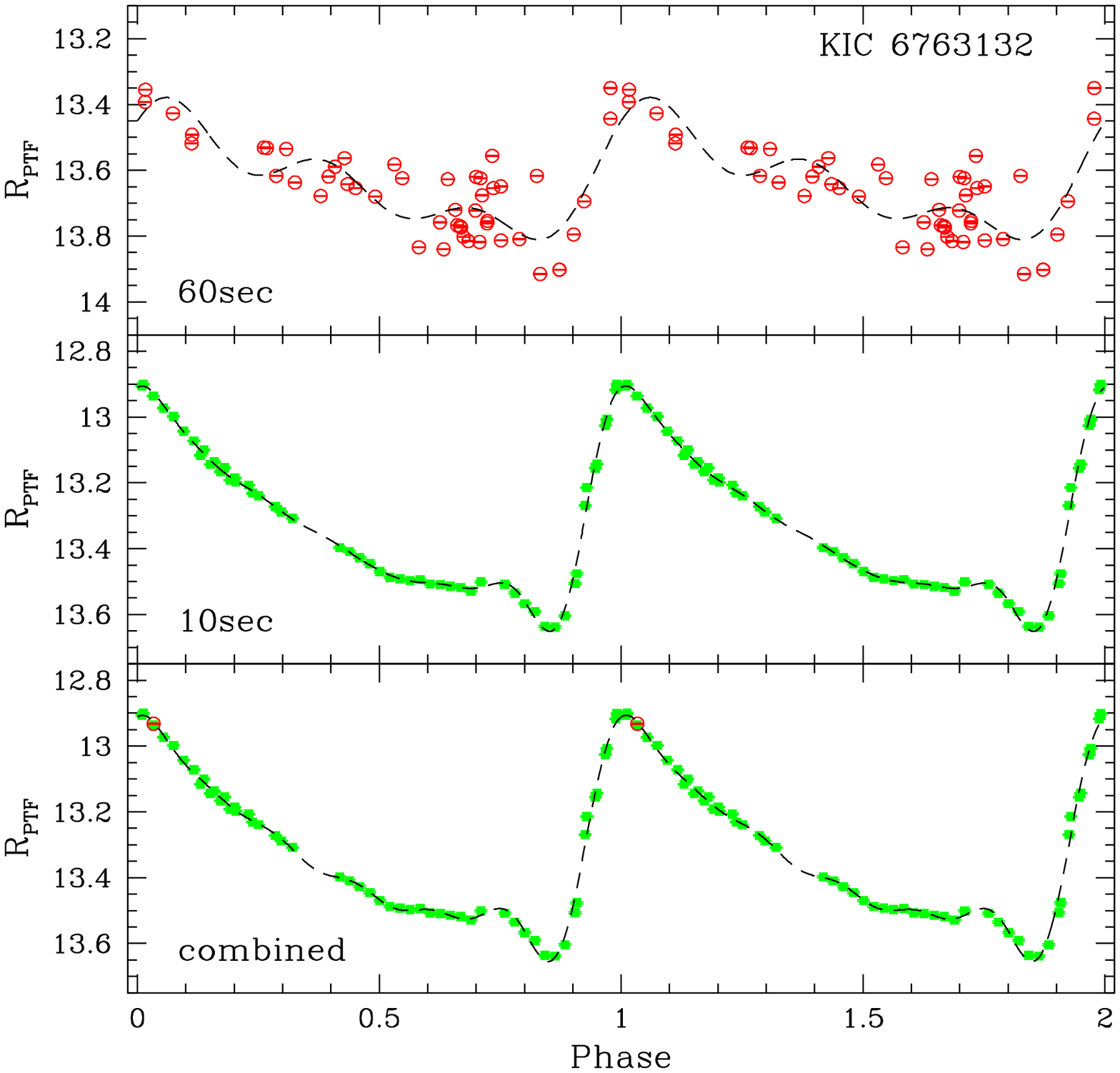} &
    \includegraphics[angle=0,scale=0.21]{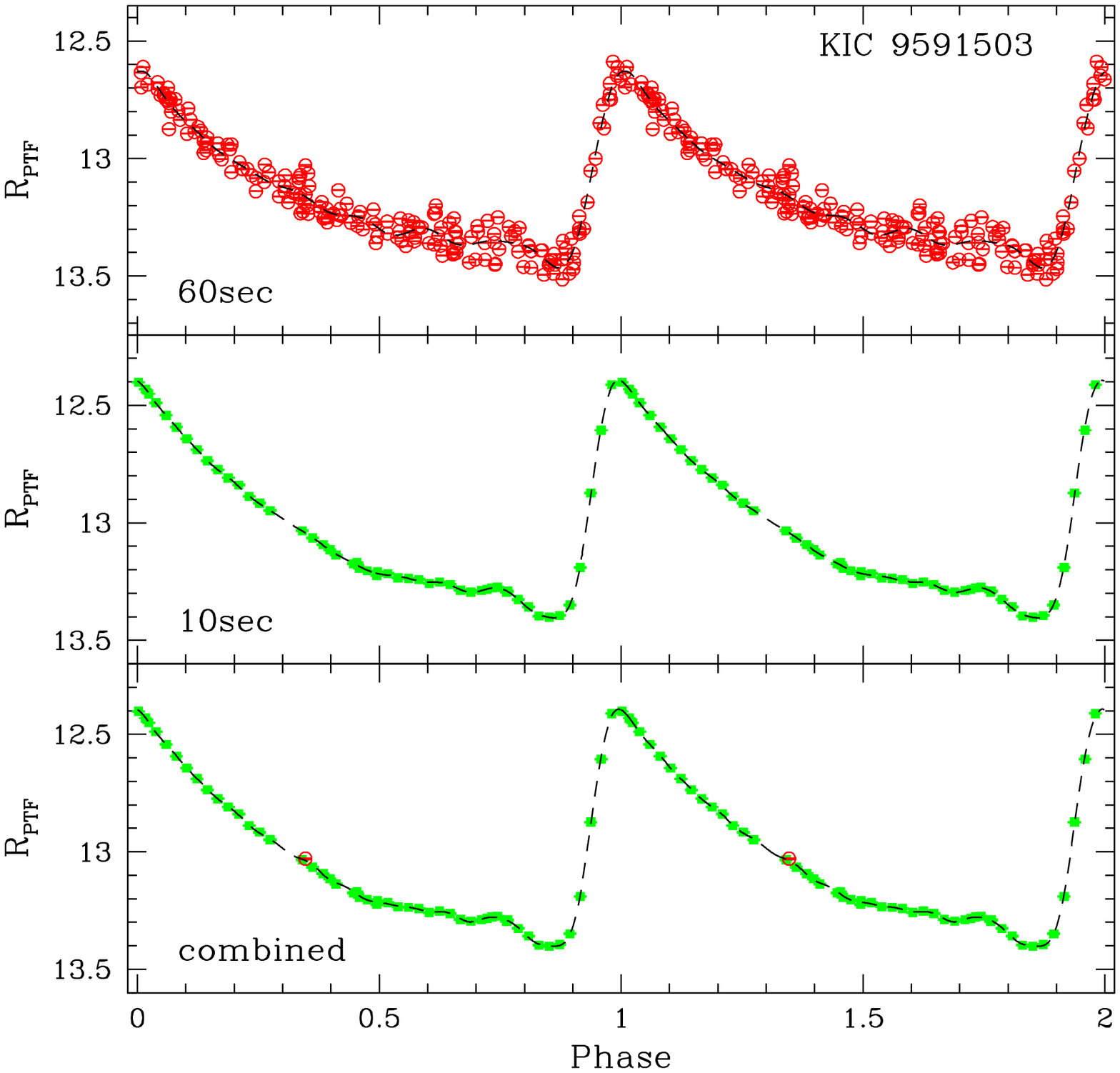} \\
    \includegraphics[angle=0,scale=0.21]{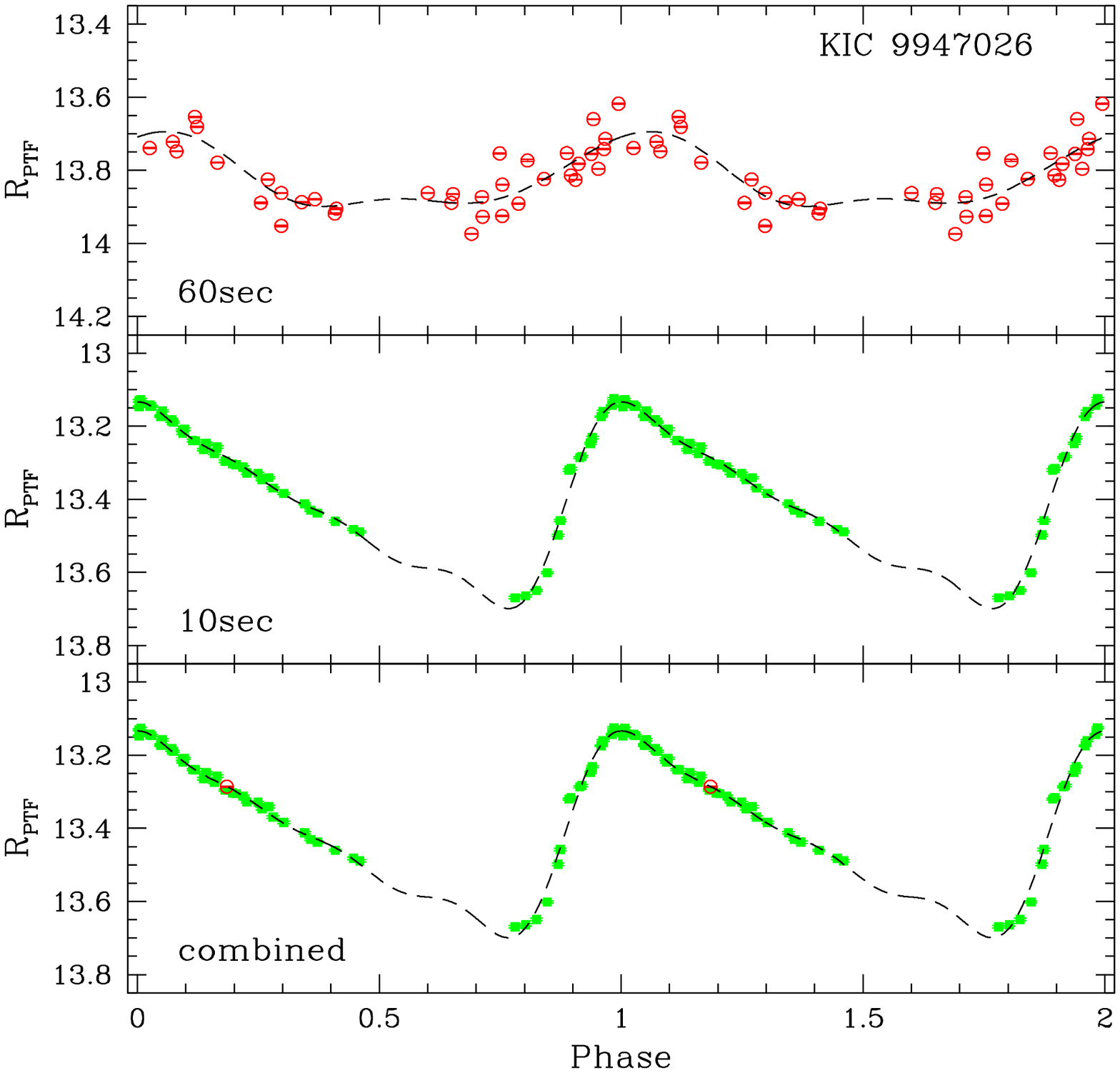} &
    \includegraphics[angle=0,scale=0.21]{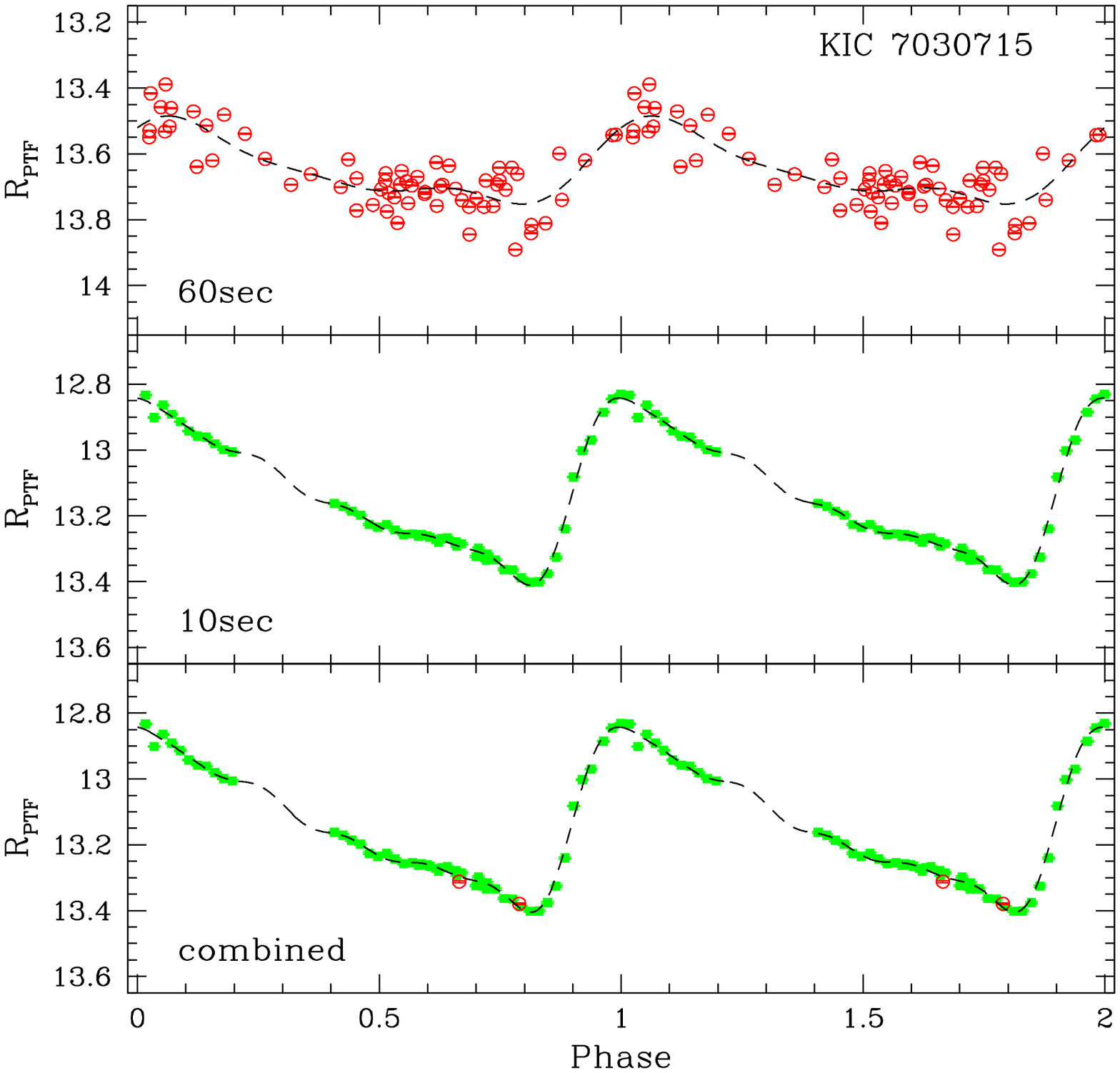} &
    \includegraphics[angle=0,scale=0.21]{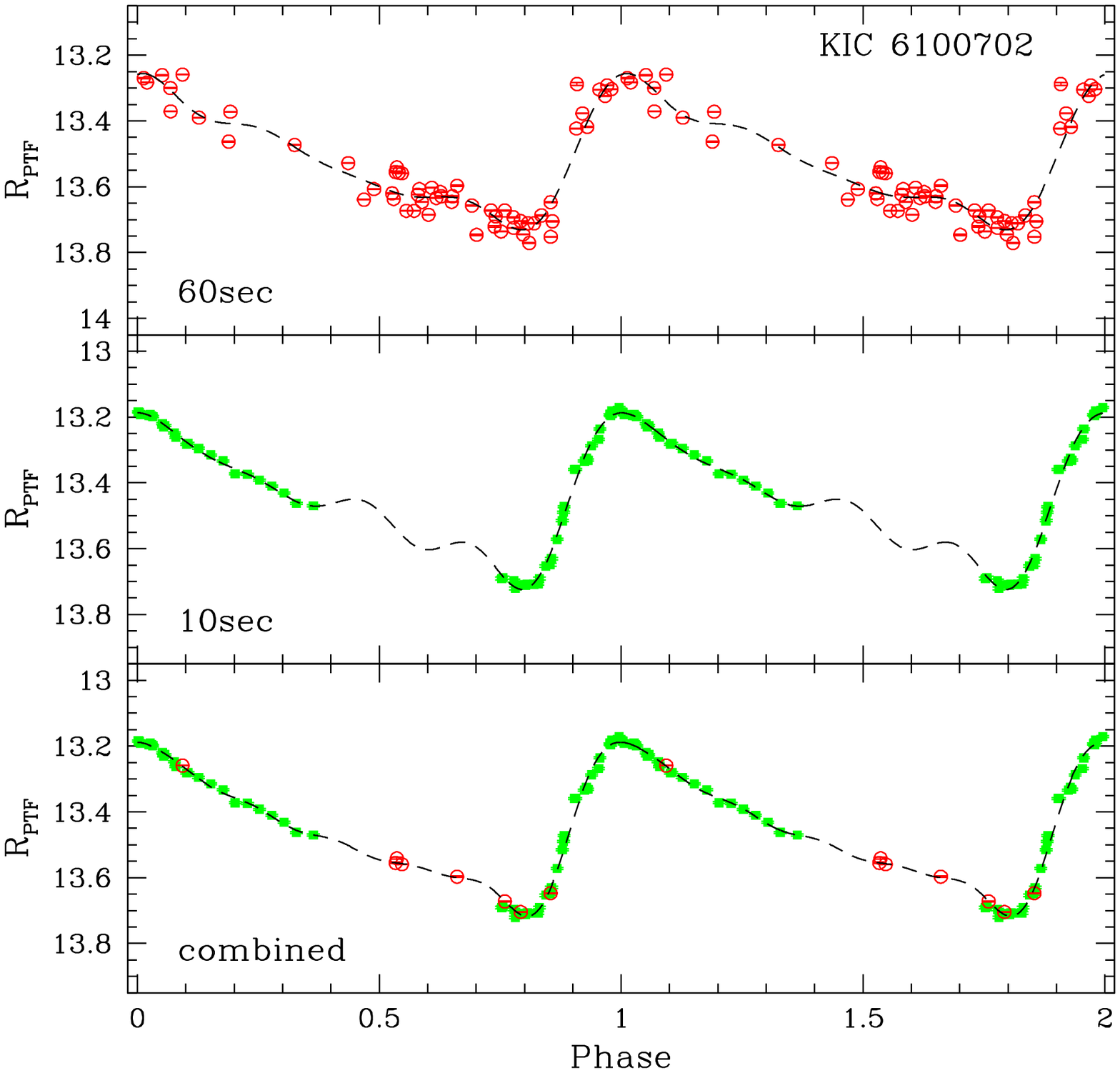} & 
    \includegraphics[angle=0,scale=0.21]{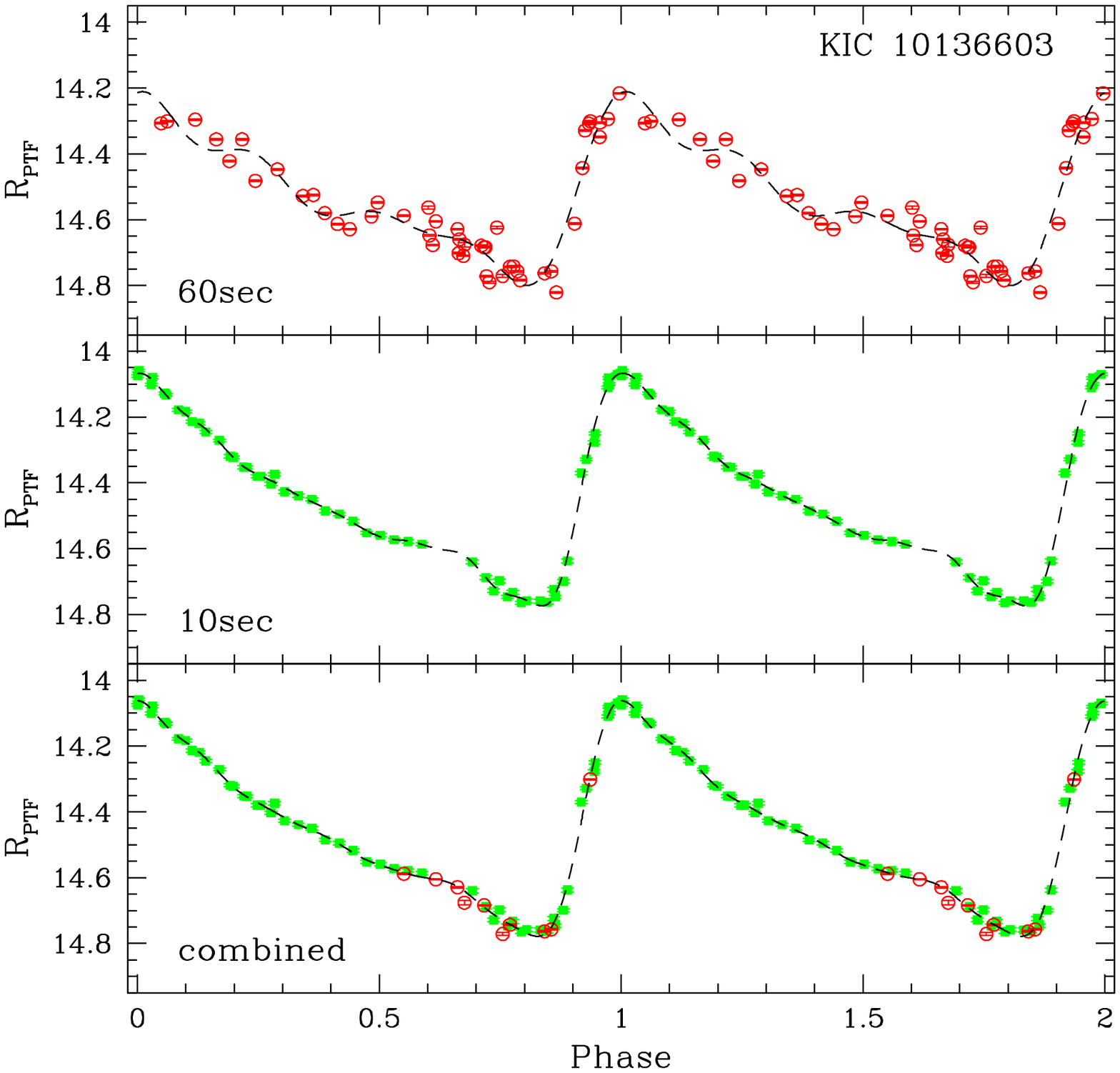} \\
    \includegraphics[angle=0,scale=0.21]{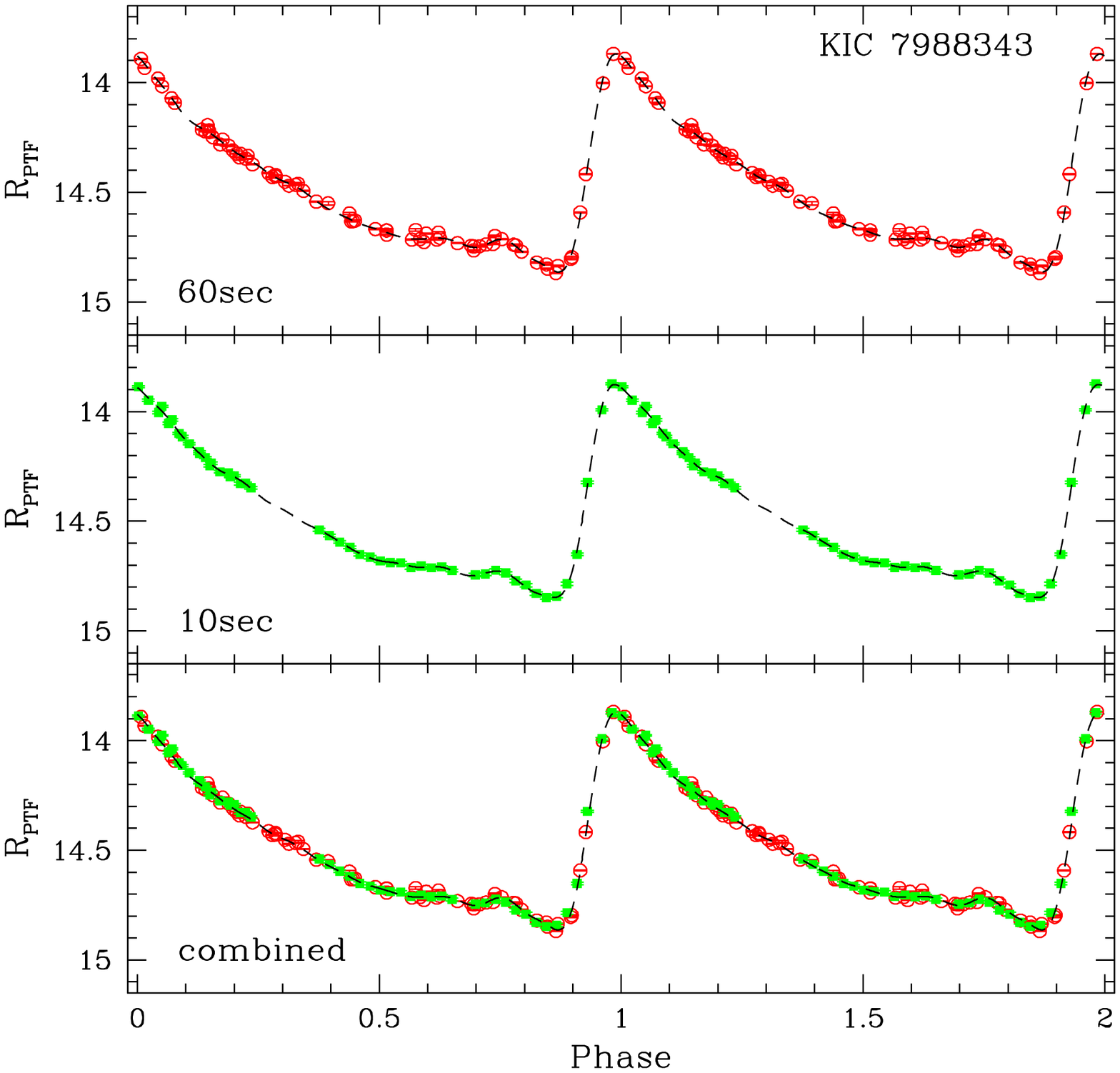} &
    \includegraphics[angle=0,scale=0.21]{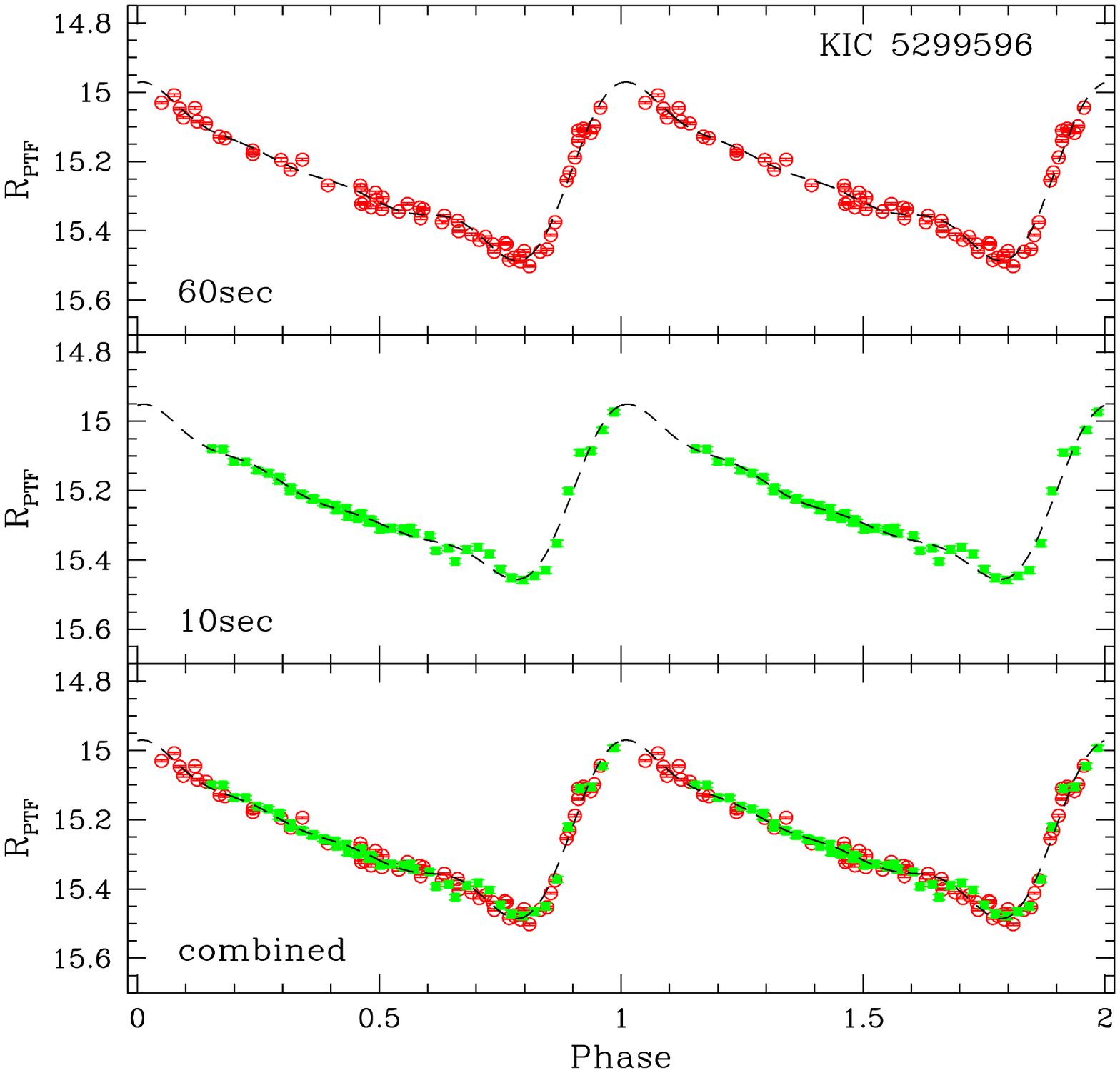} &
    \includegraphics[angle=0,scale=0.21]{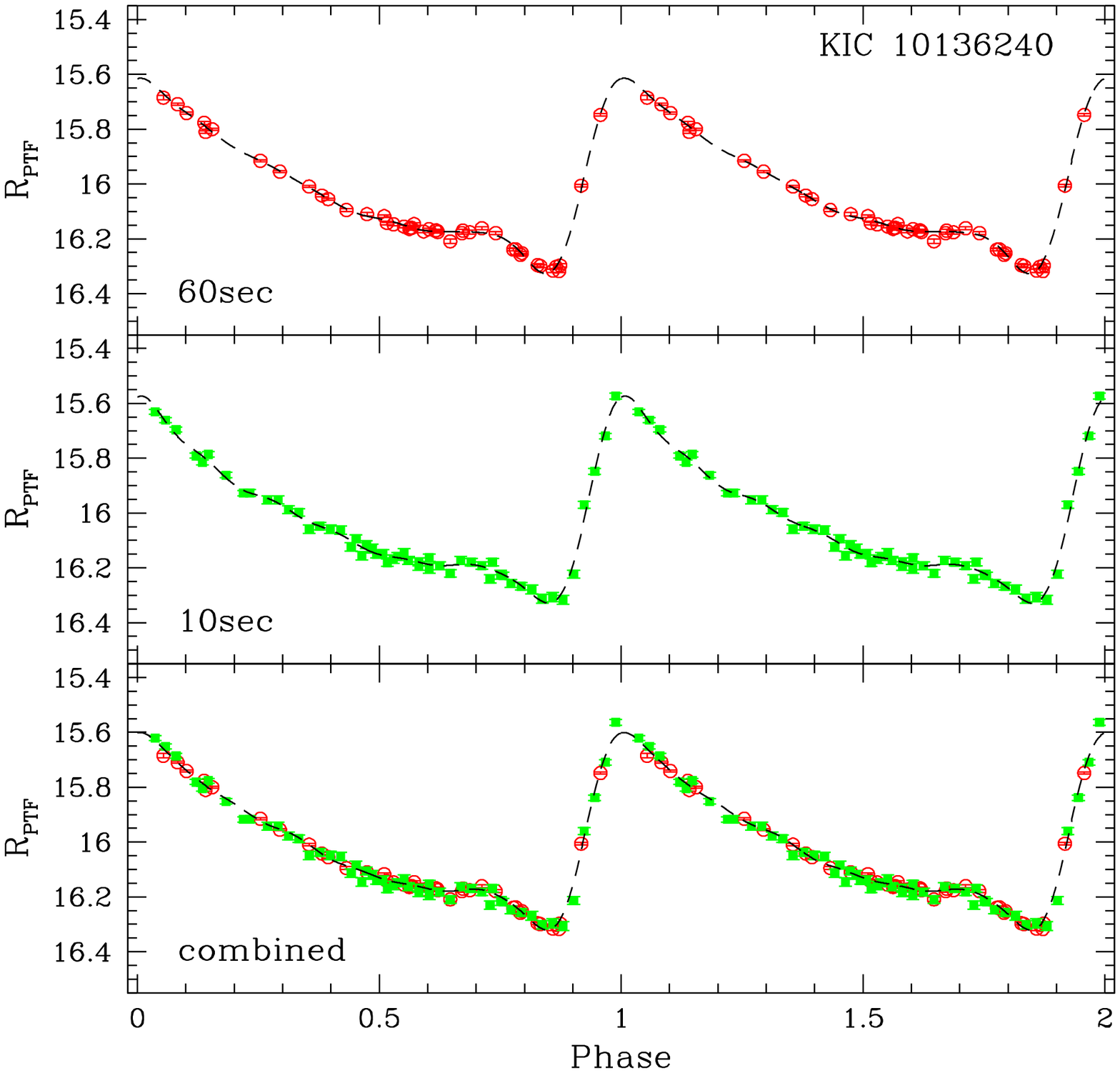} &
    \includegraphics[angle=0,scale=0.21]{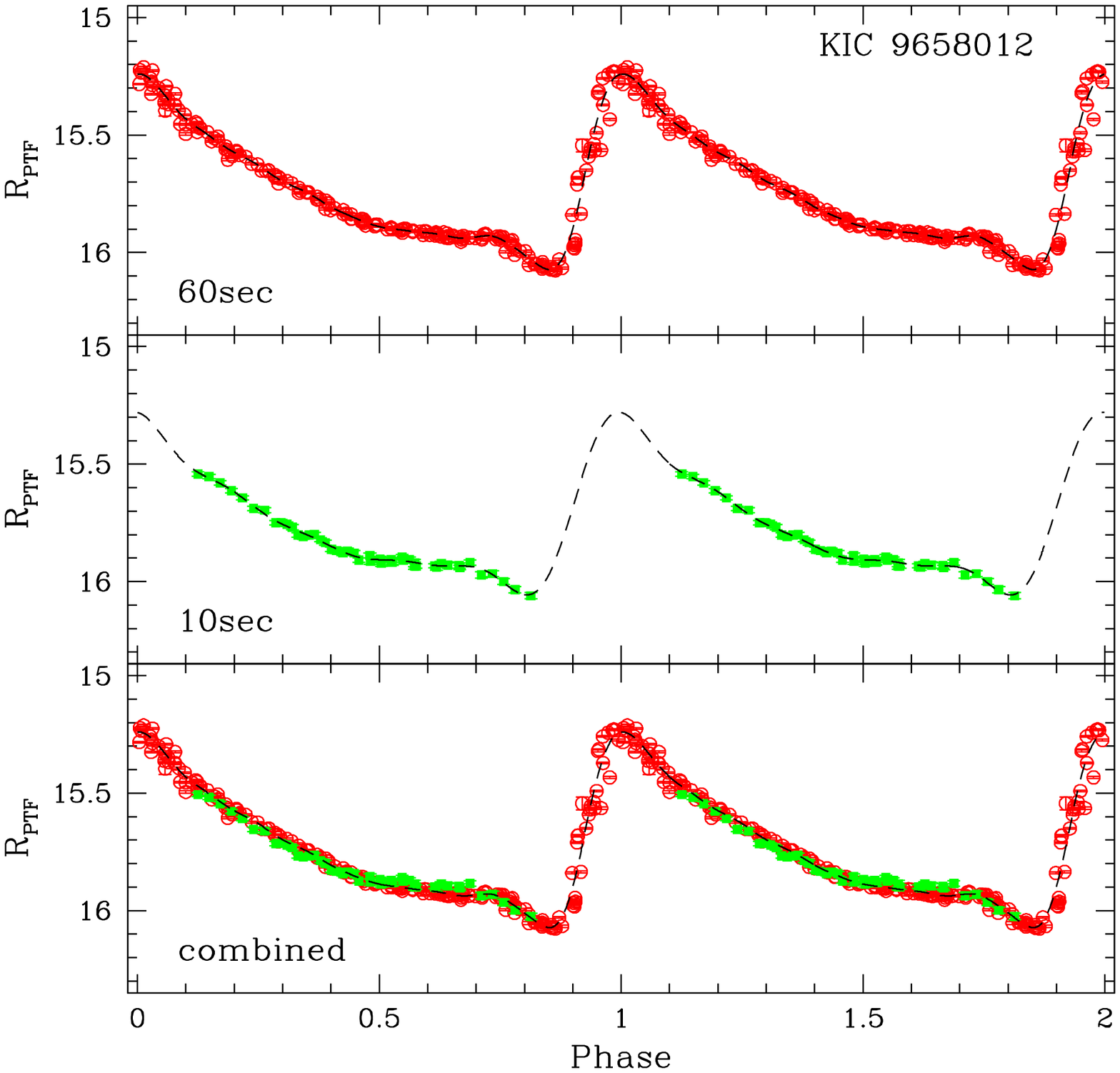} \\
    \includegraphics[angle=0,scale=0.21]{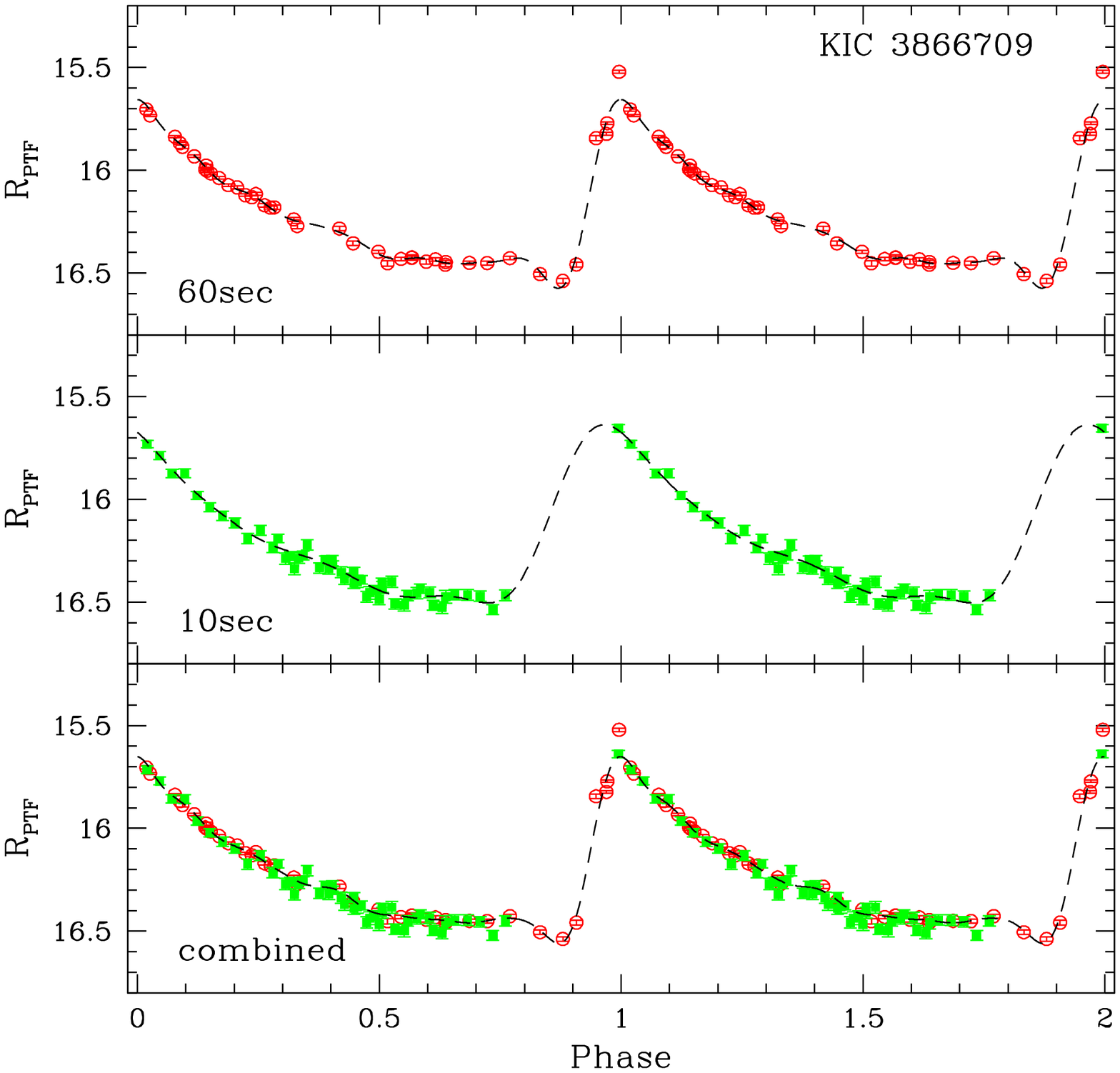} & 
    \includegraphics[angle=0,scale=0.21]{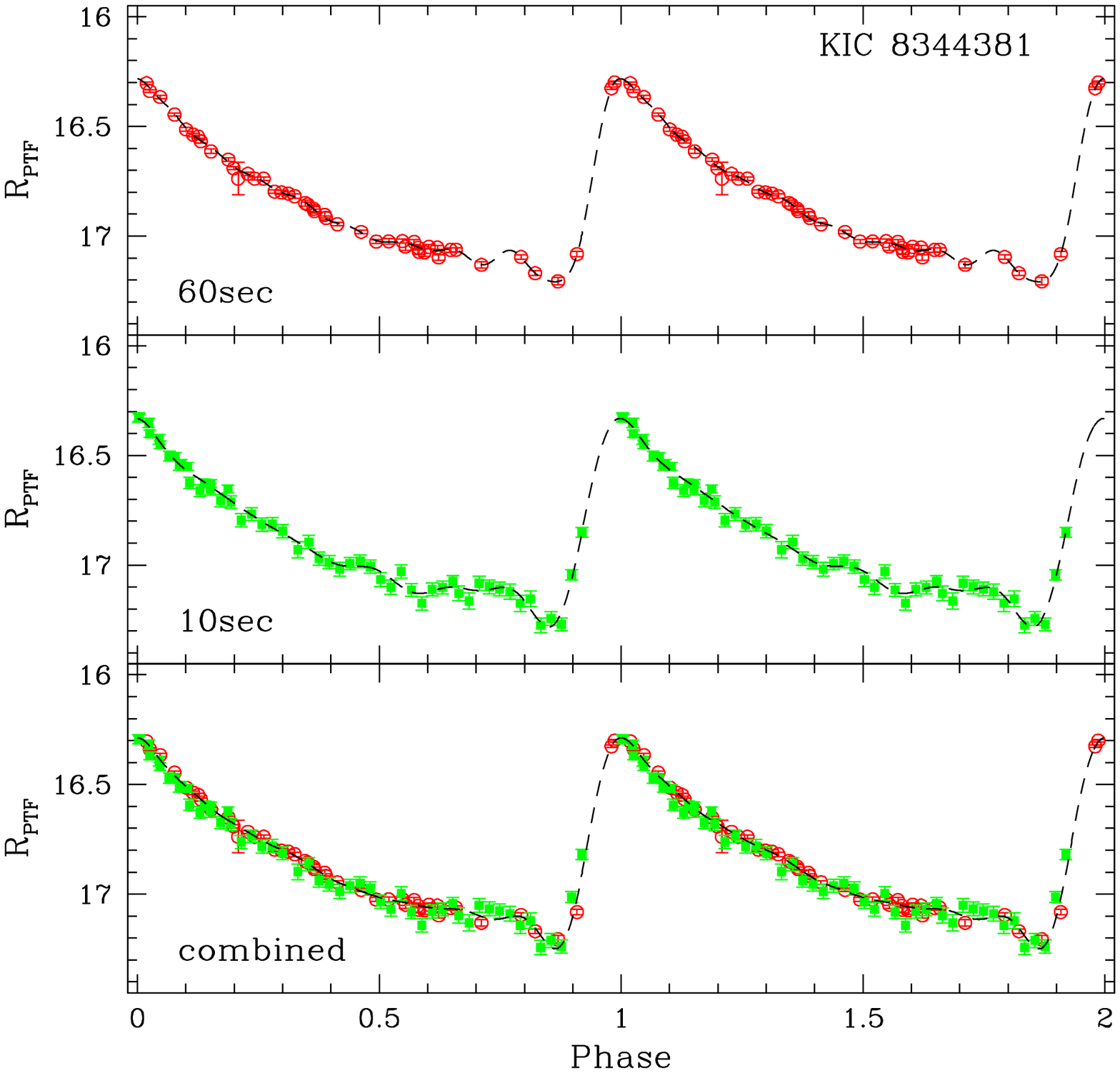} &
    \includegraphics[angle=0,scale=0.21]{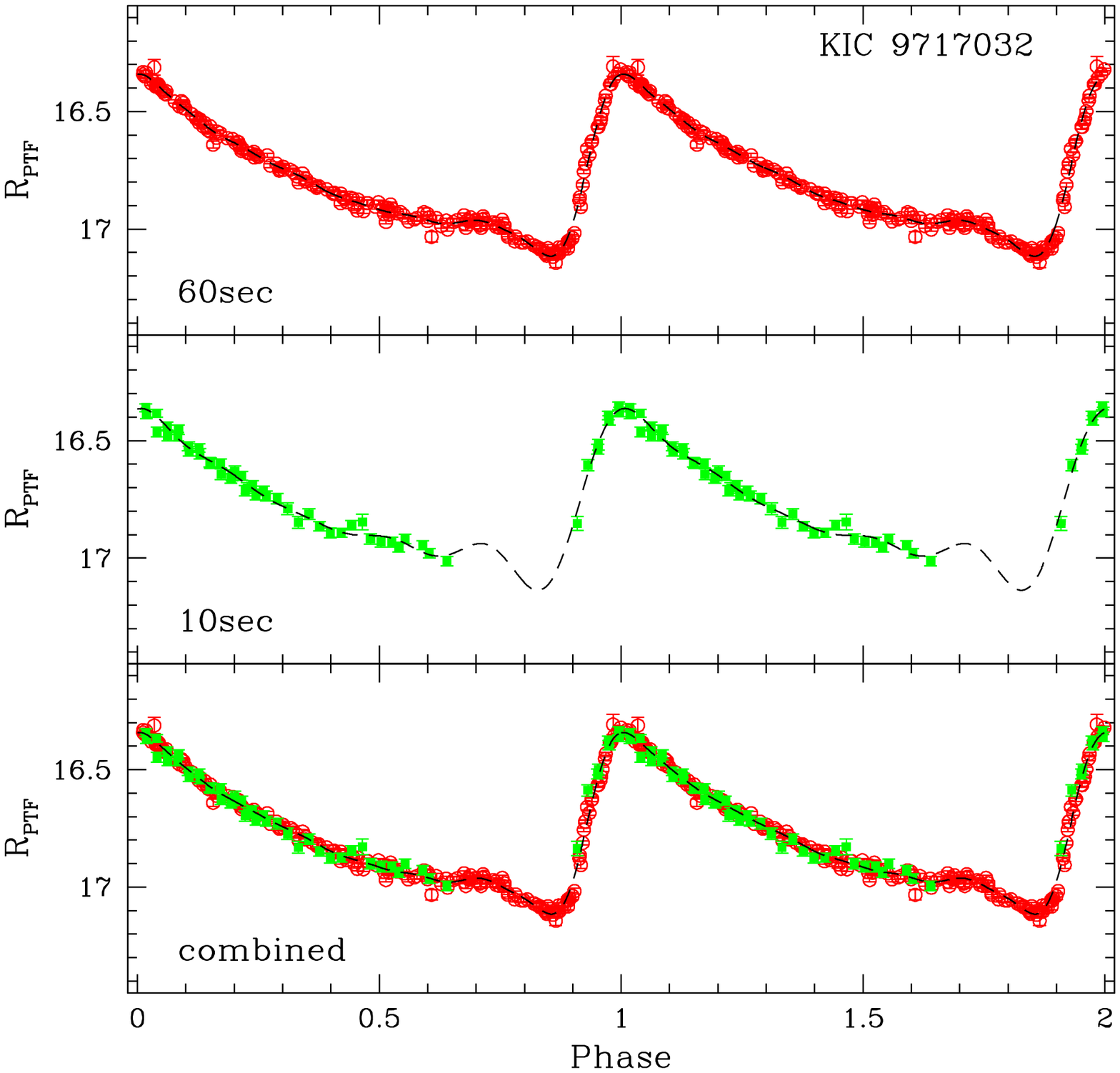} &
    \includegraphics[angle=0,scale=0.21]{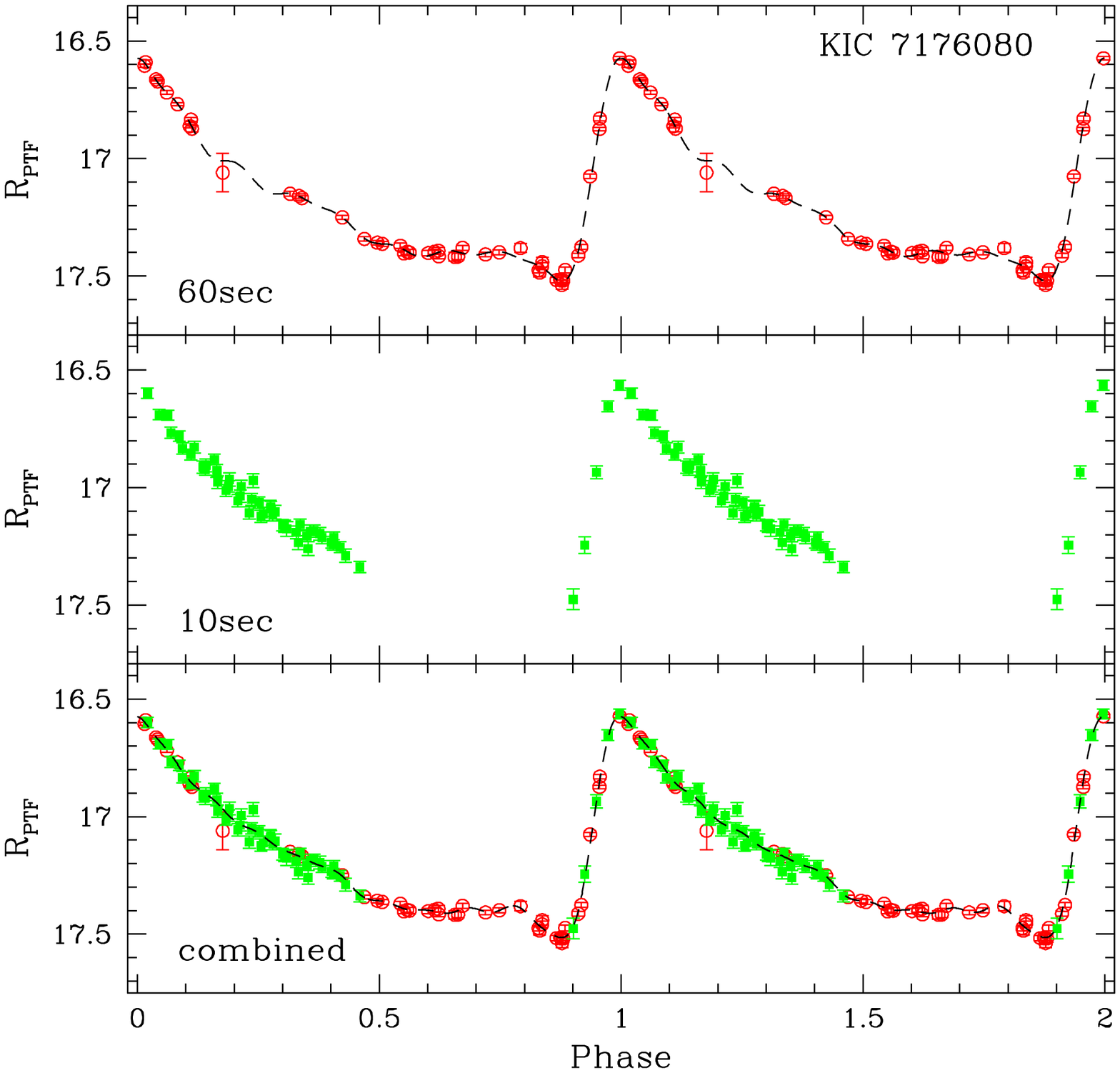} \\
  \end{array}$ 
  \caption{Differential light curves for non-Blazhko RR Lyrae stars in our sample after applying the differential photometry technique as described in the text. Red (opened) circles and green (filled) squares are for data with 60~s exposure time (from regular PTF/iPTF surveys, top panels in each sub-figure) and 10~s exposure time (from the dedicated iPTF experiment, middle panels in each sub-figures), respectively. Bottom panels in the sub-figures are the combined light curves (see the text for further details). The dashed curves are the fitted light curves using Fourier expansion as given in Equation (1).}
  \label{fig_nonblazhko_lc}
\end{figure*}

Differential light curves for the remaining 19 non-Blazhko RR Lyrae stars were fitted with the truncated Fourier decomposition as given in Equation (1). We fit the 60~s and the 10~s light curves separately, as well as to the combined light curves. The best-fit orders $n$ of the Fourier decomposition were chosen based on visual inspection of the fitted light curves. For the majority of the 10~s light curves, a relatively large gap was seen in the phased light curves, which will affected the fitted light curves when applying the Fourier decomposition technique. To remedy this, we added a data point near the mid-point of the phased gap either taken from the 60-second light curves (for faint RR Lyrae) or derived from a cubic spline interpolation function (for bright RR Lyrae). Note that we only applied this additional data point to the 10~s light curves and not to the 60~s light curves. The only exception is the 10~s light curve for KIC 7176080, at which the phased gap is too large to apply a meaningful Fourier fit.

\begin{figure}
  %\epsscale{1.17}
  \plotone{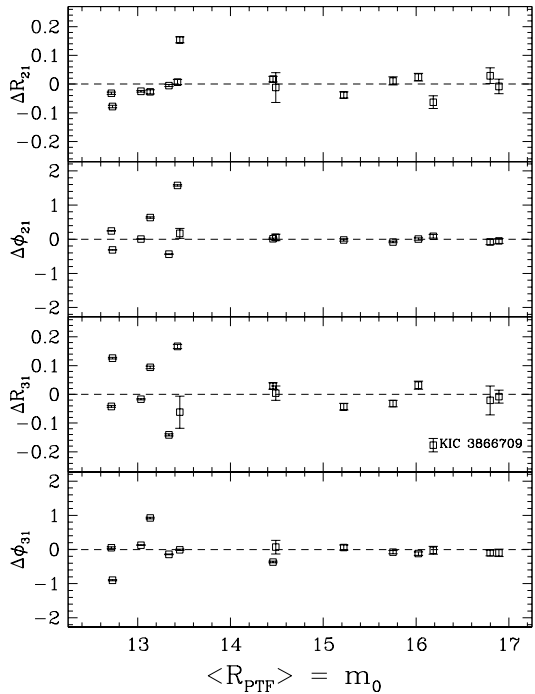}
  \caption{Difference, $\Delta$, of the lower order Fourier parameters between the 60~s and 10~s light curves, where the Fourier parameters were calculated via Equation (2). The horizontal dashed lines indicate the cases of $\Delta=0$, and not the fit to the data. The mean magnitude $m_0$ is calculated from the Fourier decomposition as given in Equation (1). The ``outlier'' in the $\Delta R_{31}$ plot is KIC 3866709, at which the fitting of Fourier decomposition to the 10~s light curves was affected by the phased gap at the ascendant branch.} 
  \label{fig_10vs60}
\end{figure}

Among the 19 non-Blazhko RR Lyrae stars, three RR Lyrae do not have 10~s light curves and their 60~s light curves are displayed in Figure \ref{fig_3nonBL}. The 60~s and 10~s phased light curves for the rest of the 16 non-Blazhko RR Lyrae were shown in the top and middle panels for each of the sub-Figures in Figure \ref{fig_nonblazhko_lc}. The dashed curves displayed in Figure \ref{fig_3nonBL} and \ref{fig_nonblazhko_lc} are the fitted light curves based on the Fourier decomposition technique. We compared the low-order Fourier parameters, $R_{21}$, $R_{31}$, $\phi_{21}$, and $\phi_{31}$, calculated with Equation (2), between the 60~s and 10~s light curves. The difference of the Fourier parameters as a function of mean $R_{PTF}$-band magnitudes is given in Figure \ref{fig_10vs60}. As can be seen from this figure, light curves with mean magnitudes fainter than $\sim14$~mag show a smaller dispersion of the difference in Fourier parameters than those with mean magnitudes brighter than $\sim14$~mag. Therefore, we merged the 60~s and 10~s light curves for the eight faint non-Blazhko RR Lyrae. For other eight bright non-Blazhko RR Lyrae, we added few data points from the 60-second light curves that have $FLAG=0$ in the {\tt SExtractor} catalogs, as these data points follow the light curve shapes defined by the 10~s light curves. We referred the merged 60~s and 10~s light curves as the combined light curves, as shown in bottom panels in each sub-figure of Figure \ref{fig_nonblazhko_lc}. The final adopted $\phi_{31}$ Fourier parameters, based on these combined light curves, for the non-Blazhko RRab stars are listed in Table \ref{tab3}.

\subsection{For Blazhko RRab Stars}

\begin{figure*}
  $\begin{array}{cccc}
    \includegraphics[angle=0,scale=0.21]{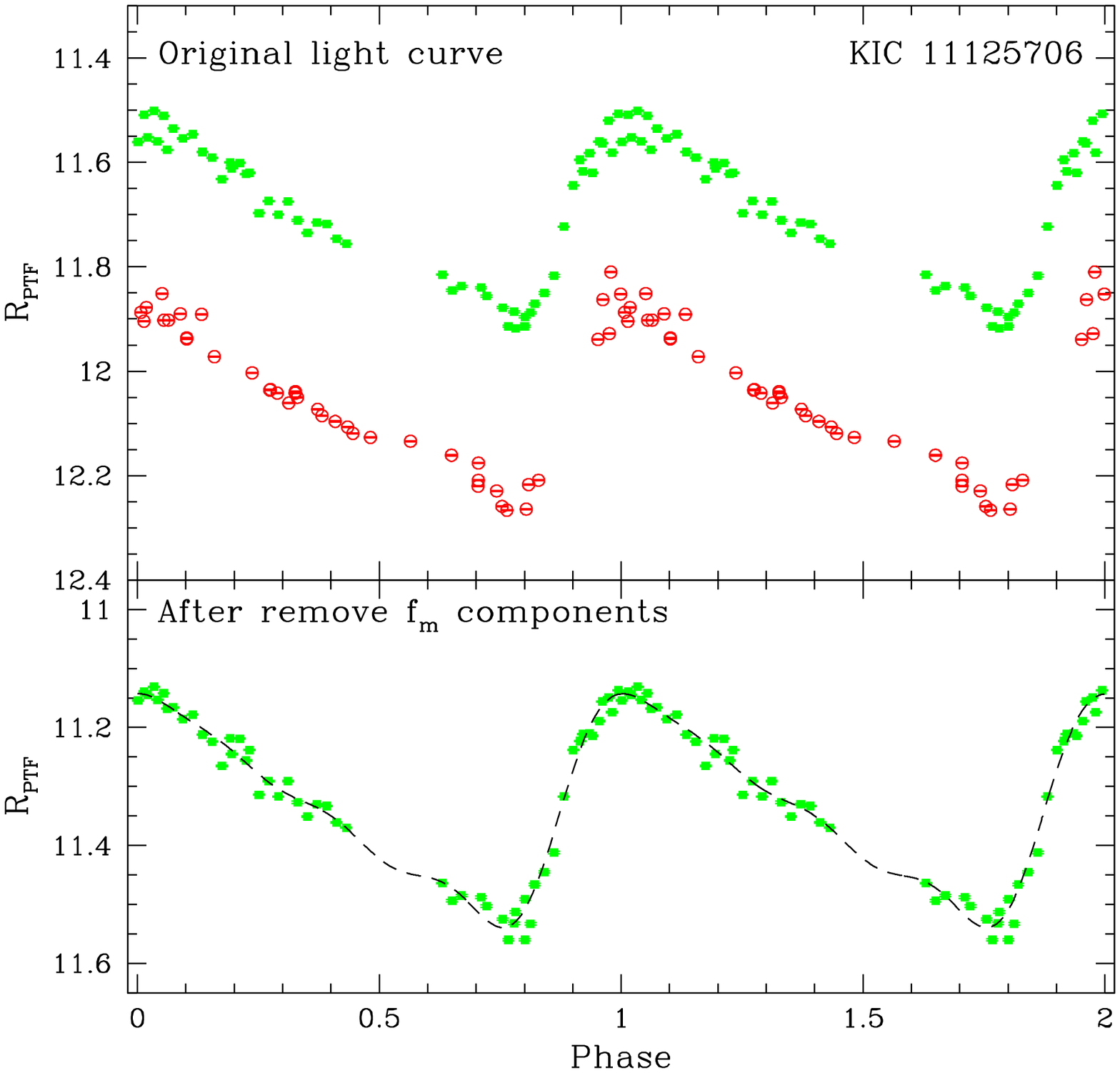} &
    \includegraphics[angle=0,scale=0.21]{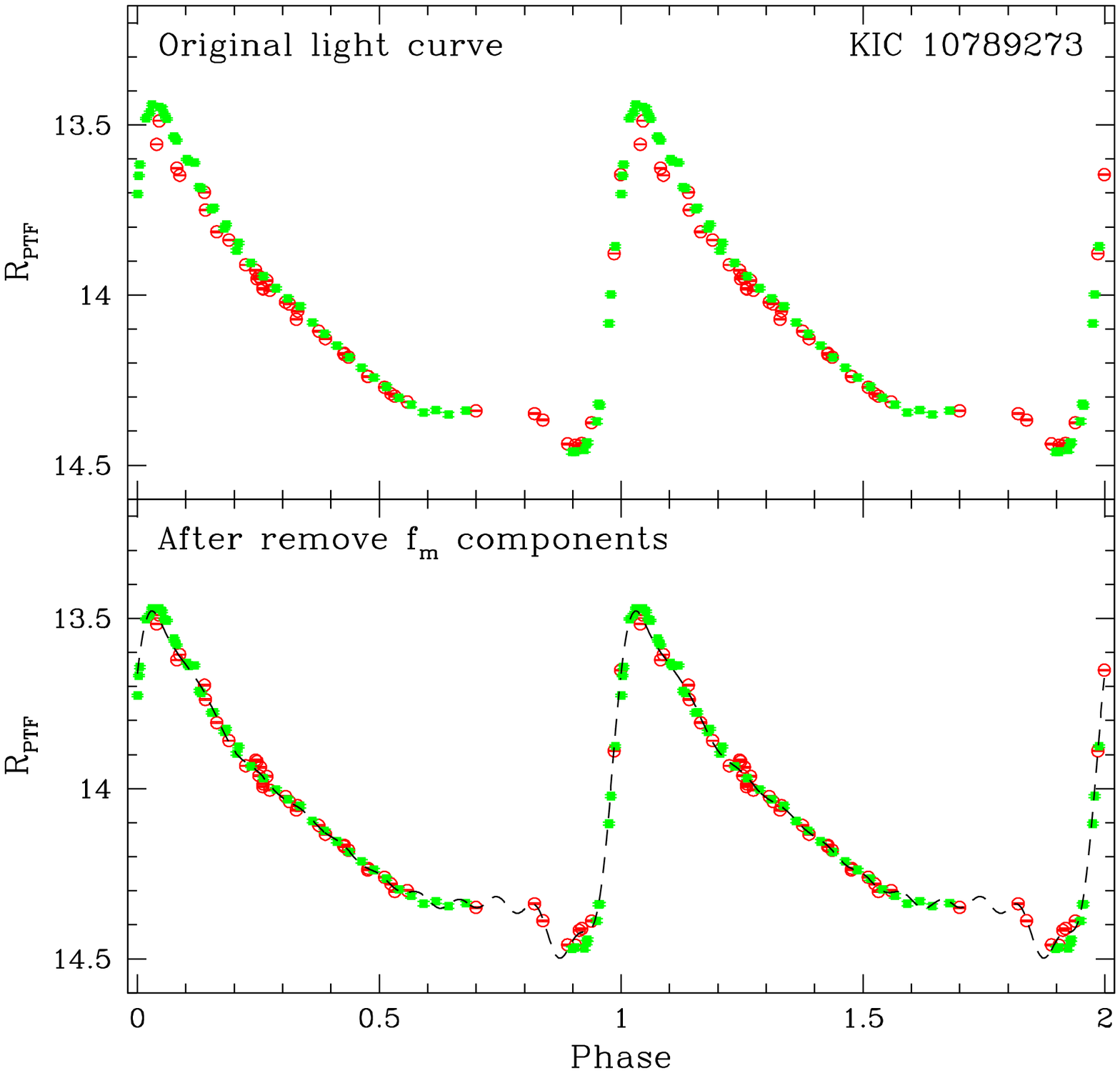} &
    \includegraphics[angle=0,scale=0.21]{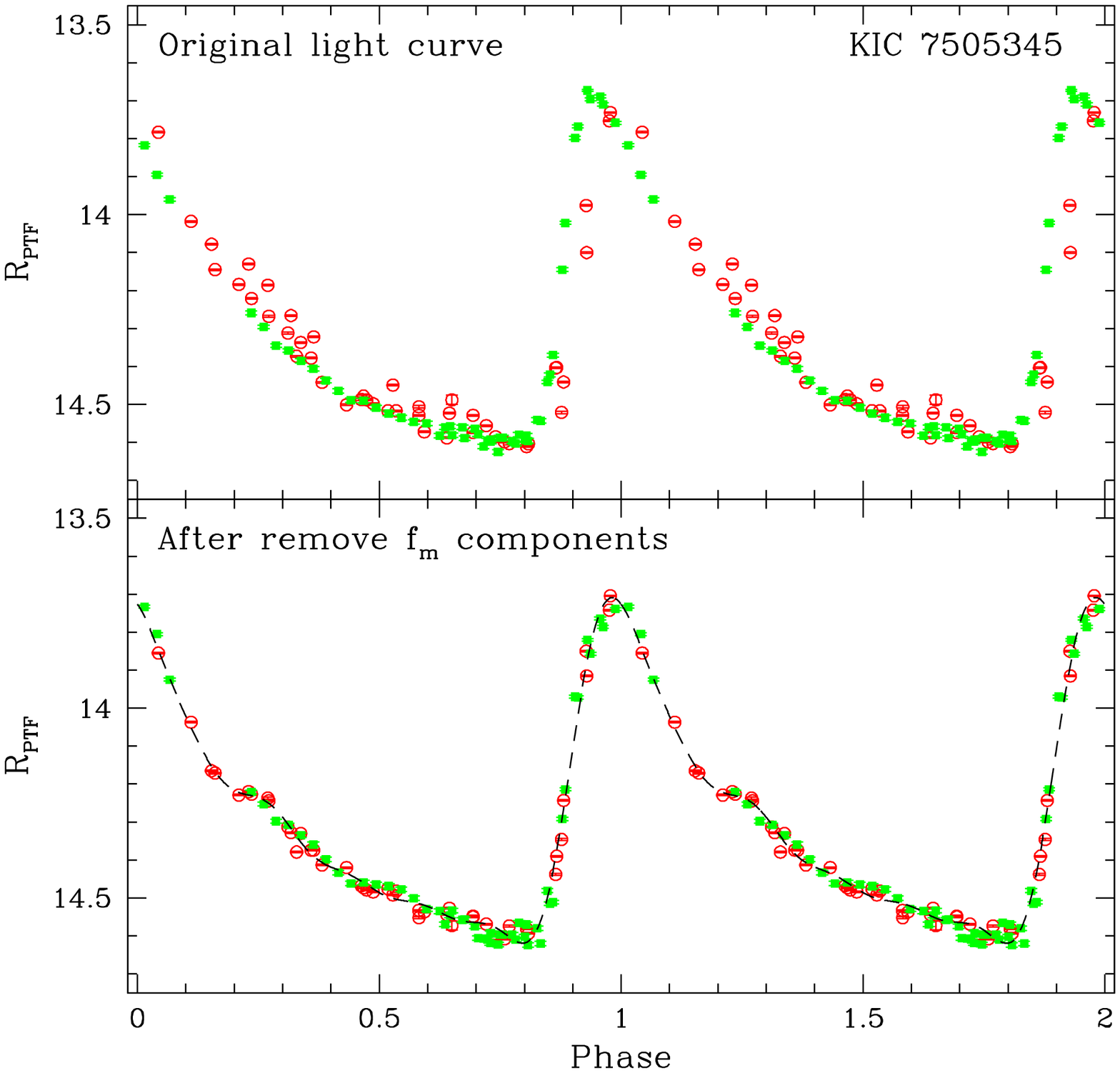} &
    \includegraphics[angle=0,scale=0.21]{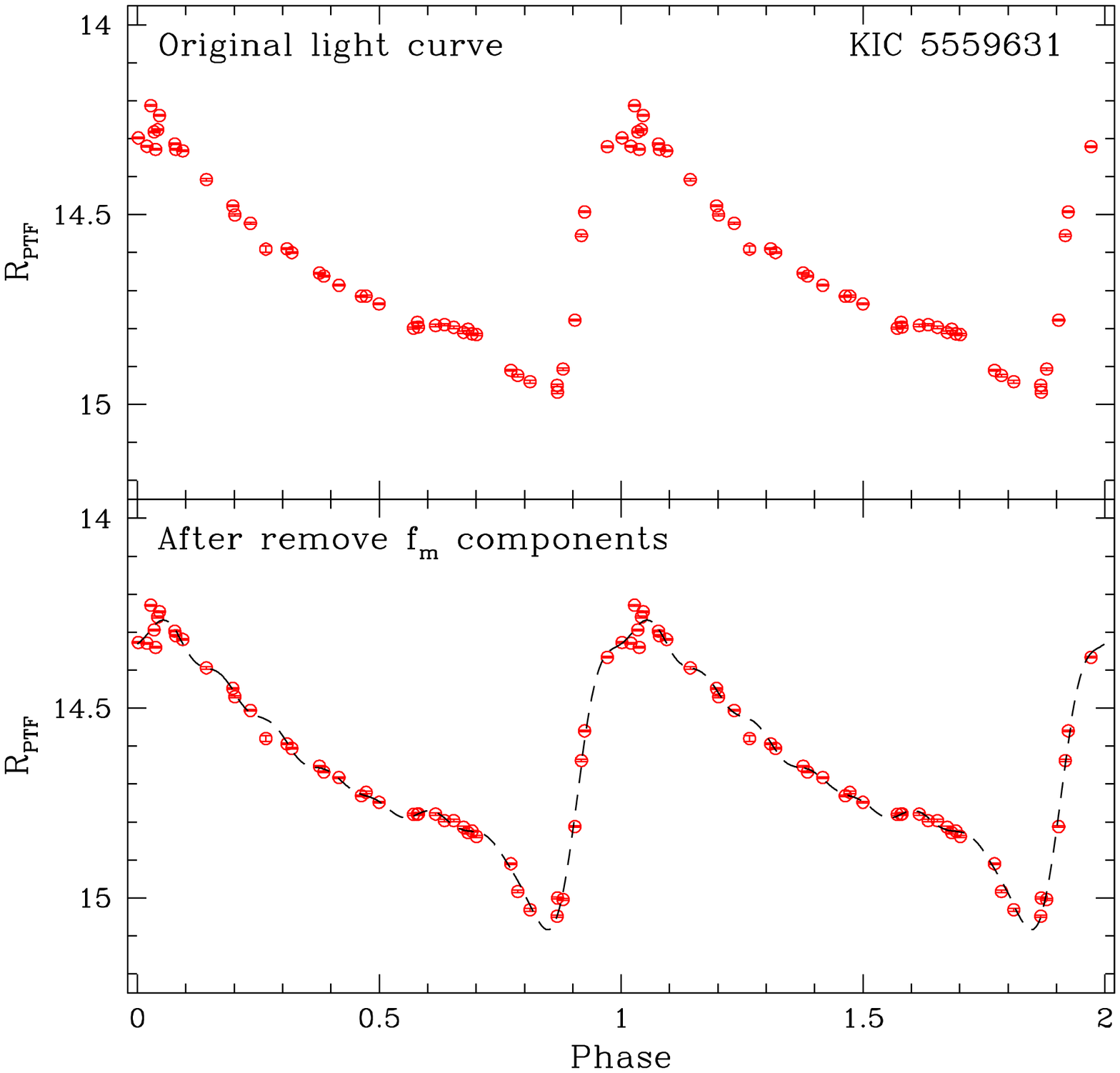} \\
    \includegraphics[angle=0,scale=0.21]{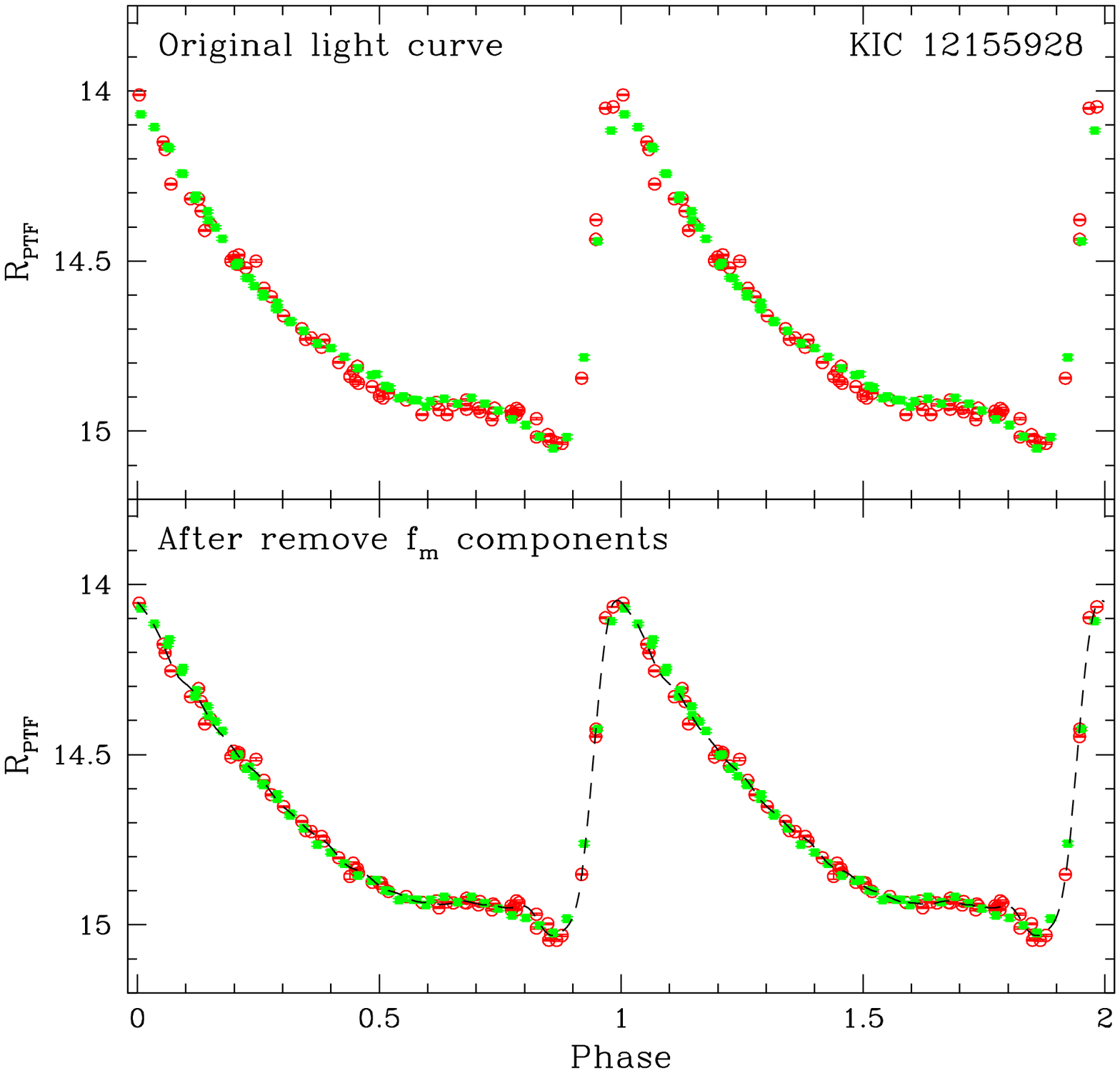} &
    \includegraphics[angle=0,scale=0.21]{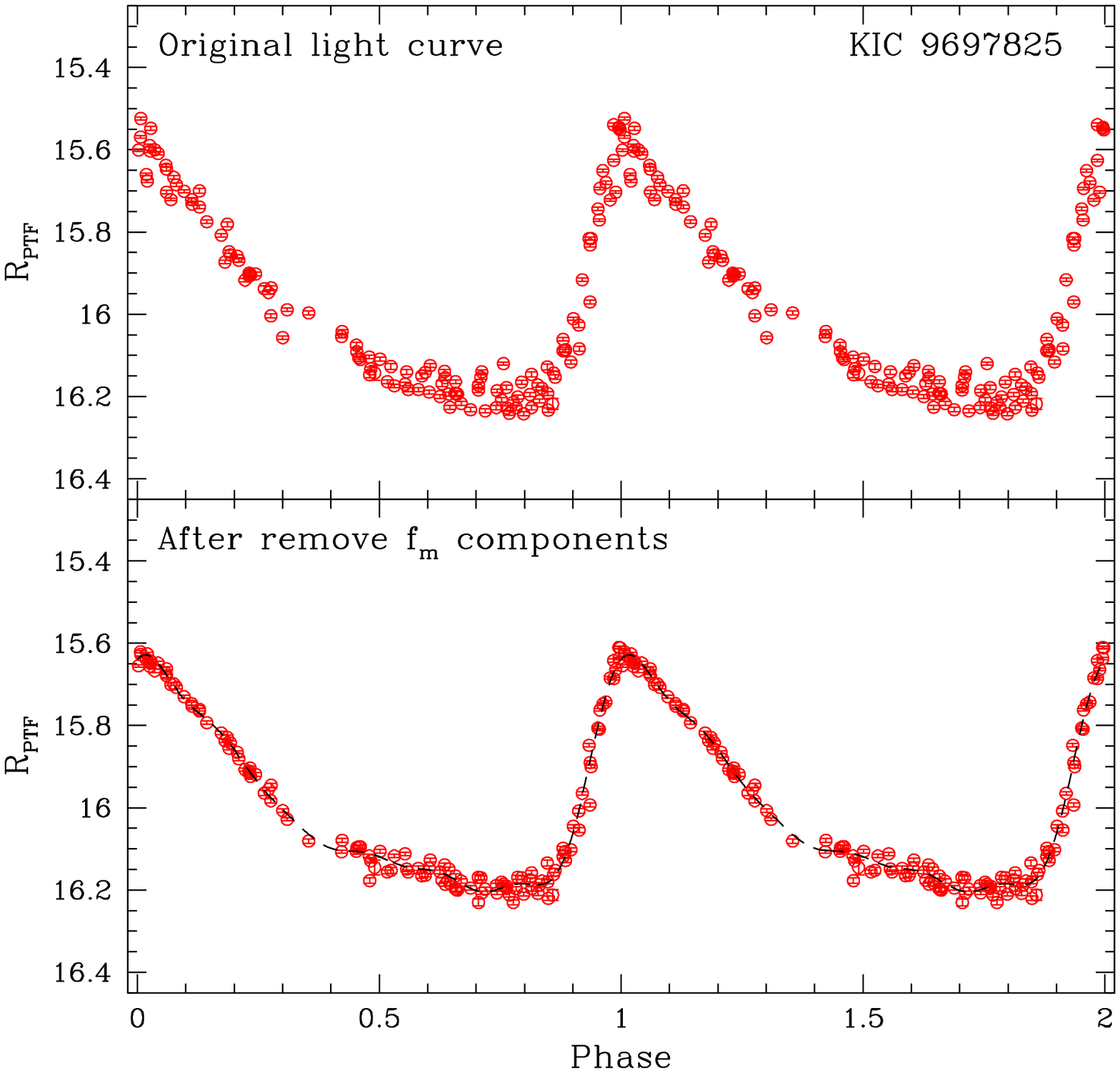} &
    \includegraphics[angle=0,scale=0.21]{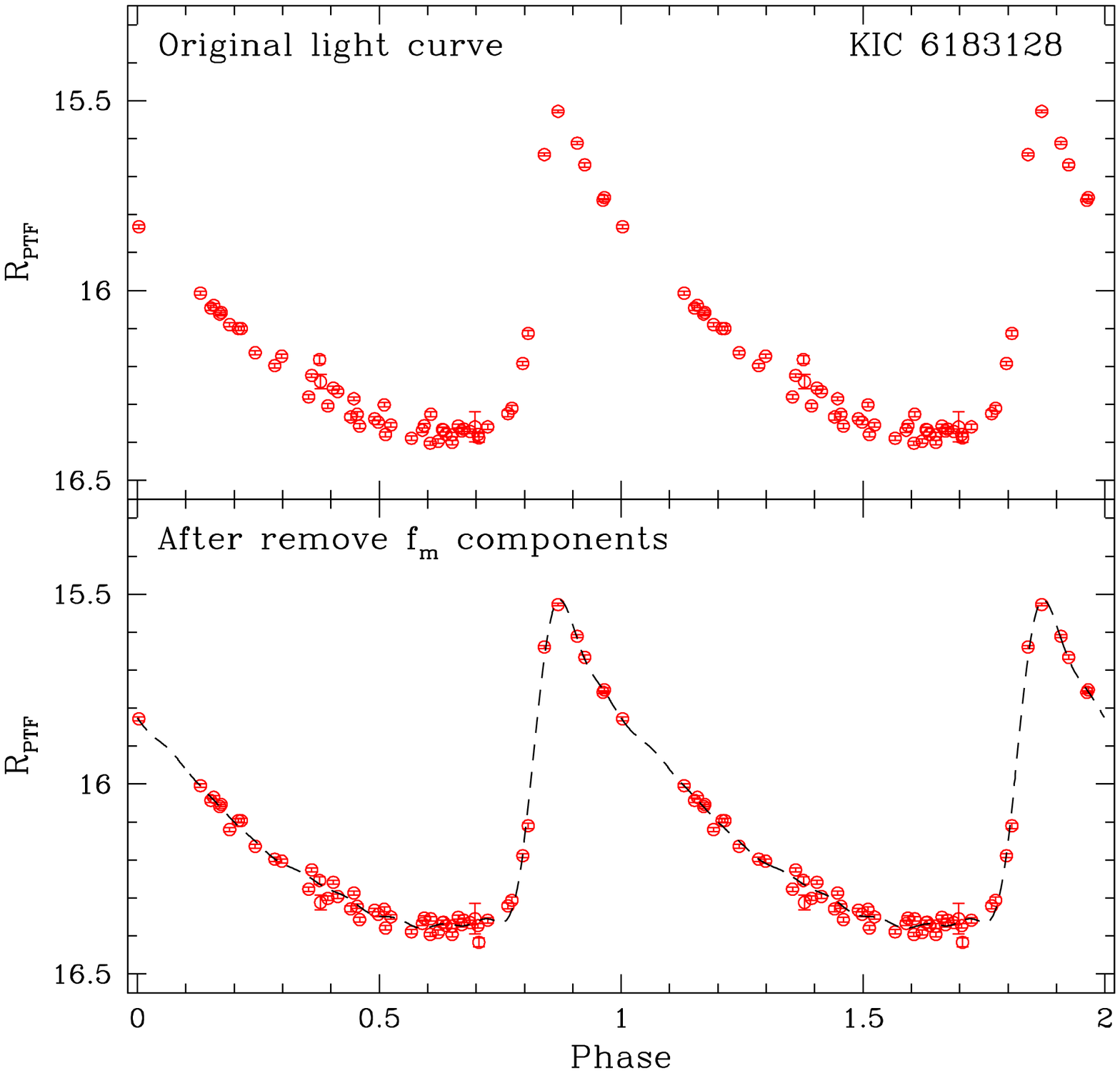} &
    \includegraphics[angle=0,scale=0.21]{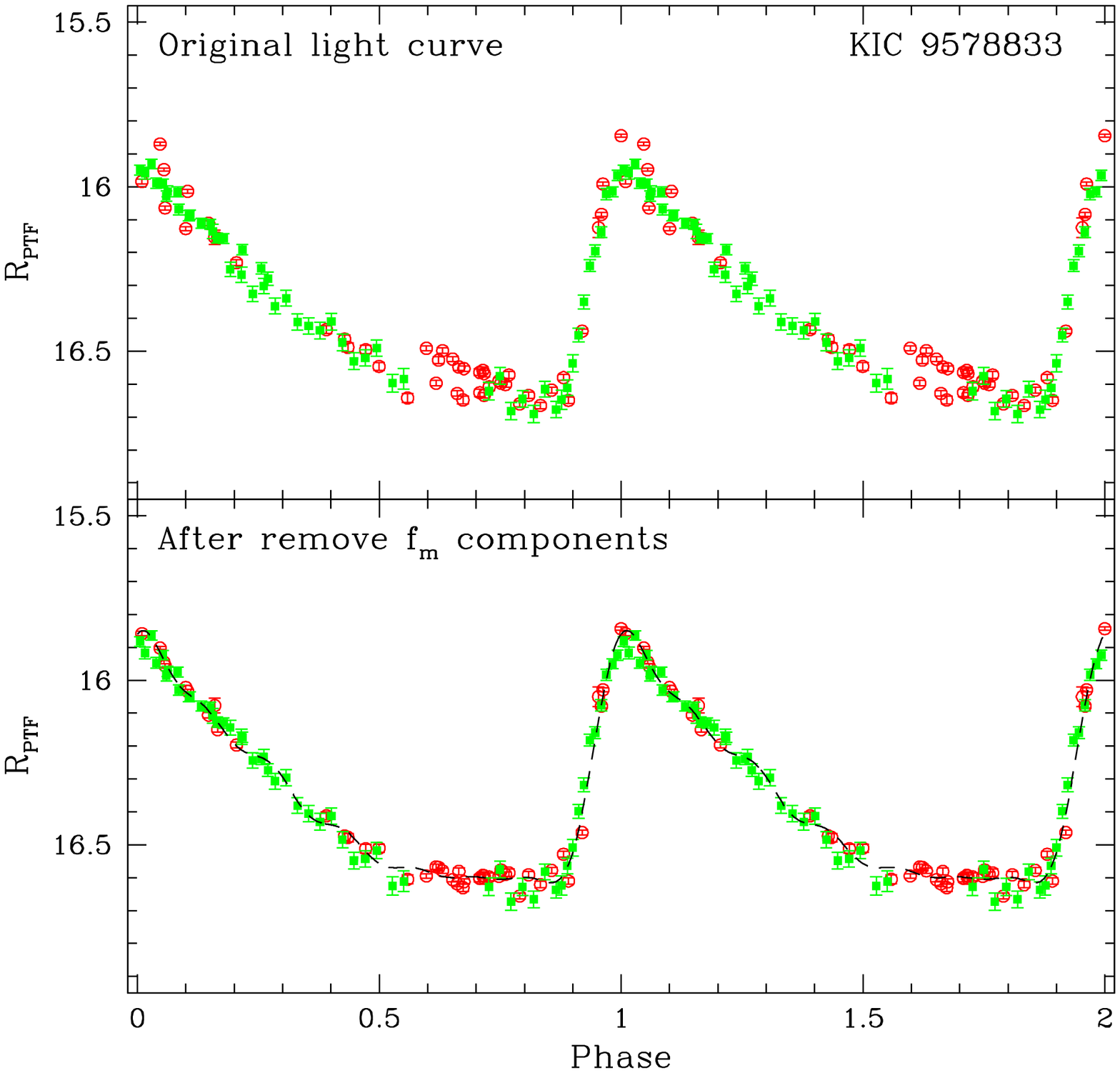} \\
    \includegraphics[angle=0,scale=0.21]{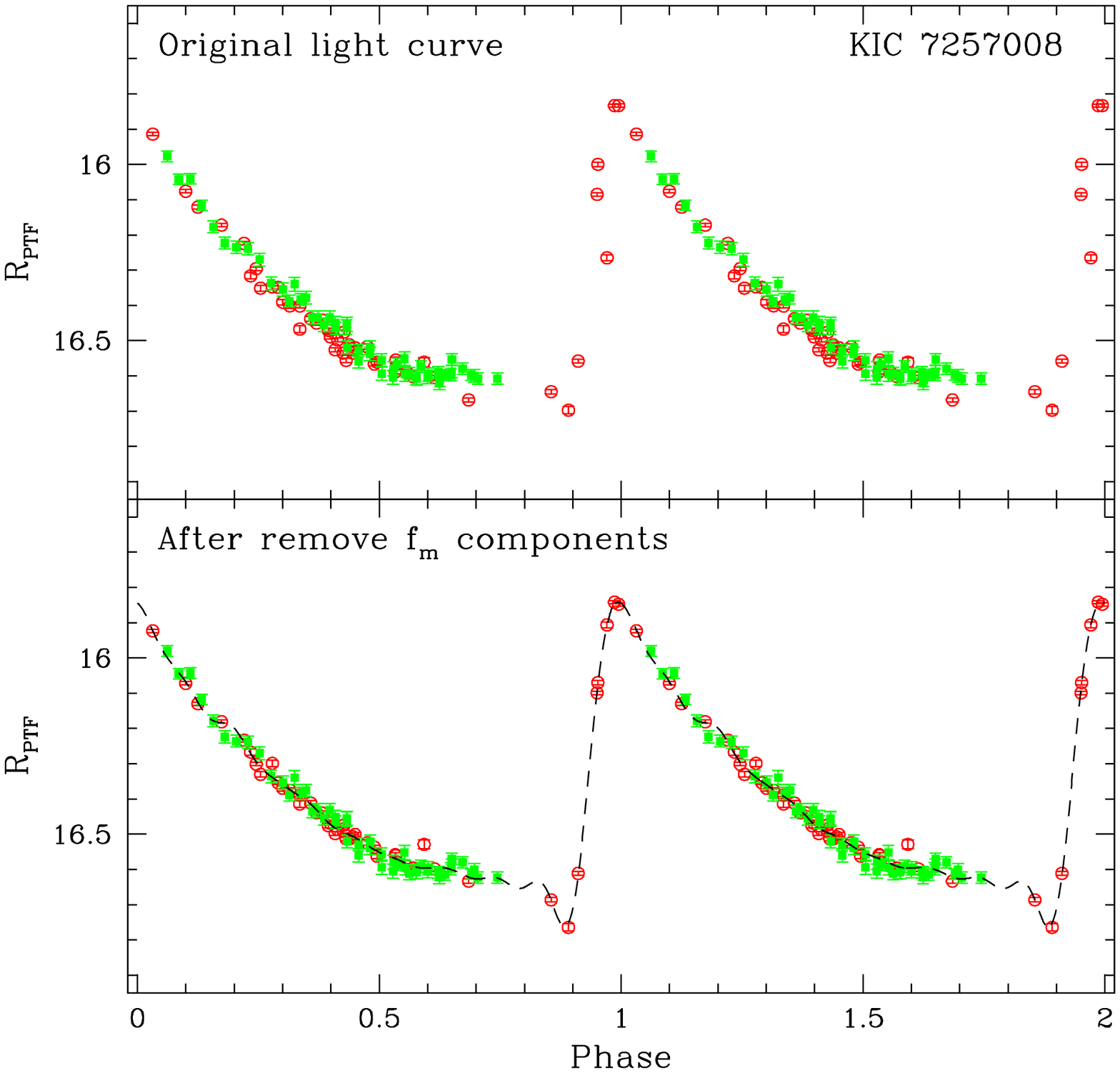} & 
    \includegraphics[angle=0,scale=0.21]{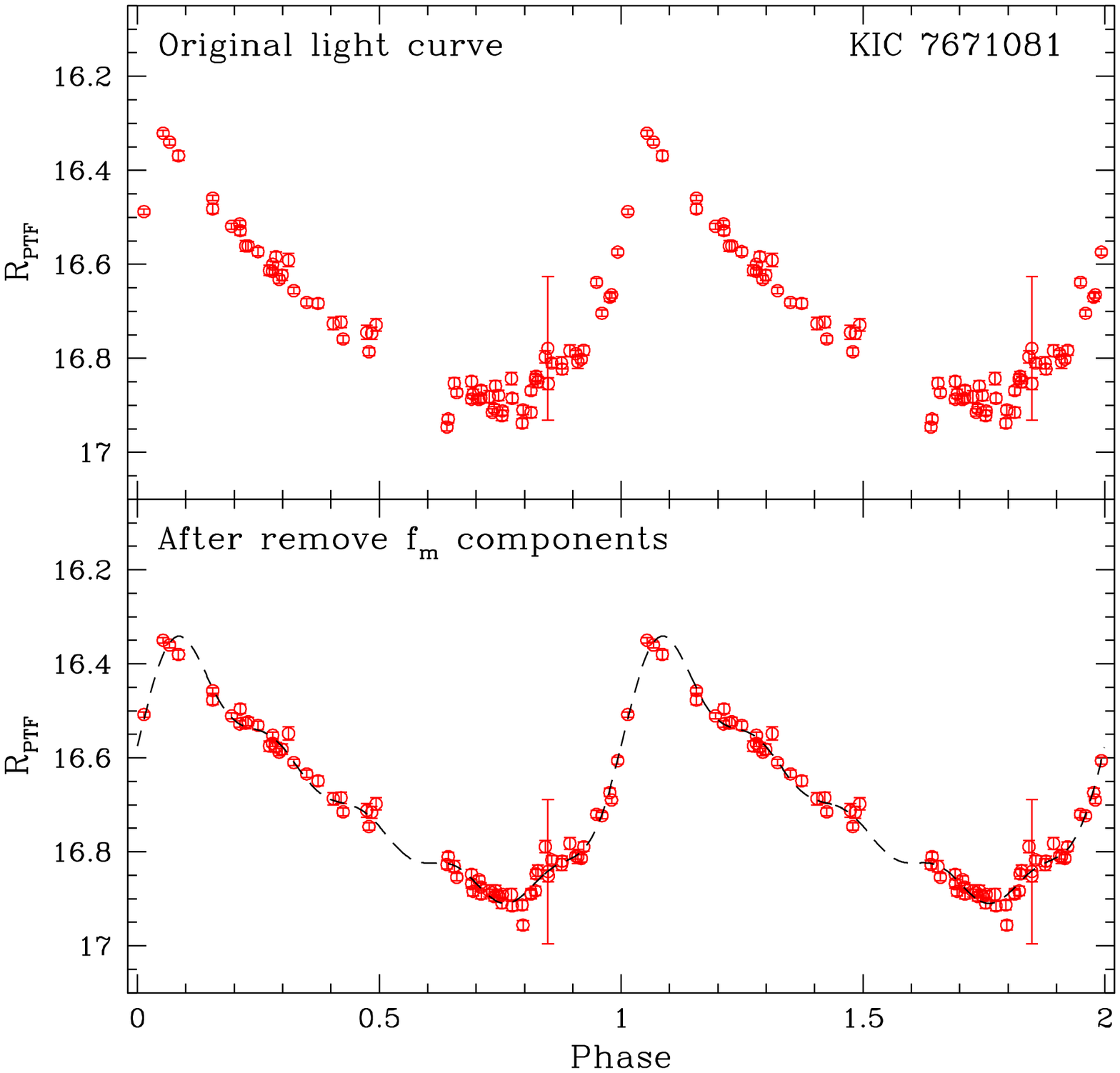} &
    \includegraphics[angle=0,scale=0.21]{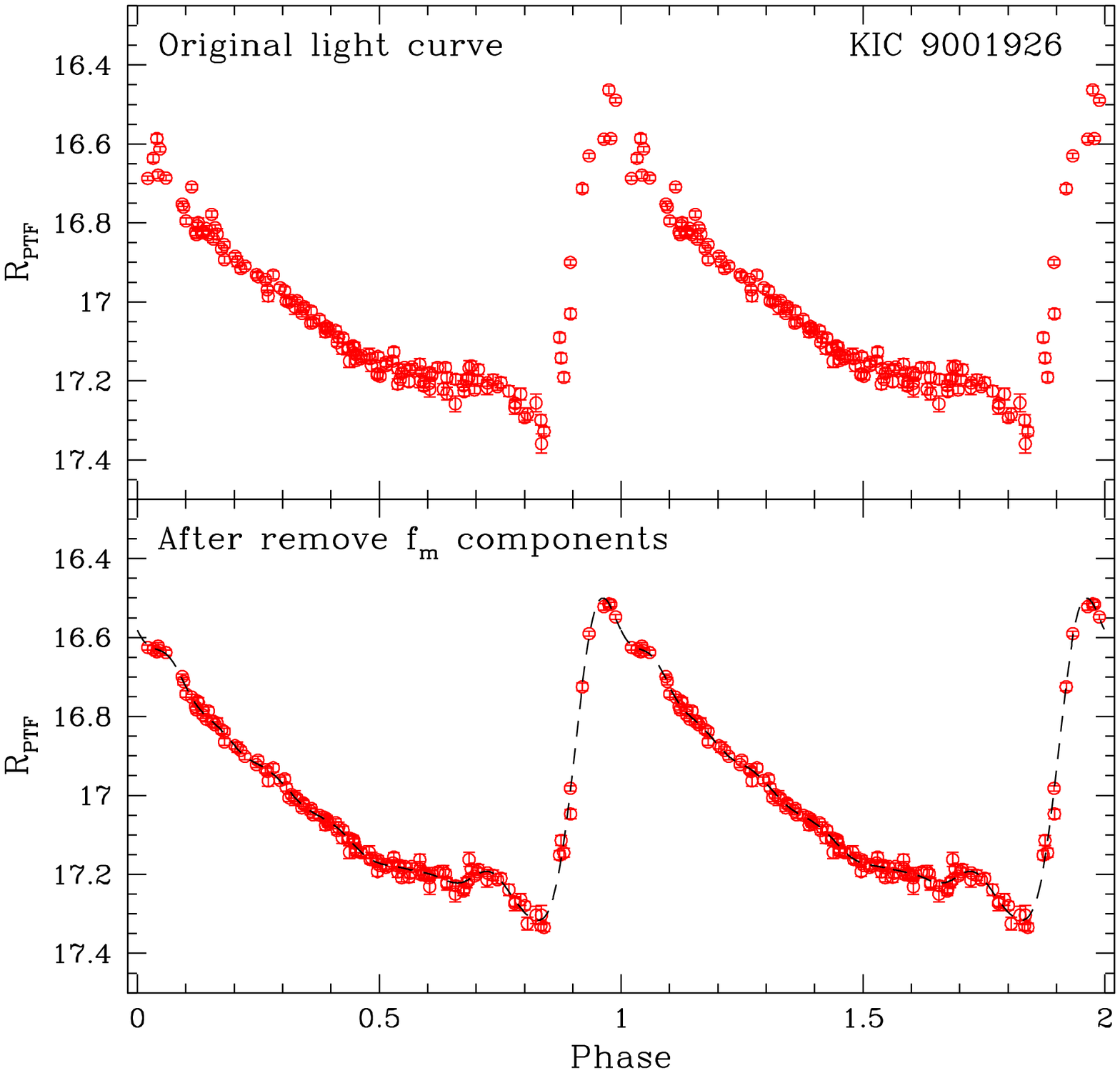} &
    \includegraphics[angle=0,scale=0.21]{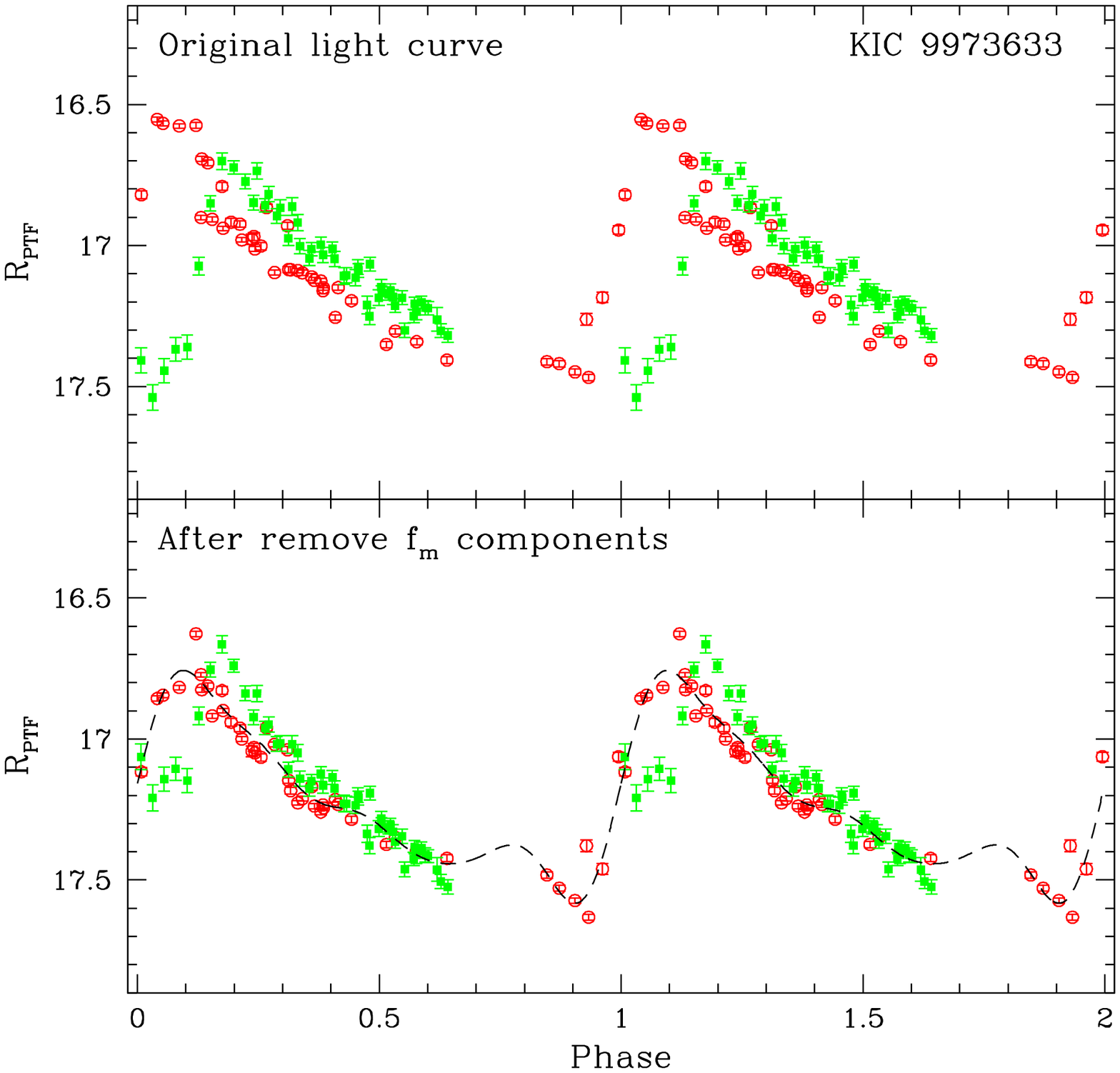} \\
  \end{array}$ 
  \caption{Upper panels of each of the sub-figures display the ``original'' differential light curves for Blazhko RR Lyrae stars in our sample after applying the differential photometry technique as described in the text. Bottom panels of each of the sub-figures are the light curves after removing the components related to the modulated (Blazhko) frequencies $f_m$, and the dashed curves are the fitted light curves using Fourier expansion as given in Equation (1). Red filled circles and green triangles are for data with 60~s exposure times (from regular PTF/iPTF surveys) and 10~s exposure times (from the dedicated iPTF experiment).}
  \label{fig_blazhko_lc}
\end{figure*}

For the remaining 12 Blazhko RRab stars in our sample, seven of them have both 10~s and 60~s light curves, and only KIC 11125706 has a mean magnitude brighter than $\sim 14$~mag. Therefore, we only fit the 10~s light curve of this star, and merged the 10~s and 60~s light curves for the other six Blazhko RRab stars. The differential light curves for these 12 Blazhko RRab stars were displayed in upper panels of Figure \ref{fig_blazhko_lc}. Due to the amplitude and/or phase modulation, these light curves are ``noisier'' than the light curves of non-Blazhko RRab stars. Based on a similar approach presented in \citet{smolec05}, these modulated light curves were fitted with the following expression \citep[for examples, see][]{kovacs95a,alcock03,benko11}:

\begin{eqnarray}
m(t) & = & F_0(t,f_0) + F_m(t,f_0,f_m),
\end{eqnarray}

\noindent where $F_0$ is the same as in Equation (1):

\begin{eqnarray}
F_0 (t,f_0) & = & m_0 + \sum_{i=1}^{n} A_i \sin \left( 2\pi i f_0 t + \phi_i \right), \nonumber
\end{eqnarray}

\noindent and $F_m$ includes the modulated components:

\begin{eqnarray}
F_m (t,f_0,f_m) & = & \sum_{j=0}^{r} A_j^m \sin \left( 2\pi j f_m t + \phi_j^m \right) \nonumber \\
               &   & + \sum_{k=0}^{q} A_k^- \sin [2\pi (kf_0 - f_m) t + \phi_k^-] \nonumber \\
               &   & + \sum_{k=0}^{q} A_k^+ \sin [2\pi (kf_0 + f_m) t + \phi_k^+]. \nonumber
\end{eqnarray}

\noindent In Equation (3), $f_0=1/P$ is the fundamental frequency, $f_m = 1/P_{BL}$ is the modulated frequency, and $A_0^m=A_0^-=A_0^+=0$. The Blazhko periods, $P_{BL}$, of these Blazhko RRab stars have already been derived in \citet{nemec13}, and we adopted their values in this work.\footnote{There are two and three $P_{BL}$ listed for KIC 9001926 and KIC 10789273, respectively. We adopted the first value given in \citet{nemec13} when fitting Equation (3) to their light curves.} Various combinations of Fourier order $(n,r,q)$ were visually inspected, and the best-fit combinations were adopted. Following an approach similar to that of \citet{smolec05}, we removed the $F_m$ components that are associated with the modulated frequency $f_m$ after fitting the light curves with Equation (3). The resulted light curves were shown in lower panels of Figure \ref{fig_blazhko_lc}, and were used to determine the $\phi_{31}$ Fourier parameters of these Blazhko RRab stars. Except for KIC 9973633, these light curves resemble the light curves for RR Lyrae stars pulsating in the fundamental frequency (and its harmonics) only. 

The differential light curve for KIC 9973633 exhibits strong amplitude and phase modulation, similar to KIC 3864443 and KIC 6186029 (as shown in Figure \ref{fig_extreme}). After experimenting with various combinations of Fourier order $(n,r,q)$ to fit the combined 10~s and 60~s light curves with Equation (3), we still could not remove the modulated components in the combined light curve (see lower panel in Figure \ref{fig_blazhko_lc}). This implies that additional frequency terms, such as $k f_0\pm l f_m$ (where $l$ is an integer), might need to be included in Equation (3), or this Blazhko star exhibits complex modulations as in the case of KIC 6186029 \citep[V445 Lyr,][]{guggenberger12}. Nevertheless, a detailed investigation of the power spectrum of KIC 9973633 is beyond the scope of this work, and it is obvious that it should be excluded from our sample.

\begin{deluxetable}{llccr}
\tabletypesize{\scriptsize}
\tablecaption{Fitted Results for Fourier Parameter $\phi_{31}$ from PTF Light Curves and the Derived Photometric Metallicity \label{tab3}}
\tablewidth{0pt}
\tablehead{
\colhead{KIC} &
\colhead{$P$ [days]\tablenotemark{a}} &
\colhead{$\phi_{31}$} & 
\colhead{$[Fe/H]_{\mathrm{spec}}$\tablenotemark{a}} &
\colhead{$[Fe/H]_{\mathrm{PTF}}$} 
}
\startdata
\cutinhead{Non-Blazhko RR Lyrae} 
6936115  & 0.52739847	& $4.796\pm0.019$ & $-1.98\pm0.09$ & $-1.82\pm0.05$ \\
11802860 & 0.6872160	& $5.644\pm0.007$ & $-1.33\pm0.09$ & $-1.91\pm0.07$ \\
6763132  & 0.5877887	& $5.207\pm0.010$ & $-1.89\pm0.10$ & $-1.74\pm0.04$ \\
9591503  & 0.5713866	& $5.178\pm0.012$ & $-1.66\pm0.12$ & $-1.66\pm0.04$ \\
9947026  & 0.5485905	& $5.946\pm0.040$ & $-0.59\pm0.13$ & $-0.51\pm0.06$ \\
7030715  & 0.68361247	& $5.971\pm0.023$ & $-1.33\pm0.08$ & $-1.47\pm0.07$ \\
6100702  & 0.4881457	& $5.803\pm0.017$ & $-0.16\pm0.09$ & $-0.25\pm0.05$ \\
10136603 & 0.4337747	& $5.764\pm0.013$ & $-0.05\pm0.14$ & $0.10\pm0.07$ \\
7988343  & 0.5811436	& $5.130\pm0.005$ & $\cdots$ & $-1.79\pm0.04$ \\
6070714  & 0.5340941	& $6.210\pm0.029$ & $-0.05\pm0.10$ & $-0.06\pm0.07$ \\
5299596  & 0.5236377	& $5.949\pm0.016$ & $-0.42\pm0.10$ & $-0.32\pm0.05$ \\
10136240 & 0.5657781	& $5.406\pm0.019$ & $-1.29\pm0.23$ & $-1.33\pm0.04$ \\
9508655  & 0.5942369	& $5.215\pm0.008$ & $-1.83\pm0.12$ & $-1.78\pm0.04$ \\
9658012  & 0.533206	& $5.250\pm0.006$ & $-1.28\pm0.14$ & $-1.29\pm0.03$ \\
7742534  & 0.4564851	& $4.886\pm0.028$ & $-1.28\pm0.20$ & $-1.19\pm0.06$ \\
3866709  & 0.47070609	& $4.876\pm0.033$ & $-1.13\pm0.09$ & $-1.31\pm0.06$ \\
8344381  & 0.5768288	& $5.260\pm0.033$ & $\cdots$ & $-1.59\pm0.05$ \\
9717032  & 0.5569092	& $5.340\pm0.016$ & $-1.27\pm0.15$ & $-1.35\pm0.04$ \\
7176080  & 0.5070740	& $4.954\pm0.039$ & $\cdots$ & $-1.47\pm0.06$ \\
\cutinhead{Blazhko RR Lyrae} 
11125706 & 0.6132200 	& $6.098\pm0.021$ & $-1.09\pm0.08$ & $-0.79\pm0.05$ \\
10789273 & 0.48027971 	& $5.068\pm0.015$ & $-1.01\pm0.10$ & $-1.13\pm0.04$ \\
7505345  & 0.4737027 	& $5.027\pm0.005$ & $-1.14\pm0.17$ & $-1.13\pm0.04$ \\
5559631  & 0.62070001 	& $5.967\pm0.011$ & $-1.16\pm0.11$ & $-1.01\pm0.05$ \\
12155928 & 0.43638507 	& $4.983\pm0.005$ & $-1.23\pm0.15$ & $-0.92\pm0.05$ \\
9697825  & 0.5575765 	& $4.947\pm0.016$ & $-1.50\pm0.29$ & $-1.85\pm0.05$ \\
6183128  & 0.561691 	& $5.137\pm0.019$ & $-1.44\pm0.16$ & $-1.64\pm0.04$ \\
9578833  & 0.5270283 	& $5.069\pm0.043$ & $-1.16\pm0.09$ & $-1.47\pm0.07$ \\
7257008  & 0.51177516 	& $5.163\pm0.051$ & $\cdots$ & $-1.24\pm0.07$ \\
7671081  & 0.5046123 	& $5.134\pm0.050$ & $-1.51\pm0.12$ & $-1.22\pm0.07$ \\
9001926  & 0.5568016 	& $5.220\pm0.020$ & $-1.50\pm0.20$ & $-1.50\pm0.04$ 
\enddata
\tablenotetext{a}{Values are taken from \citet{nemec13}.}
\end{deluxetable}

\section{The Metallicity-Light Curve Relation}

The $\phi_{31}$ Fourier parameters derived in the previous section were listed in the third column of Table \ref{tab3}. Among these RRab stars, 26 of them have $[Fe/H]$ values, listed in the forth column of Table \ref{tab3}, based on high-resolution spectroscopic observations taken from the 3.6m Canada-France-Hawaii Telescope and the 10m Keck I Telescope \citep{nemec13}. To derive the $R_{PTF}$-band metallicity-light curve relation, we adopted the well established regression function as in \citet{jurcsik96} and \citet{wu06}: $[Fe/H] = b_0 + b_1 P + b_2 \phi_{31}$. We did not adopt the five-parameter regression function from \citet{nemec13} because it did not improve the dispersion ($\sigma$) of the fitted relation. The initial fit to the 26 RRab stars given in Table \ref{tab3} returns a $\sigma\sim 0.20$~dex, which is much higher than the typical dispersion based on this technique \citep[$\sim0.13$~dex to $\sim0.14$~dex,][]{jurcsik96,wu06,ngeow15}. After removing outliers that deviate more than $2\times 0.13$~dex from the regression, we derived the following relation,\footnote{The actual regression fitting was done via the {\tt kmpfit} package, available at {\tt https://github.com/josephmeiring/kmpfit}, because errors are presented in both of the independent variable $\phi_{31}$ and the dependent variable $[Fe/H]$.} in the native $R_{PTF}$-band photometric system:

\begin{eqnarray}
[Fe/H]_{PTF} & = & -4.089 (\pm0.339) - 7.346 (\pm0.439) P  \nonumber \\
            &   & + 1.280 (\pm0.062) \phi_{31},
\end{eqnarray}

\noindent with a dispersion of $\sigma=0.118$~dex. Uncertainty on the photometric $[Fe/H]$ based on the above Equation, which incorporates the covariance matrix, can be calculated with the following expression.

\begin{eqnarray}
\sigma^2_{[Fe/H]} & = & 0.115 + 0.193 P^2 + 0.004 \phi_{31}^2 - 0.103 P  \\ \nonumber 
                &   & - 0.031 \phi_{31} - 0.020 P \phi_{31} + 1.638 \sigma_{\phi_{31}}^2 + 53.965 \sigma_P^2.
\end{eqnarray}

\noindent The fifth column in Table \ref{tab3} listed the photometric $[Fe/H]$ in the native $R_{PTF}$-band and the associated uncertainties calculated from Equation (4) and (5). Note that we assume $\sigma_P = 0$ for the RRab stars in the {\it Kepler} field as their very precise and accurate periods were determined from nearly continuous {\it Kepler} light curves \citep{nemec13}. 

\begin{figure}
  %\epsscale{0.7}
  \plotone{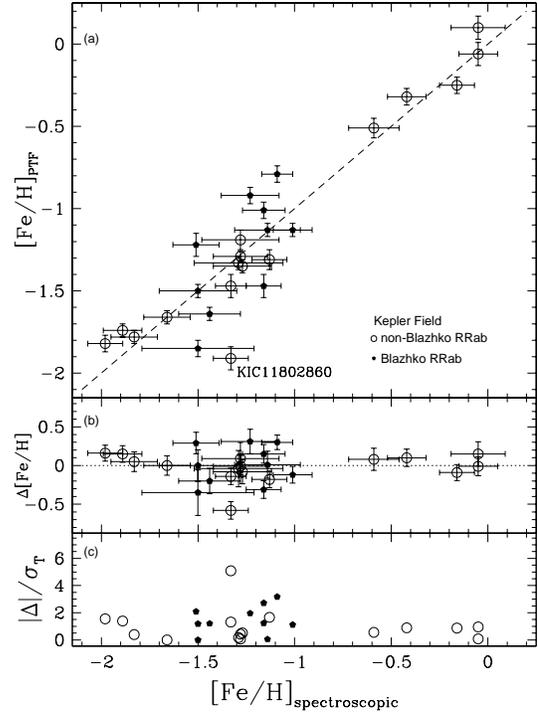}
  \caption{{\bf (a):} Comparison of the photometric $[Fe/H]$, calculated from Equation (4), in native $R_{PTF}$-band to the spectroscopic $[Fe/H]$ as presented in \citet{nemec13} for the common 26 RRab stars in the {\it Kepler} field. The dashed line indicates $y=x$ and not the fit to the data. A clear outlier, KIC 11802860, is also marked in the plot. {\bf (b):} Difference between the photometric and spectroscopic $[Fe/H]$ as a function of spectroscopic metallicity. The dotted lines are for $\Delta=0$ and not the fit to the data. {\bf (c):} Ratio of the absolute difference and $\sigma_T$ as a function of spectroscopic metallicity, where $\sigma_T$ is the quadrature sum of the uncertainties of photometric and spectroscopic metallicities. Open and filled symbols represent the non-Blazhko and Blazhko RRab stars, respectively.} 
  \label{fig_feh_compare}
\end{figure}

In Figure \ref{fig_feh_compare}, we compare the photometric metallicities derived from Equation (4) to the spectroscopic metallicities given in \citet{nemec13}. A clear outlier, KIC 11802860, can be seen from the top panel of this figure. We will discuss this outlier further in the next section. Figure \ref{fig_feh_compare}(b) shows the difference between the photometric and spectroscopic metallicities ($\Delta$) as a function of spectroscopic metallicity. Excluding KIC 11802860, these RRab stars have a $|\Delta|$ less than $\sim 0.35$~dex (or $3\sigma$, where $\sigma=0.118$ is the dispersion of Equation [4]), with a mean $\Delta$ of $-0.028$~dex. Out of the 26 RRab stars, 20 and 13 of them fall within the $2\sigma$ and $1\sigma$ boundaries, respectively. When comparing two quantities, it is customary in astronomy for these quantities to be considered in agreement if their absolute difference is within two to three times of the quadrature sum of their errors. In Figure \ref{fig_feh_compare}(c), we present the ratios of absolute difference of the two metallicities and their quadrature sum errors, at which the majority of them fall within the ratio of approximately three.   

The Blazhko RRab stars in the {\it Kepler} field provide an opportunity to test the applicability of using their modulated light curves in estimating the $\phi_{31}$ Fourier parameter and hence the photometric metallicity. For the 11 Blazhko RRab stars in our {\it Kepler} sample (excluding KIC 9973633), we fit the light curves of the Blazhko RRab stars as presented in the upper panels of Figure \ref{fig_blazhko_lc} using Equation (1) only, without removing the modulated components (i.e. the $F_m$ term as done in Section 5.2). For differential comparison, we adopted the same order $n$ in the Fourier decomposition as in the case of including the modulated components. We found that several Blazhko RRab stars show a small difference in the $\phi_{31}$ Fourier parameter (KIC~9001926: $0.001$; KIC~10789273: $0.013$; KIC~12155928: $0.015$) with and without removing the modulated components, while few others the differences are much larger (KIC~5559631: $0.233$; KIC~11125706: $0.295$; KIC~7671081: $0.666$). The averaged difference of $0.102$ translates to a difference of $0.13$~dex in $[Fe/H]_{PTF}$ from Equation (4), which is comparable to the dispersion of the metallicity-light curve relation. Excluding KIC~7671081 the averaged difference in the $\phi_{31}$ Fourier parameter is reduced to $0.045$ or a difference of $0.06$~dex in $[Fe/H]_{PTF}$. Therefore, except for a few cases, our test suggested the Blazhko RRab stars can be included in the estimation of photometric $[Fe/H]_{PTF}$, given that their Blazhko periods can be well determined.

In the following sub-sections, we test and verify our metallicity-light curve relation using the new spectra data from a low-resolution spectroscopic observation of few RRab stars in the {\it Kepler} field (Section 6.1), as well as several publicly available data taken from the literature (Section 6.2 - 6.3).

\subsection{P200 Observations on Selected RR Lyrae in the {\it Kepler} Field}

\begin{figure*}
  \epsscale{1.1}
  \plotone{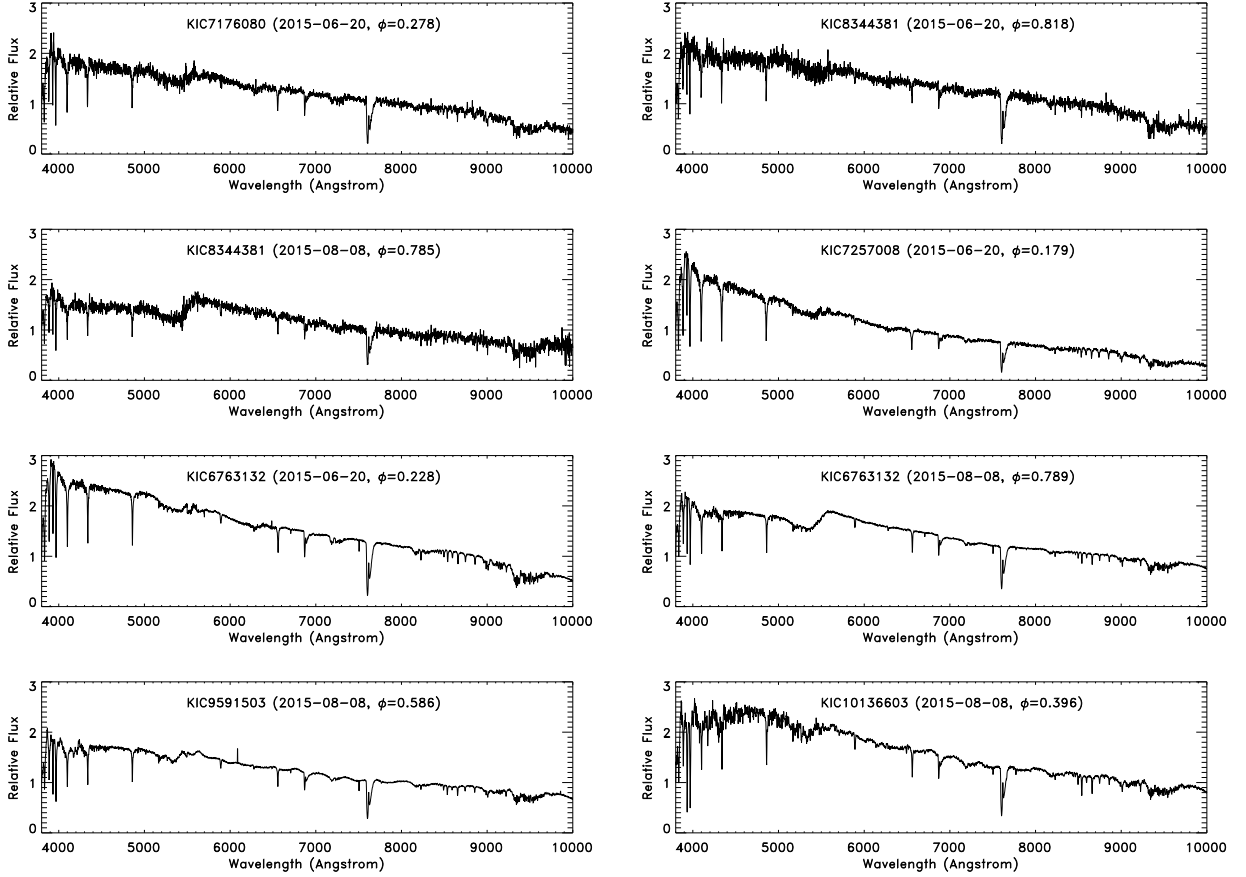}
  \caption{P200/DBSP low-resolution spectra for the RRab stars listed in Table \ref{tab4}. Flux is normalized to the flux at $3800\AA$ of each target. $\phi$ represents the pulsation phases when the spectra were taken.} 
  \label{fig_spectra}
\end{figure*}

\begin{deluxetable*}{llccccccr}
\tabletypesize{\scriptsize}
\tablecaption{Metallicity for Selected RRab Stars in the {\it Kepler} Field with P200/DBSP Observations \label{tab4}}
\tablewidth{0pt}
\tablehead{
\colhead{KIC} &
\colhead{Type\tablenotemark{a}} &
\colhead{Observed Date} &
\colhead{Exposure time\tablenotemark{b}} &
\colhead{$[Fe/H]_{S12}$} & 
\colhead{$[Fe/H]_{S13}$} & 
\colhead{$[Fe/H]_{\mathrm{spec}}$\tablenotemark{c}} &
\colhead{$[Fe/H]_{Kp}$\tablenotemark{c}} &
\colhead{$[Fe/H]_{\mathrm{PTF}}$} 
}
\startdata
6763132  & RRab-NB & 2015-06-20 & 240 & $-1.48$ & $-1.54$ & $-1.89\pm0.10$ & $-1.81\pm0.03$ & $-1.74\pm0.04$ \\
6763132  & RRab-NB & 2015-08-08 & 300 & $-1.83$ & $-1.90$ & $-1.89\pm0.10$ & $-1.81\pm0.03$ & $-1.74\pm0.04$ \\
9591503  & RRab-NB & 2015-08-08 & 300 & $-1.39$ & $-1.48$ & $-1.66\pm0.12$ & $-1.74\pm0.03$ & $-1.66\pm0.04$ \\
10136603 & RRab-NB & 2015-08-08 & 300 &  $0.72$ &  $0.43$ & $-0.05\pm0.14$ & $-0.06\pm0.05$ &  $0.10\pm0.07$ \\
7176080  & RRab-NB & 2015-06-20 & 300 &  $0.37$ &  $0.15$ & $\cdots$ & $-1.63\pm0.04$ & $-1.47\pm0.06$ \\
8344381  & RRab-NB & 2015-06-20 & 300 & $-1.65$ & $-1.75$ & $\cdots$ & $-1.82\pm0.03$ & $-1.59\pm0.05$ \\ 
8344381  & RRab-NB & 2015-08-08 & 300 & $-1.28$ & $-1.40$ & $\cdots$ & $-1.82\pm0.03$ & $-1.59\pm0.05$ \\ 
7257008  & RRab-B  & 2015-06-20 & 300 &  $0.29$ &  $0.31$ & $\cdots$ & $-1.02\pm0.03$ & $-1.24\pm0.07$ \\
\cutinhead{Averaged $[Fe/H]_{S12}$ and $[Fe/H]_{S13}$ values for KIC 6763132 and KIC 8344381}
6763132  & RRab-NB & $\cdots$ & $\cdots$ & $-1.66$ & $-1.72$ & $-1.89\pm0.10$ & $-1.81\pm0.03$ & $-1.74\pm0.04$ \\
8344381  & RRab-NB & $\cdots$ & $\cdots$ & $-1.47$ & $-1.58$ & $\cdots$ & $-1.82\pm0.03$ & $-1.59\pm0.05$  
\enddata
\tablenotetext{a}{Types are the same as in Table \ref{tab1}.}
\tablenotetext{b}{Exposure time in seconds for the P200/DBSP observations.}
\tablenotetext{c}{Adopted from \citet{nemec13}.}
\end{deluxetable*}

Spectra of several RRab stars listed in Table \ref{tab3} were obtained with the P200 Telescope, using the available low-resolution spectrograph DBSP \citep[the Double-Beam Spectrograph;][with $R\sim 1360$]{oke82}, on 2015 June 20 and August 08. These spectra were reduced using a {\tt pyRAF}-based reduction pipeline\footnote{Available at {\tt https://github.com/ebellm/pyraf-dbsp}} tailored for the P200/DBSP spectrograph \citep{bellm16s,bellm16}, including the bias subtraction, flat-field correction, spectral extraction, and wavelength calibration. Spectroscopic metallicity were then measured on these reduced spectra using the pseudo-equivalent widths of Balmer lines and [Ca II] K lines, following the procedures outlined in \citet[][hereafter S12]{sesar12} and \citet[][hereafter S13]{sesar13}. To avoid line broadening due to velocity gradients and shock waves, \citet{nemec13} only took the spectra at pulsational phases between $\sim 0.2$ and $\sim 0.5$. Since only a small fraction of the P200/DBSP spectra were fall within this range of pulsational phases, we adopted the criterion given in \citet{sesar13} to retain those spectra taken at the pulsational phases between 0.10 and 0.85 in order to increase usable P200/DBSP spectra in our sample. The six RRab stars and their measured metallicities from the P200/DBSP observations were listed in Table \ref{tab4}, and their spectra are presented in Figure \ref{fig_spectra}. Two of the six RRab stars were observed on both nights. Also, three of the RRab stars have spectroscopic metallicities obtained from high-resolution spectra \citep{nemec13}, which can be used to compare to the spectroscopic metallicities taken from P200/DBSP.

Since the coefficients in the transformation of pseudo-equivalent widths were slightly different in S12 and S13, we measured the metallicity using both transformations. They are listed in columns 5 and 6 in Table \ref{tab4} as $[Fe/H]_{S12}$ and $[Fe/H]_{S13}$, respectively. We assume an uncertainty of $\sim 0.15$~dex on these metallicities, which is a typical value based on the method of using pseudo-equivalent widths \citep{sesar12}. As shown in Table \ref{tab4}, the two prescriptions give consistent spectroscopic metallicities (except for KIC 10136603, with the largest difference of $0.29$~dex), with $[Fe/H]_{S13}$ being more metal-poor than $[Fe/H]_{S12}$ by $\sim 0.1$~dex on average. When compared to the metallicity from high-resolution spectra ($[Fe/H]_{\mathrm{spec}}$), excellent agreement was found for KIC 6763132 taken on August 08, followed by marginal agreements for KIC 9591503 and KIC 6763132 taken on June 20. For KIC 10136603, even though the values for both $[Fe/H]_{S12}$ and $[Fe/H]_{\mathrm{spec}}$ indicate a metal-rich RRab star, they are not in agreement with each others. In contrast, value from $[Fe/H]_{S13}$ is closer to $[Fe/H]_{\mathrm{spec}}$ for this RRab star. In short, values from $[Fe/H]_{S13}$ are in better agreement with those from $[Fe/H]_{\mathrm{spec}}$ than those from $[Fe/H]_{S12}$, with a mean difference of $\sim 0.25$~dex and $\sim0.38$~dex, respectively.

\begin{figure}
  \epsscale{1.0}
  \plotone{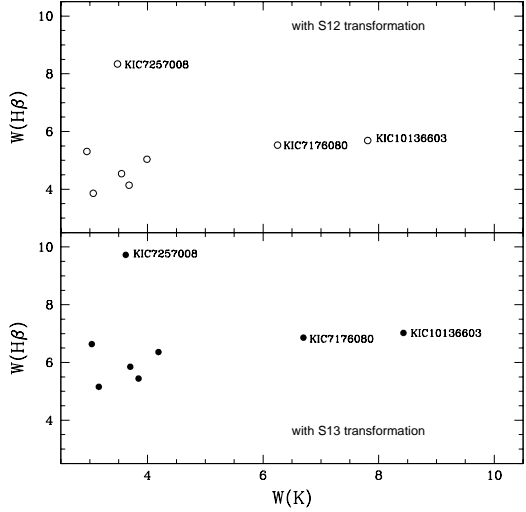}
  \caption{Comparison of the transformed pseudo-equivalent widths for [Ca II] K lines and $H\beta$ lines from the eight spectra taken from P200/DBSP observations (see Table \ref{tab4}), with transformations given in S12 (upper panel) and S13 (lower panel).}
  \label{fig_W}
\end{figure}

When comparing the metallicities from P200/DBSP spectra to photometric metallicities based on PTF/iPTF light curves ($[Fe/H]_{\mathrm{PTF}}$), we found that good agreements can be seen in three RRab stars: KIC 6763132, KIC 9591503, and KIC 8344381. Based on the five measurements of low-resolution spectroscopic metallicities for these three RRab stars, the $[Fe/H]_{S13}$ values are again in better agreement with $[Fe/H]_{\mathrm{PTF}}$ than those from $[Fe/H]_{S12}$, the mean difference becomes $\sim0.05$~dex and $\sim0.14$~dex, respectively. For KIC 10136603, $[Fe/H]_{S12}$ disagrees with $[Fe/H]_{\mathrm{PTF}}$ and yet $[Fe/H]_{S13}$ marginally agrees with the latter value. Finally, both KIC 7176080 and KIC 7257008 show large discrepancies between the metallicities from P200/DBSP spectra and PTF/iPTF light curves, by more than $1.5$~dex. We suspected that this might be caused by noisier spectra toward the short-wavelength ends. Figure \ref{fig_W} compares the transformed pseudo-equivalent widths for the eight spectra, those from KIC 7176080, KIC 7257008, and KIC 10136603 appeared to be outliers in this figure. Another possibility is that the observations of KIC 7176080 and KIC 7257008 were affected by weather because their spectra were taken within 10 minutes of each other on June 20. A similar result and conclusion can also be found when comparing $[Fe/H]_{S12/S13}$ to $[Fe/H]_{Kp}$.

For the two RRab stars, KIC 6763132 and KIC 8344381, that have two P200/DBSP observations, the arithmetic average of the $[Fe/H]_{S12/S13}$ values are listed in the last two rows of Table \ref{tab4}. These averaged values were in good agreement with those from $[Fe/H]_{\mathrm{PTF}}$, in particular, excellent agreements can be seen between the values of $[Fe/H]_{S13}$ and $[Fe/H]_{\mathrm{PTF}}$. Note that the difference of the measured spectroscopic metallicities at the two different pulsation phases is $\sim \pm0.35$~dex for these two RRab stars.

\begin{figure}
  %\epsscale{0.8}
  \plotone{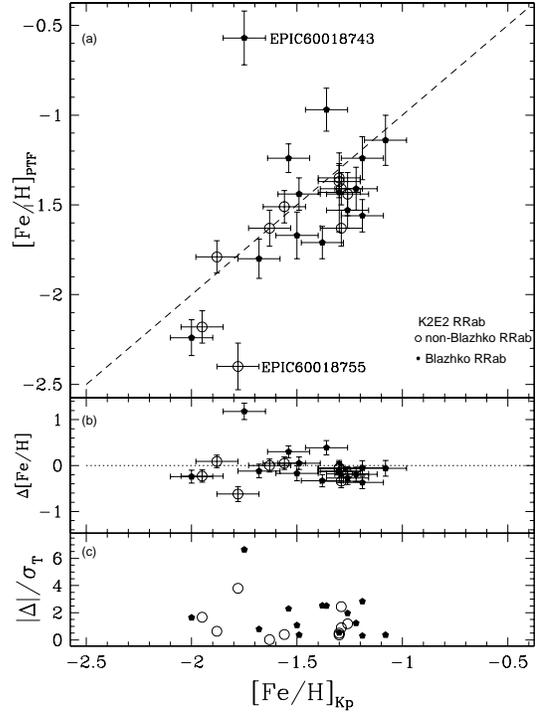}
  \caption{Same as in Figure \ref{fig_feh_compare}, but for RRab stars in the K2E2 Field. Two obvious outliers are labeled in the plot. We adopted $\pm0.1$~dex for the uncertainties on $[Fe/H]_{Kp}$ \citep{molnar15}.}
  \label{fig_feh_compare_k2e2}
\end{figure}

\begin{figure*}
  \epsscale{1.0}
  \plottwo{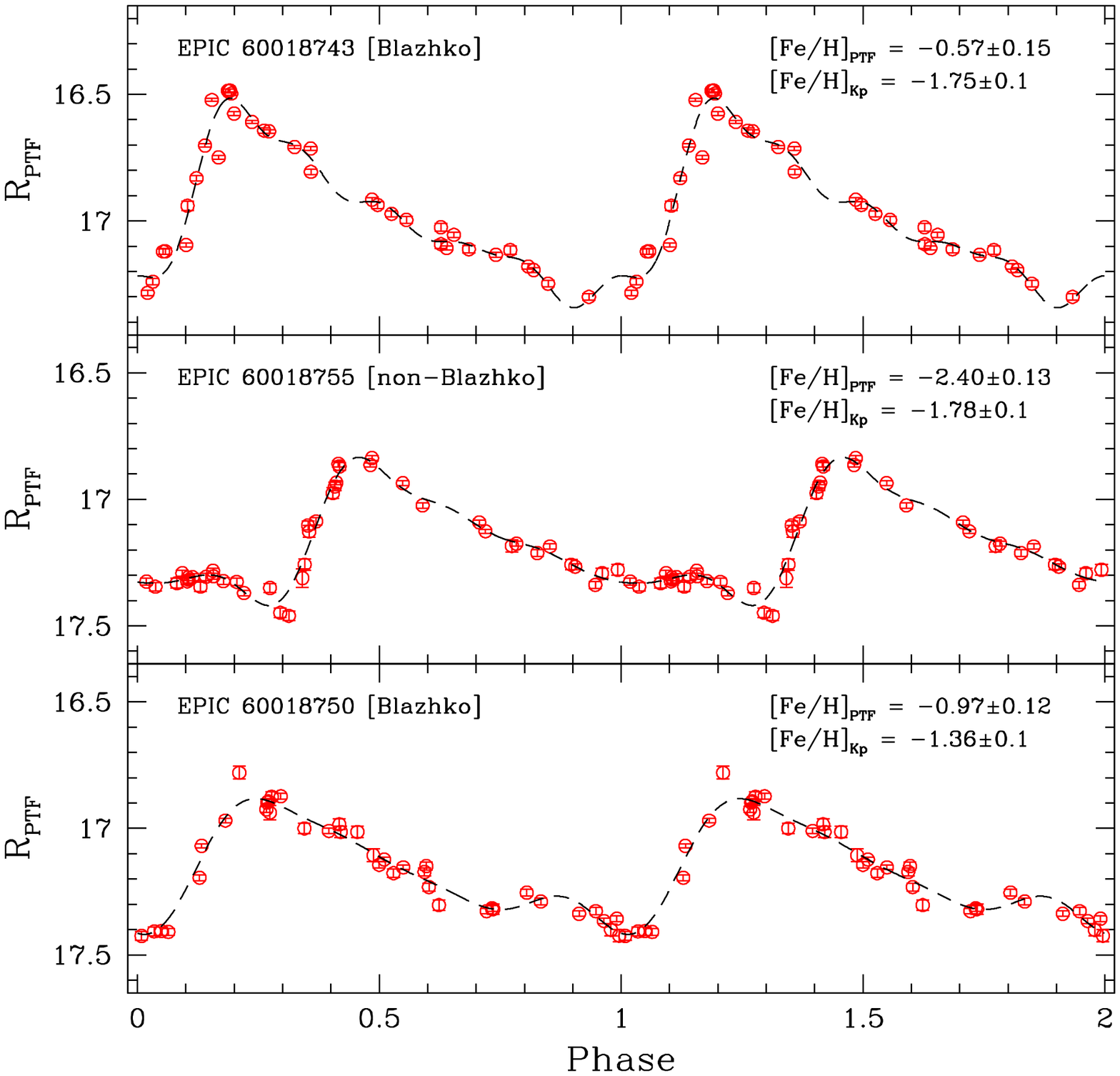}{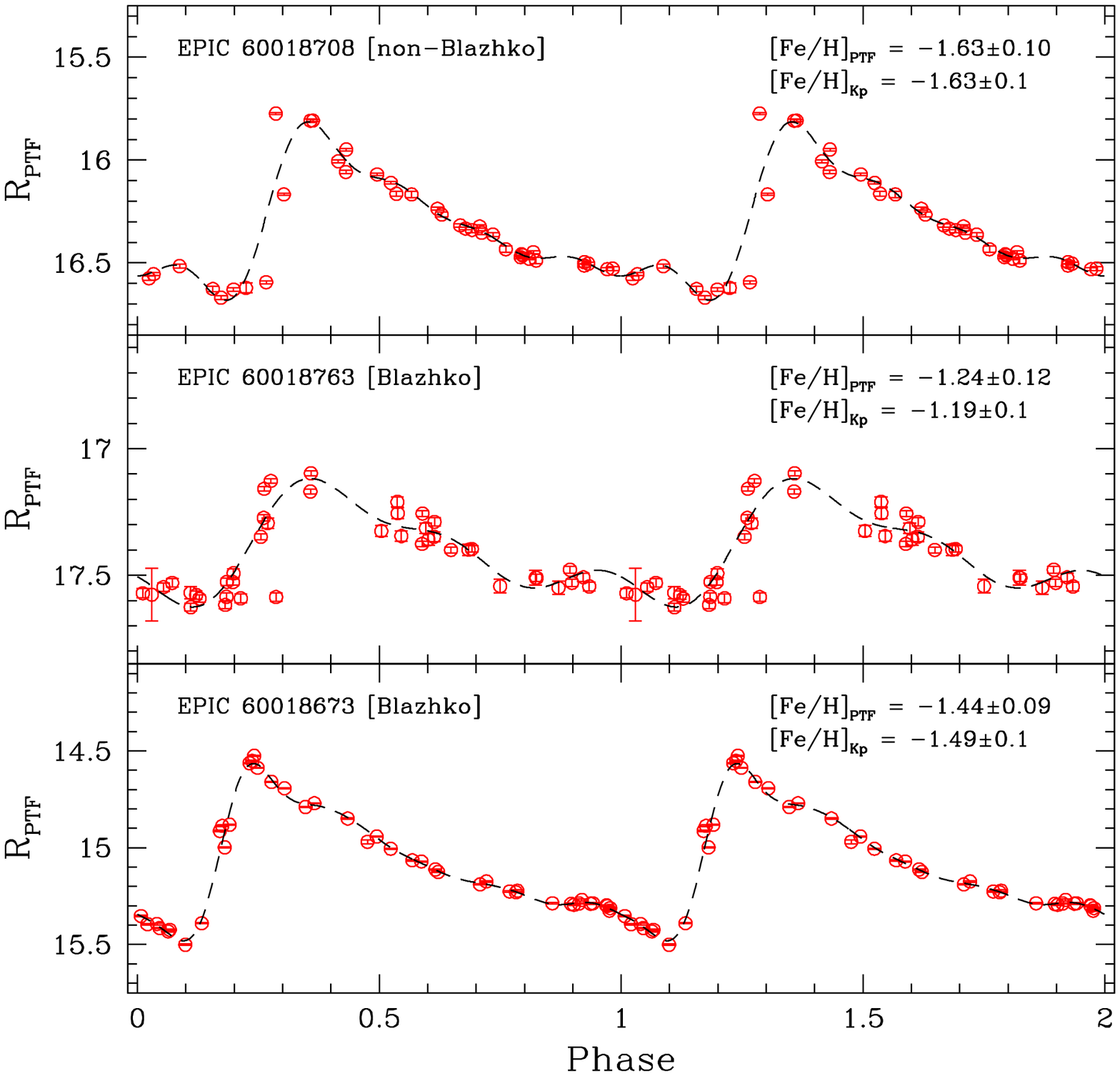}
  \caption{$R_{PTF}$-band light curves for RRab stars in K2E2 Field that display the largest (left panels) and smallest (right panels) deviation of the derived photometric metallicities. Dashed curves are the best-fit light curves using Equation (1).} 
  \label{fig_k2e2_lc}
\end{figure*}

\subsection{RR Lyrae Samples in K2E2 Field}

\citet{molnar15} presented the light-curve analysis for 27 RRab stars toward Pisces. These light-curve data were taken under the K2 Two-wheel Engineering Test (hereafter K2E2) after the failure of the second reaction wheel on board of {\it Kepler}. Based on the 8.9-days light curves, \citet{molnar15} classified 13 of them as non-Blazhko RRab stars, and the remaining 14 of them are Blazhko RRab stars.\footnote{Even though three of them are possible modulated RRab stars, for simplicity, we grouped them into the Blazhko RRab stars.} These authors also derived the photometric metallicity $[Fe/H]_{Kp}$ based on the K2E2 light curves with the relation presented in \citet{nemec13}. Therefore, the RRab stars in the K2E2 Field provide a sizable sample to test our metallicity-light curve relation. We retrieved and constructed the PTF light curves for 24 of these RRab stars using the same approaches as described in Section 4, the remaining three of them (EPIC 60018663, 60018669, and 60018779) were either fell on the inoperational CCD 03 or on the gap between the CCD chips. Number of data points per light curves for these RRab star ranges from $\sim30$ to $\sim 240$. Similar to Section 5, these light curves were fitted with Equation (1) to determine the Fourier parameters $\phi_{31}$ and hence the photometric metallicity $[Fe/H]_{PTF}$ with Equation (4). We did not remove the $F_m$ components for the Blazhko RRab stars because the majority of them do not have modulated period determined in \citet{molnar15}. Nevertheless, this also provides an opportunity to test our metallicity-light curve relation in the absence of modulated period. Finally, pulsation periods are taken from \citet{molnar15}, and assume that $\sigma_P \sim 0$.

Similar to Figure \ref{fig_feh_compare}, we compare the photometric metallicities for these RRab stars based on PTF light curves and K2K2 light curves in Figure \ref{fig_feh_compare_k2e2}. After removing the two clear outliers shown in Figure \ref{fig_feh_compare_k2e2}(a), the remaining 22 RRab stars show a mean $\Delta$ of $-0.093$~dex, which is consistent with the accuracy of this technique \citep{kocacs05}. Separating the sample into non-Blazhko RRab stars and Blazhko RRab stars, we obtained a mean $\Delta$ of $-0.094$~dex and $-0.092$~dex, respectively, suggesting that similar results can be achieved without removing the modulated component for Blazhko RRab stars. Besides the two outliers, there are only two other RRab stars with $|\Delta| > 0.35$~dex. For the remaining 22 RRab stars, there are 15 and 8 located within the $2\sigma$ and $1\sigma$ boundaries, respectively, as shown in Figure \ref{fig_feh_compare_k2e2}(b). Similarly, Figure \ref{fig_feh_compare_k2e2}(c) demonstrates that all of the RRab stars have a $|\Delta|/\sigma_T<3$, except for the two extreme outliers. Our test suggested that Equation (4) can be used to provide reliable metallicity estimation for the majority of RRab stars. Figure \ref{fig_k2e2_lc} displays three examples of the $R_{PTF}$-band light curves with the best (right panel) and the worst (left panel) agreements of the photometric metallicities, respectively.

\subsection{RR Lyrae Samples from Sesar et al (2012, 2013)}

\begin{figure*}
  $\begin{array}{ccc}
    \includegraphics[angle=0,scale=0.28]{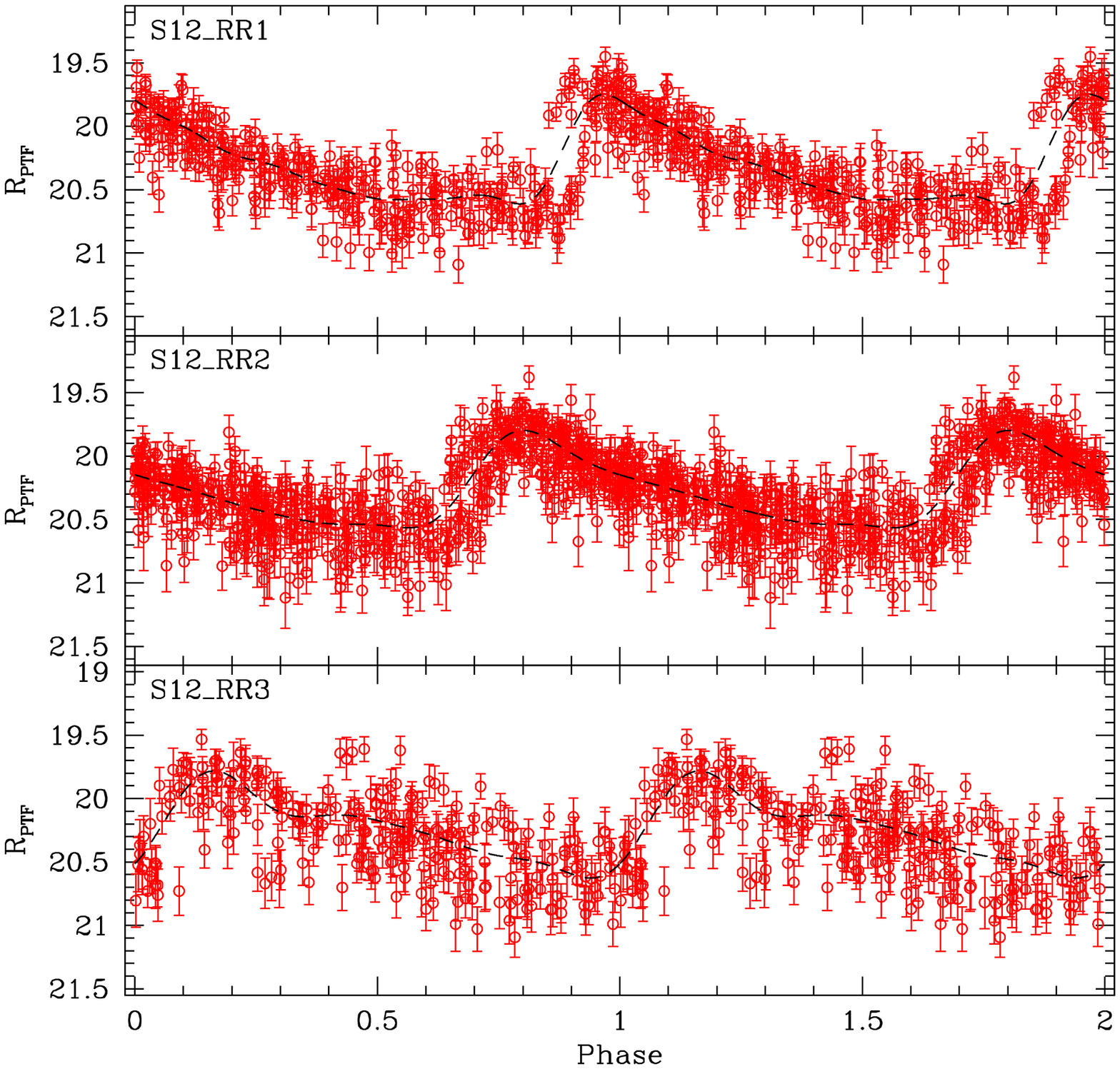} & 
    \includegraphics[angle=0,scale=0.28]{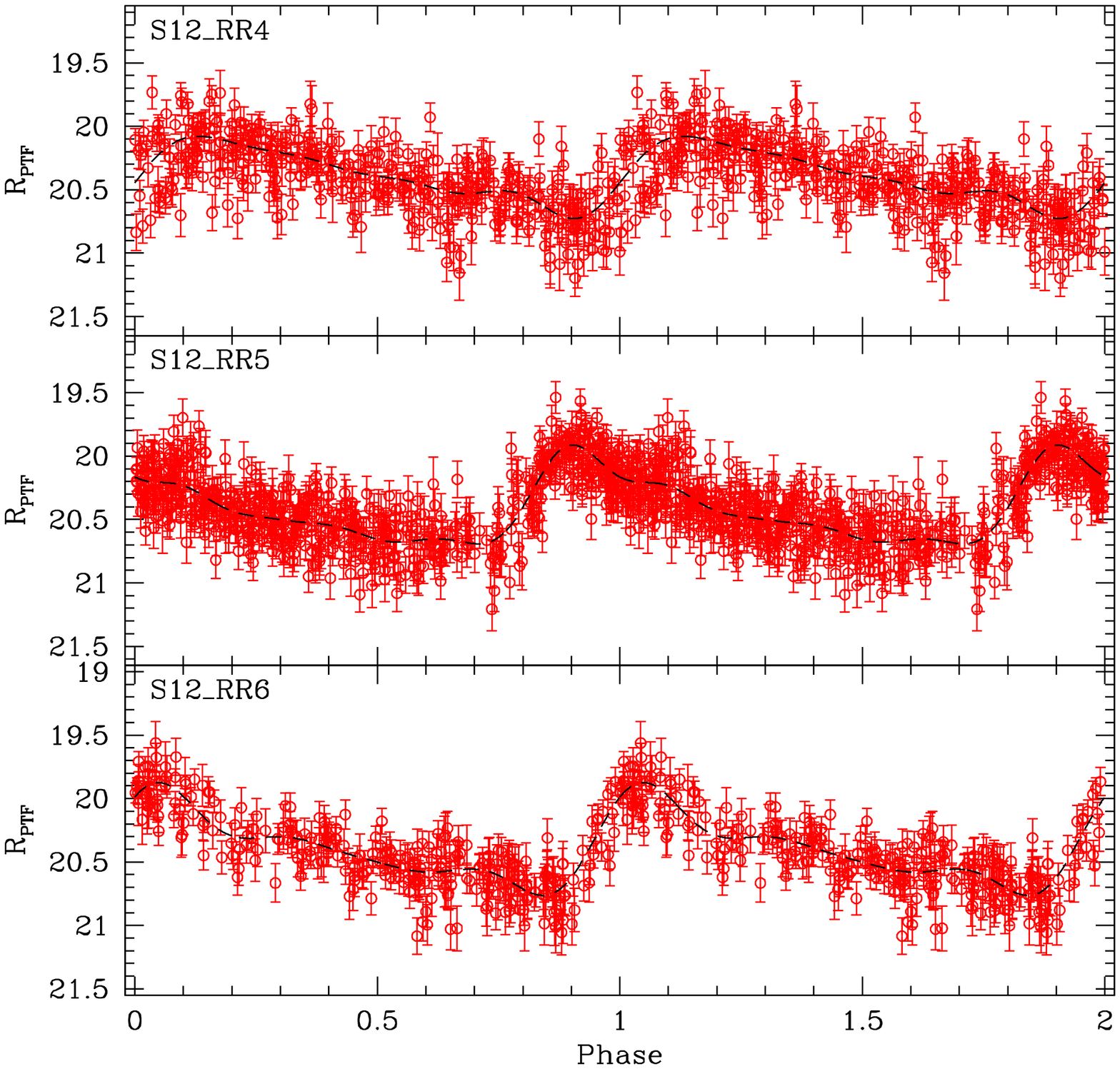} &
    \includegraphics[angle=0,scale=0.28]{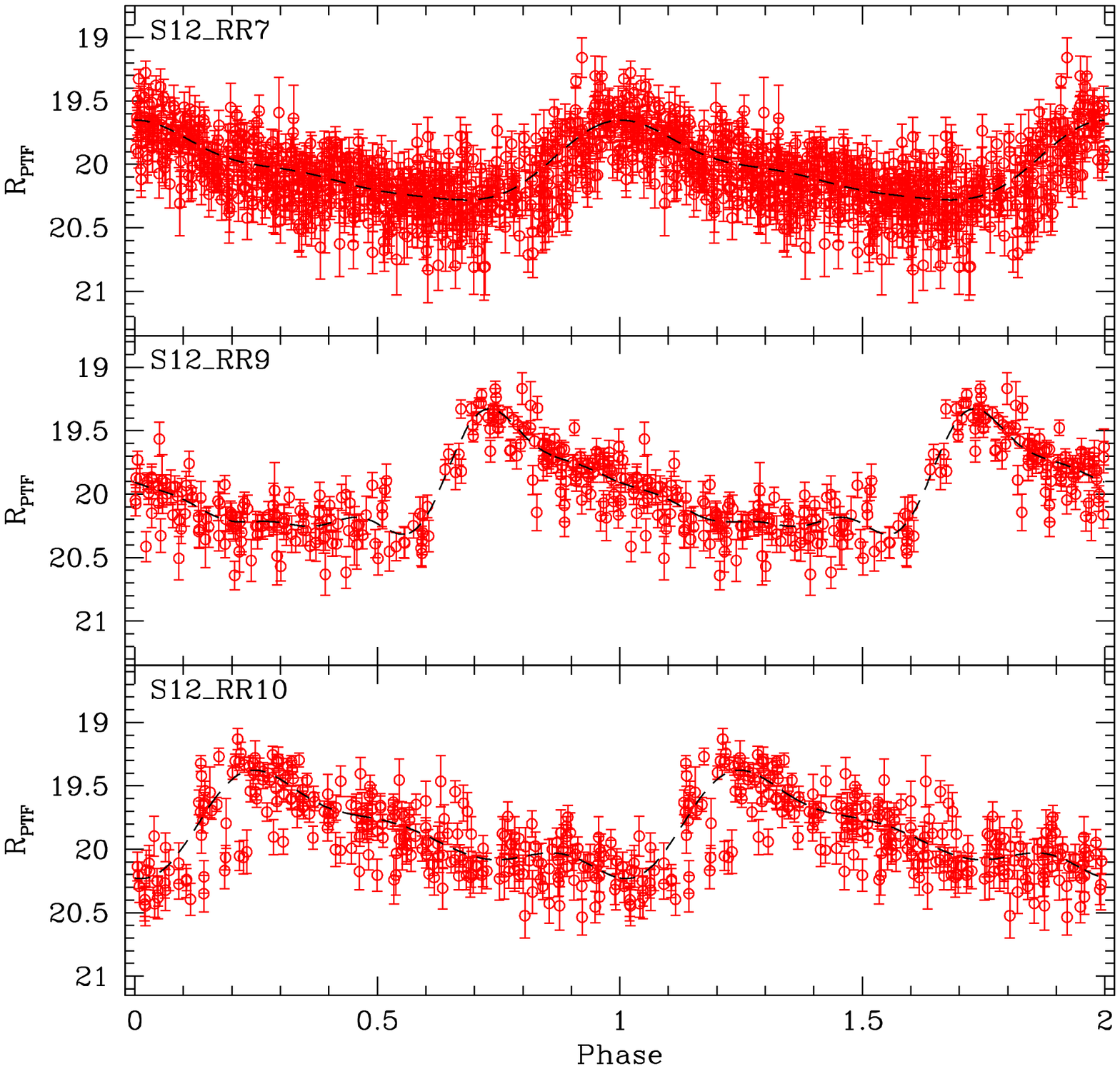} \\
  \end{array}$ 
  \caption{$R_{PTF}$-band light curves for nine faint RRab stars taken from \citet{sesar12}. The dashed curves are fitted light curves using Equation (1). We excluded S12\_RR8 because its PTF/iPTF light curve (with 192 data points) is too scattered that it does not display a typical RRab-like light curve, in contrast to those presented in this figure.}
  \label{fig_s12lc}
\end{figure*}

\begin{figure}
  %\epsscale{0.7}
  \plotone{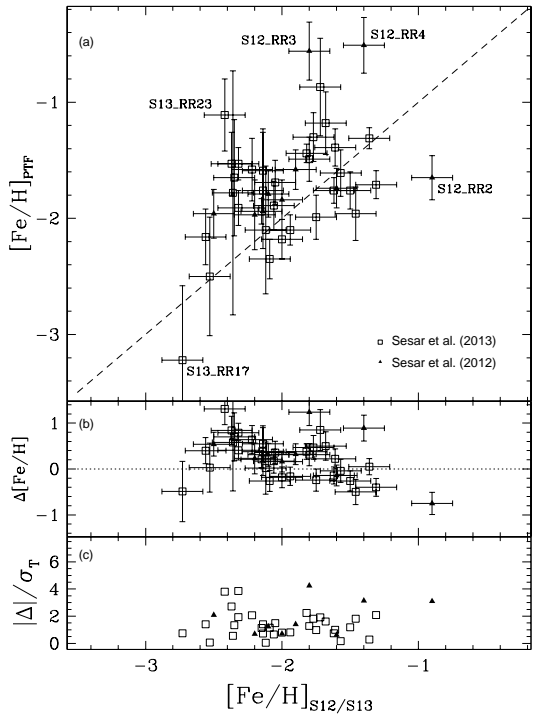}
  \caption{Same as in Figure \ref{fig_feh_compare}, but for RRab stars taken from \citet{sesar12} and \citet{sesar13}. Five obvious outliers are labeled in the plot. We adopted $\pm0.15$~dex for the uncertainties on $[Fe/H]_{S12/S13}$ \citep{sesar12}, where $S12/S13$ represents the metallicity measured from low-resolution spectra \citep{sesar12,sesar13}. We took an arithmetic mean if an RRab star has more than one spectroscopic metallicities listed in \citet{sesar12}.}
  \label{fig_feh_compare_sesar}
\end{figure}

As part of the study to search for substructures and tidal streams in the Galactic halo, S12 and S13 searched for the distant RRab stars using the PTF and other survey data. S12 reported the finding of 10 RR Lyrae with spectroscopic follow-up observations using two low-resolution spectrographs, the P200/DBSP and the Low Resolution Imaging Spectrometer \citep[LRIS;][with $R\sim 1760$, equipped on the Keck I Telescope]{oke95}. S13 presented 94 RRab stars of which only 50 of them have spectroscopic observations with P200/DBSP. Spectroscopic metallicities of these RRab stars were then measured from the low-resolution spectra. We retrieved PTF/iPTF light curves for the majority of these RRab stars and fit with Equation (1) following the procedures described in Section 4 and 5. Pulsation periods of these RRab stars were adopted from S12 and S13.

Since 2012, more data have become available from the iPTF project for the 10 RRab stars listed in S12. The number of data points per light curve increased from merely $\sim1\%$ (from 271 to $\sim290$ for S12\_RR6) to $\sim180\%$ (from 82 to $\sim235$ for S12\_RR9) as compared to S12, with a range from $\sim190$ to $\sim660$. Besides a larger number of data points, these 10 RRab stars are also much fainter than those found in the K2E2 Field shown in the previous subsection, with mean magnitudes fainter than $\sim 19.5$~mag. Consequently, their light curves exhibit a much larger scatter than the K2E2 RRab stars, as displayed in Figure \ref{fig_s12lc}. Nevertheless, these RRab stars provide an opportunity to test our metallicity-light curve relation for the faint RRab stars. Comparison of the photometric metallicities and spectroscopic metallicities of the nine RRab stars shown in Figure \ref{fig_s12lc} are given in Figure \ref{fig_feh_compare_sesar} as filled symbols. The mean $\Delta$ of these nine RRab stars is $0.31$~dex, after removing the three outliers in Figure \ref{fig_feh_compare_sesar} this mean value drops to $0.24$~dex. Five out of the six remaining faint RRab stars have a $|\Delta|$ value less than the $3\sigma$ (i.e. $\sim0.35$~dex) boundary. Hence, the performance of our metallicity-light curve relation is acceptable given the faintness, large scatter of light curves, and small number of RRab stars in this sample.
 
\begin{figure*}
  $\begin{array}{ccc}
    \includegraphics[angle=0,scale=0.28]{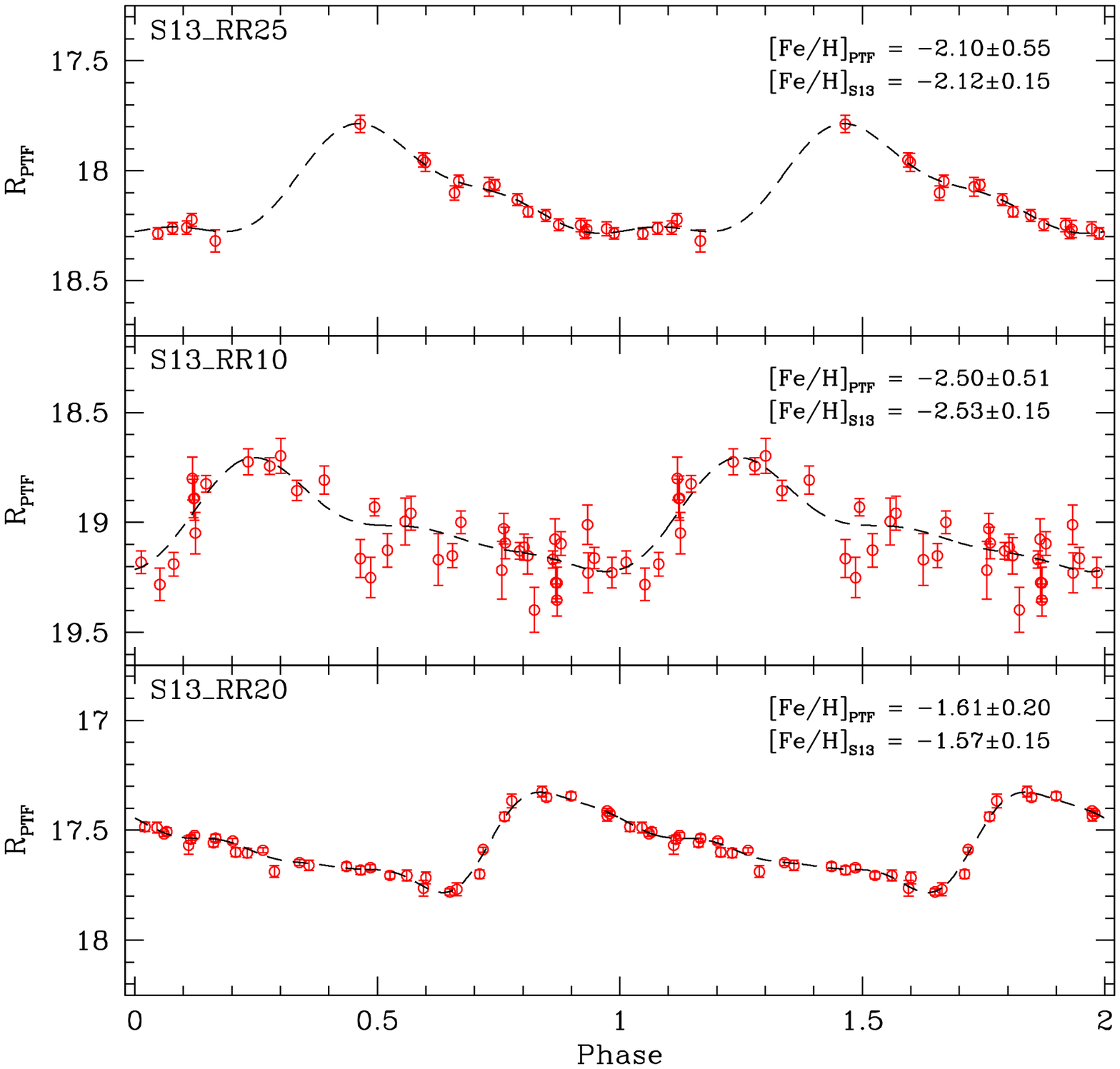} & 
    \includegraphics[angle=0,scale=0.28]{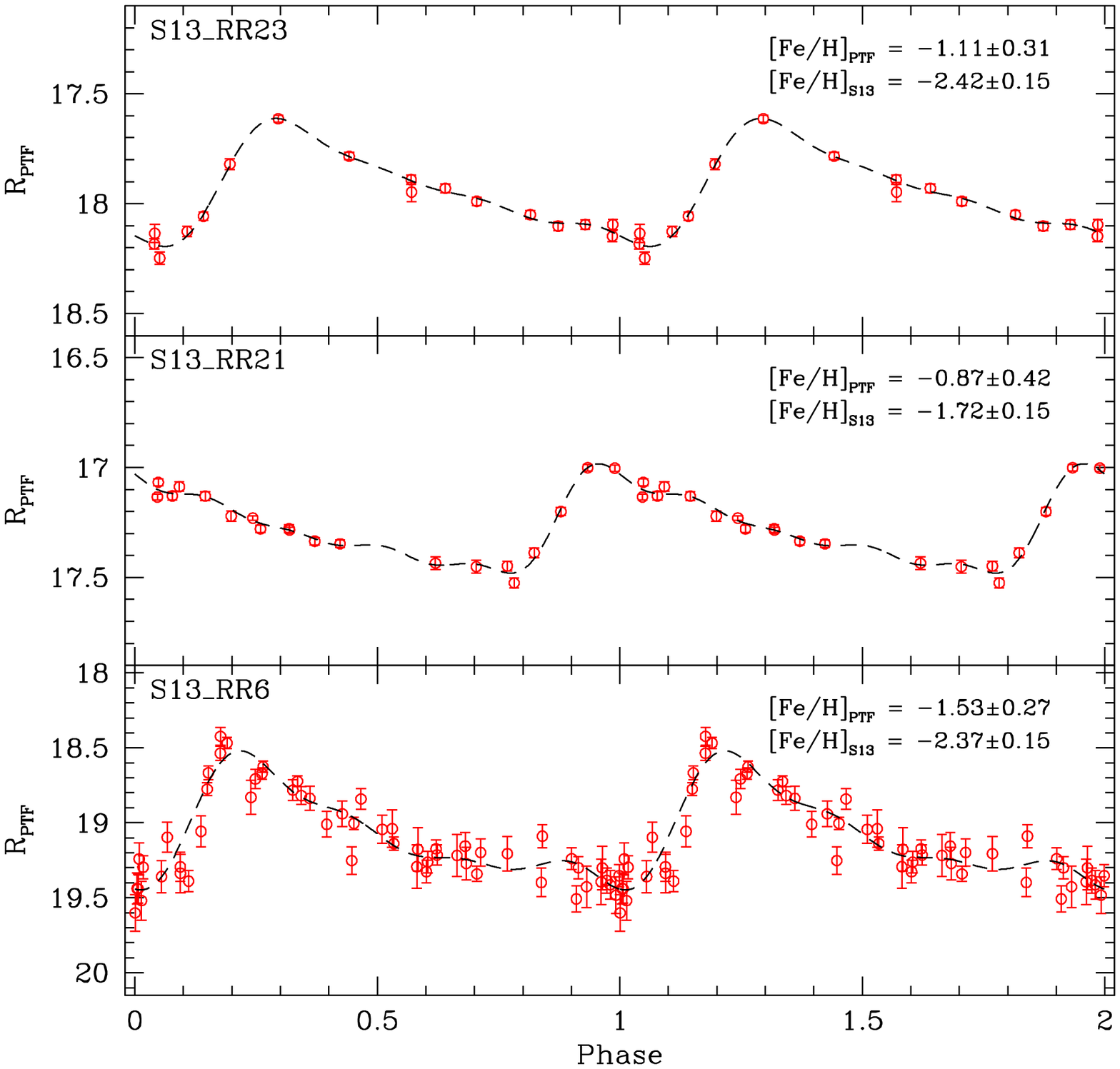} &
    \includegraphics[angle=0,scale=0.28]{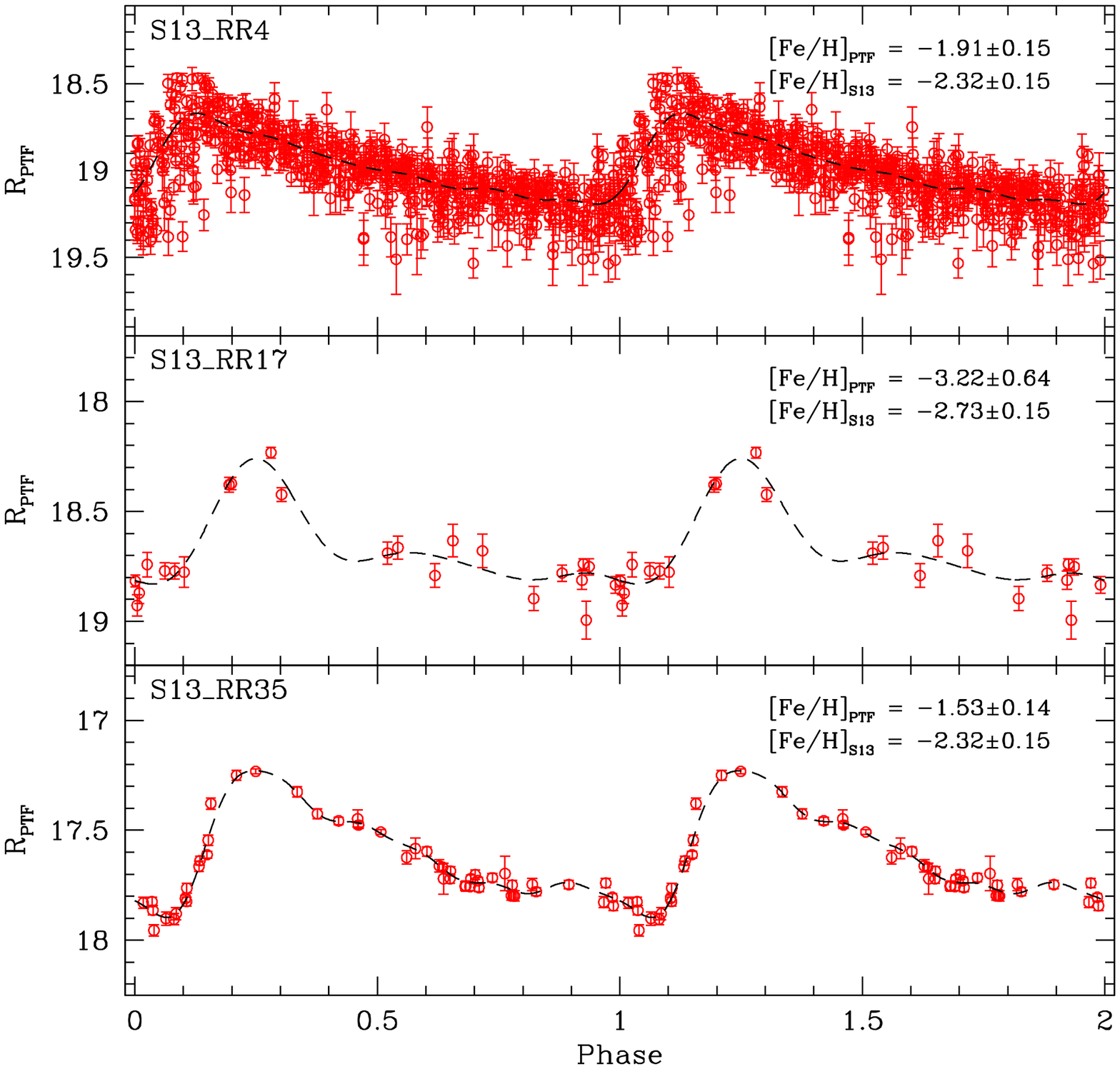} \\
  \end{array}$ 
  \caption{{\bf Left Panel:} example $R_{PTF}$-band light curves for three RRab stars with the {\it smallest} deviation between photometric and spectroscopic metallicities. {\bf Middle Panel:} example $R_{PTF}$-band light curves for three RRab stars with the {\it largest} deviation between photometric and spectroscopic metallicities. {\bf Right Panel:} example $R_{PTF}$-band light curves for three ``unusual'' RRab stars. The upper-right panel shows the light curve for the only RRab star with $\sim580$ data points in its light curve. The middle-right panel shows the light curve for the RRab star that has a $[Fe/H]_{PTF}< -3$~dex. The lower-right panel shows the light curve for the RRab star with the largest ratio of absolute difference of metallicities and the quadrature sum errors, $|\Delta|/\sigma_T \sim 3.85$, in the sample. The dashed curves are fitted light curves using Equation (1).}
  \label{fig_s13lc}
\end{figure*}

For the 50 RRab stars in S13, we only retained the PTF/iPTF light curves for 32 of them, the rest of the RRab stars either do not have data in PTF/iPTF (i.e. fall into the CCD 03), with only a small number of data points per light curve (less than 15 data points), or the light curves do not exhibit RRab-like light curves. The number of data points for these 32 RRab stars include $\sim20$ to $\sim70$ for 26 of them, $\sim100$ to $\sim300$ for 5 of them, and $\sim580$ for 1 of them. These RRab stars are brighter than those in S12, with mean magnitudes ranging from $\sim17$~mag. to $\sim19$~mag. Figure \ref{fig_feh_compare_sesar} compares the photometric and spectroscopic metallicities for these 32 RRab stars as open symbols, which exhibit a much larger scatter than the RRab stars in the K2E2 Fields (Figure \ref{fig_feh_compare_k2e2}). Without removing any outliers, the mean value of $\Delta$ for this sample is $0.23$~dex.

Figure \ref{fig_s13lc} presents $R_{PTF}$-band light curves for the selected RRab stars in this sample. Left panels of Figure \ref{fig_s13lc} display examples of three light curves that give a good agreement between the photometric and spectroscopic metallicities. In contrast, the middle panels of Figure \ref{fig_s13lc} show the light curves of three RRab stars with the largest $\Delta$, ranging from $0.84$~dex (for S13\_RR6) to $1.31$~dex (for S13\_RR23). However, these light curves do not show any ``abnormality'' when compared to those on the left panels. The upper-right panel of Figure \ref{fig_s13lc} is the light curve for the only RRab star with $\sim580$ data points in this sample, which also display a large scatter as those in the S12 sample (see Figure \ref{fig_s12lc}). The middle-right panel of Figure \ref{fig_s13lc} is the light curve for the most metal-poor RRab star within the 50 RR Lyrae in S13 sample, with a measured spectroscopic metallicity of $-2.73$~dex. We obtained a $[Fe/H]_{\mathrm{PTF}}=-3.22\pm0.64$~dex that is consistent with the spectroscopic metallicity. Finally, the lower-right panel of Figure \ref{fig_s13lc} shows the light curve of an RRab star with the largest $|\Delta|/\sigma_T$ ratio. The errors on metallicities for this RRab star are $\sim0.15$~dex, hence $\sigma_T\sim 0.21$~dex. However, this RRab star has a $\Delta$ of $\sim 0.79$~dex, and a resulting a large value of $|\Delta|/\sigma_T$ ratio. After removing the five outliers (three in the middle panel of Figure \ref{fig_s13lc}, and the two in middle-right and lower-right panel of Figure \ref{fig_s13lc}), the mean $\Delta$ reduces to $\sim0.15$~dex for this sample, which is comparable to the accuracy of both methods.

\section{Discussion and Conclusion}

\begin{figure}
  \epsscale{1.0}
  \plotone{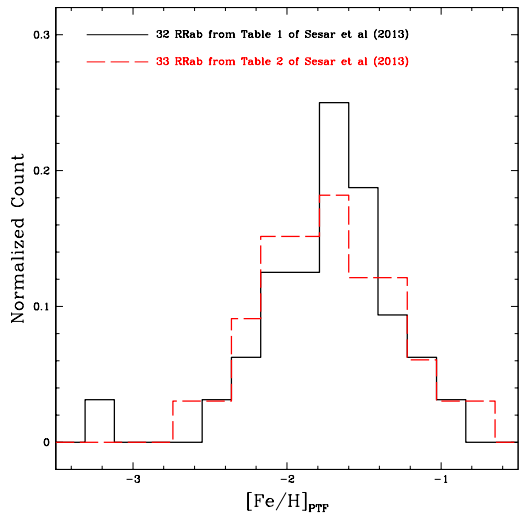}
  \caption{Distributions of the photometric metallicities based on available PTF/iPTF light curves for RRab stars in Tables 1 and 2 from \citet{sesar13}.} 
  \label{fig_hist}
\end{figure}

In this work, we derived the metallicity-light curve relation in the native $R_{PTF}$-band photometric system using the RRab stars found in the {\it Kepler} field. The main reasons for selecting this sample of RRab stars include the availability of accurate pulsation periods (based on {\it Kepler} light curves) and spectroscopic metallicities derived from high-resolution spectra \citep{nemec13}. Since about half of the RRab stars in the {\it Kepler} field are brighter than the saturation limit of $R_{PTF}\sim 14$~mag, we re-observed a number of them with a 10~s exposure time from a dedicated iPTF experiment. Our derived metallicity-light curve relation is presented in Equation (4). When we tested our metallicity-light curve relation for six RRab stars in the {\it Kepler} field with low-resolution P200/DBSP observations, we obtained mixed results with good agreements and discrepancies. The later cases might be due to problems in observed spectra rather than our relation. We further tested our relation with three samples taken from the literature and obtained overall good agreements with our derived $[Fe/H]$ to the published values. Specifically, after removing outliers, we obtained the mean difference between our photometric metallicities and published metallicities with the following values: $\sim -0.09$~dex for the K2E2 RRab stars, $\sim 0.15$~dex for the halo RRab stars in S13, and $\sim 0.24$~dex for a few faint RRab stars in S12 (mainly due to the large scatters of the light curves). When applied our relation to the only RRab star in the Bo\"{o}tes 3 dSphs galaxy, we derived $[Fe/H]_{\mathrm{PTF}}=-2.15\pm0.28$~dex, which is consistent with the spectroscopic measurement given in \citet[][$-2.0\pm0.1$~dex]{sesar14}.

As a demonstration of the applicability of our metallicity-light curve relation, we derived the photometric metallicities for the RRab stars listed in Table 2 of \citet{sesar13} that did not have spectroscopic observations. Out of the 44 RRab stars listed in that table, we could only retrieved the PTF/iPTF light curves for 33 of them (for the same reasons as those mentioned in Section 6.3). The distribution of $[Fe/H]_{\mathrm{PTF}}$ for this sample is similar to the 32 RRab stars taken from Table 1 of \citet[][for which their photometric metallicities have already been derived in Section 6.3]{sesar13}, as demonstrated in Figure \ref{fig_hist}. In the near future, we can use our relation to select RR Lyrae candidates in the Galactic halo found from the Zwicky Transient Facility \citep[ZTF,][]{bellm14,simth14} before requesting spectroscopic observations with large aperture telescopes for confirmation.\footnote{For example, a faint RR Lyrae could be either a distant halo star or a highly extinct field star. If the derived photometric metallicity indicates a metal-poor RR Lyrae, then it is most likely a halo star and is worth the spectroscopic follow-up observations, and vice versa.} The ZTF project is using the same P48 Telescope and almost the same $R$-band filter as the PTF/iPTF projects, but with an upgraded mosaic CCD camera that fills out the focal plane of the P48 Telescope. With much improved survey rates, ZTF can accumulate a much larger number of data points per light curves for the RR Lyrae candidates, and their $\phi_{31}$ Fourier parameters and hence the photometric metallicity can be better constrained. 

\begin{figure*}
  $\begin{array}{cc}
    \includegraphics[angle=0,scale=0.42]{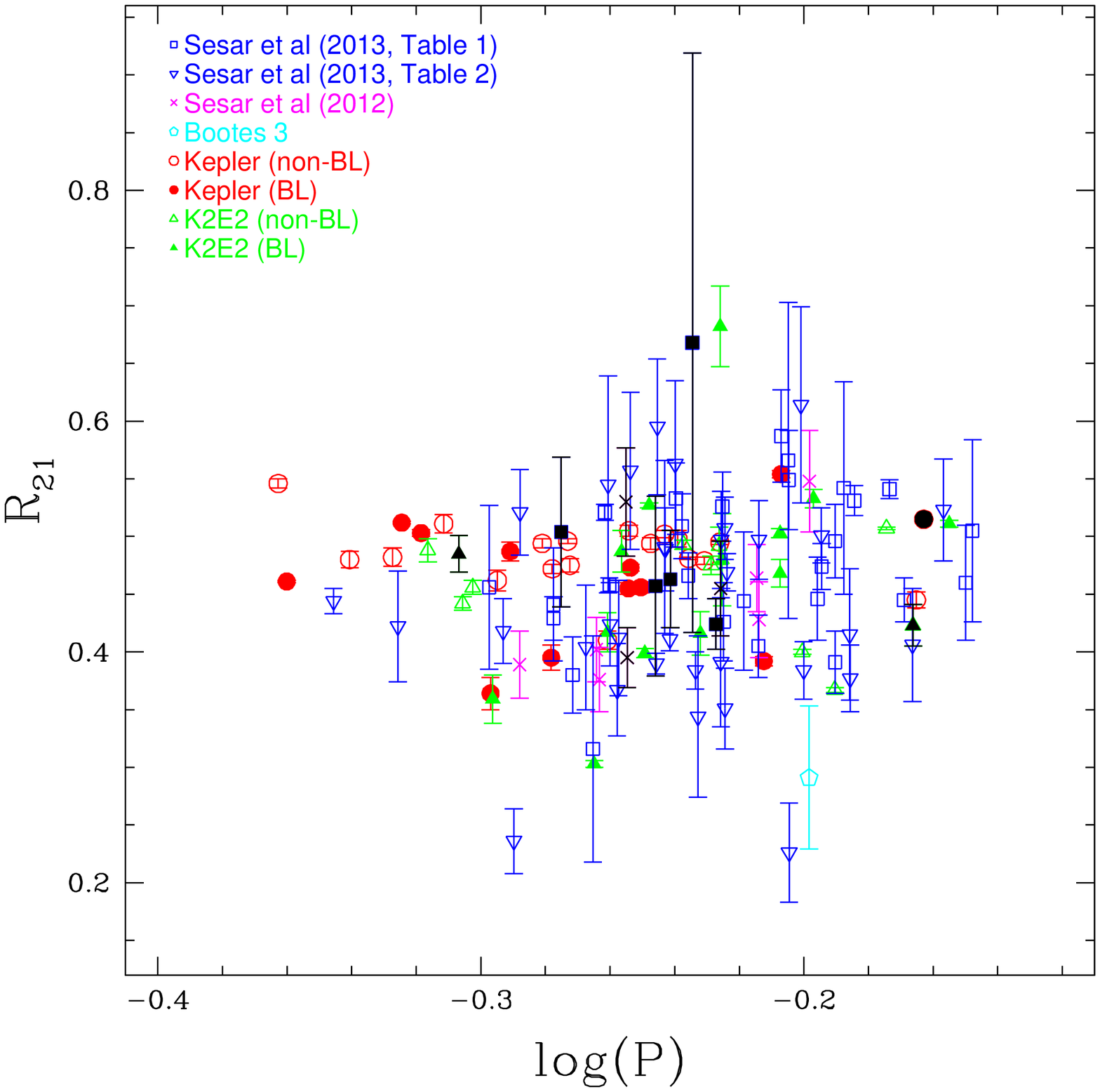} & 
    \includegraphics[angle=0,scale=0.42]{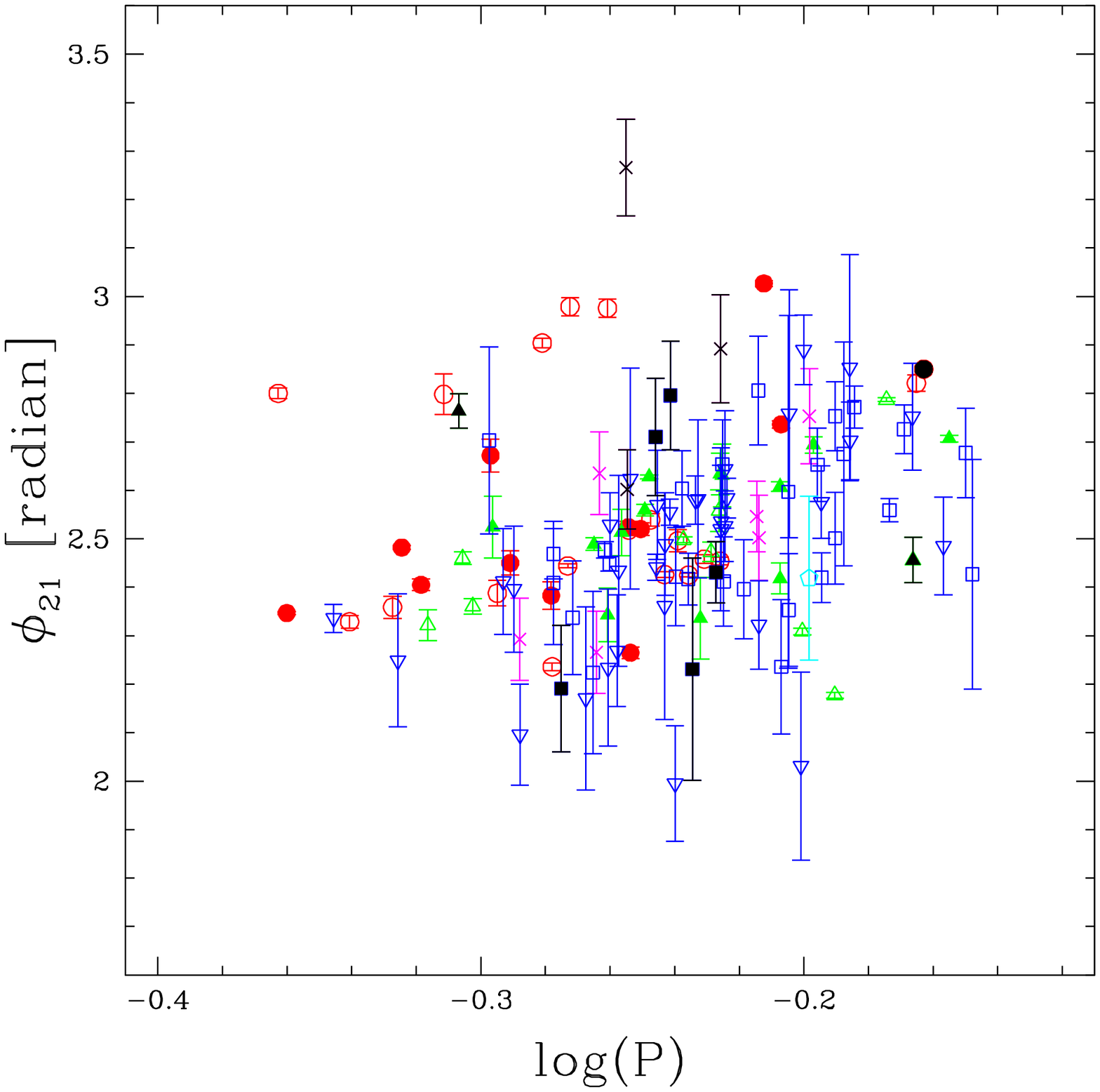} \\
    \includegraphics[angle=0,scale=0.42]{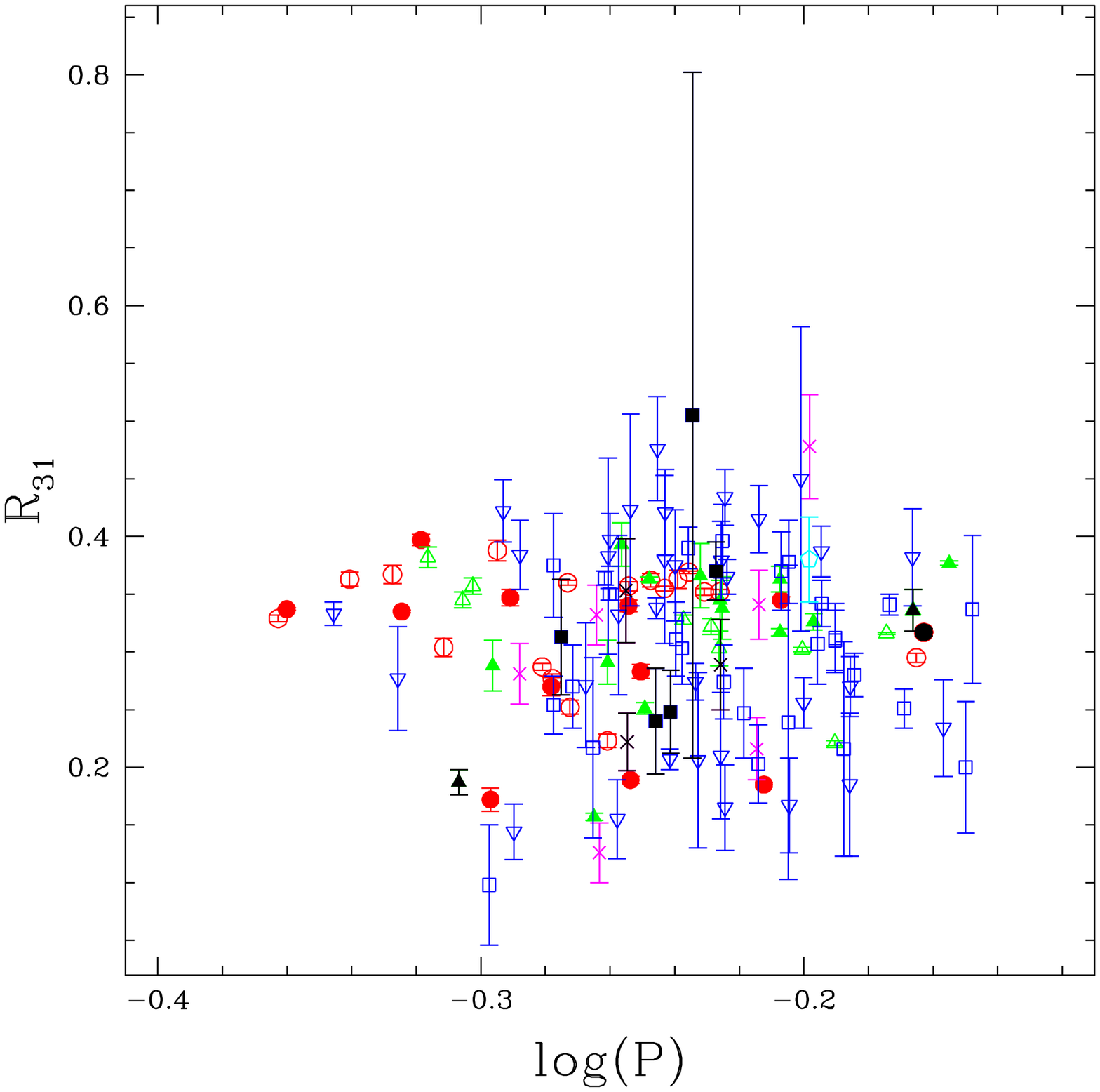} & 
    \includegraphics[angle=0,scale=0.42]{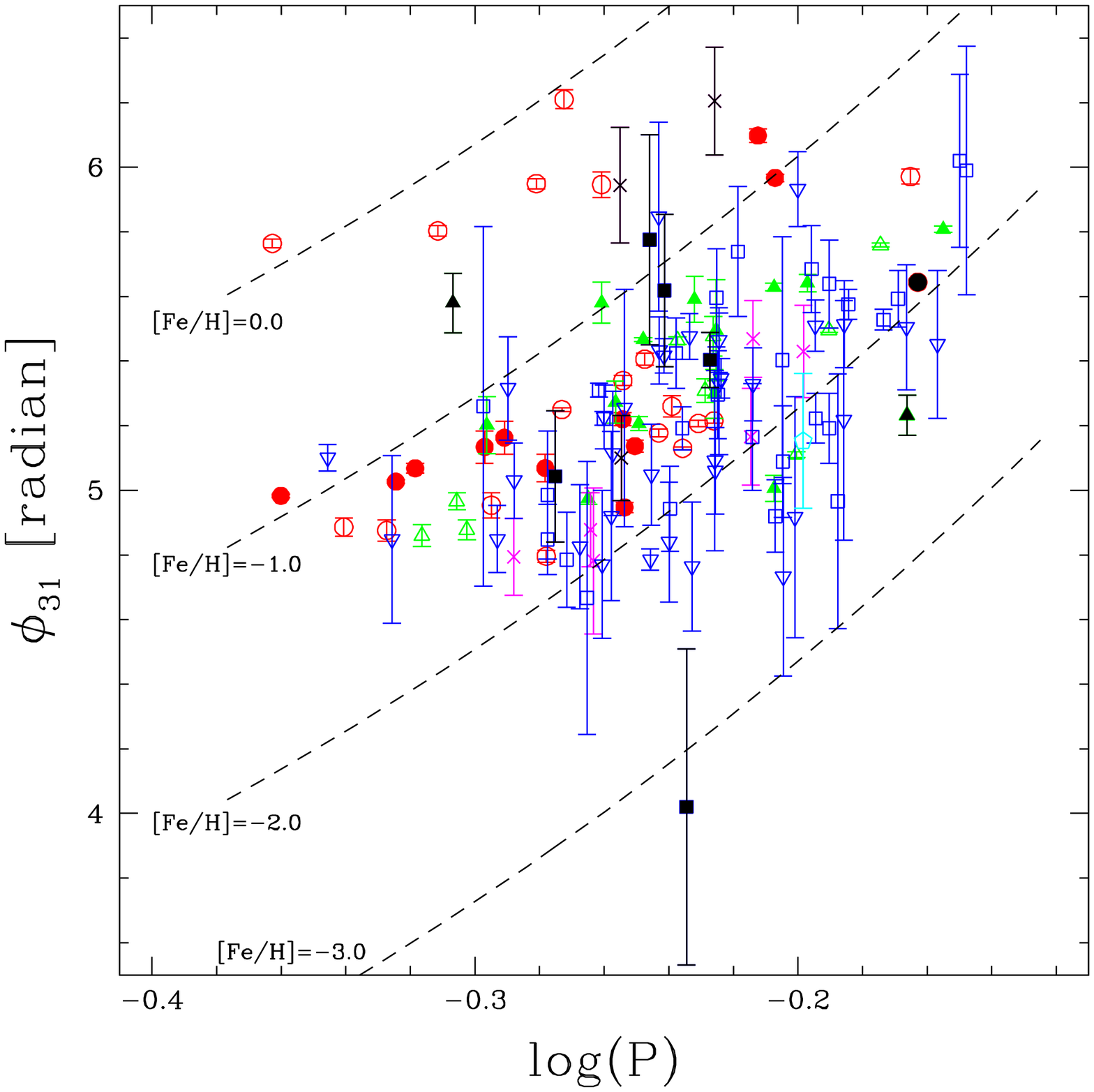} \\
  \end{array}$ 
  \caption{Low-order Fourier parameters, as defined in Equation (2), as a function of periods for all of the 129 RRab stars considered in this work. The different colored symbols represent various samples: red circles are for the RRab stars in the {\it Kepler} field, separated by filled symbols for Blazhko (BL) stars and opened symbols for non-Blazhko (non-BL) stars; upward triangles are for RRab stars from K2E2 Fields (again separated by Blazhko and non-Blazhko stars); crosses are for RR Lyrae taken from \citet{sesar12}; squares and downward triangles are those from Tables 1 and 2 of \citet{sesar13}, respectively. The filled symbols in black are the 11 rejected outliers as mentioned in Section 6. The filled squares in black color with large errors bar are for S13\_RR17. The black cross in the upper-left panel with $\phi_{21}\sim3.3$ is for S12\_RR3. The dashed curves in lower-right panel, for the $\phi_{31}$ Fourier parameters, show the expected $\phi_{31}$ at four different metallicities. These curves are constructed by inverting Equation (4).}
  \label{fig_fourier}
\end{figure*}

Figure \ref{fig_fourier} presents the low-order Fourier parameters for all of the RRab stars that have PTF/iPTF light curves and have been studied in this work. This figure also includes those outliers, shown in black symbols, when comparing the derived $[Fe/H]_{PTF}$ to published metallicities (see Section 6). In terms of Fourier parameters, these outliers are mostly confined within the parameter space defined by other RRab stars. Therefore, these outliers do not exhibit abnormality in terms of the PTF/iPTF light curves. We note that in previous studies, outliers were seen in the comparison of metallicities from (low-resolution) spectroscopic measurements and from the metallicity-light curve relation \citep[see, for example,][]{kocacs05,wu06}. The causes of the majority of the outliers can be traced back to various reasons, including, stars exhibiting Blazhko modulation (without removing the modulated components), problems in the photometry and/or light curves (e.g. noisy light curves, gaps in phased light curves, issues due to photometric calibrations, and etc), inaccurate spectroscopic metallicity, or even the wrong pulsational period being adopted. However, there still exist few outliers that cannot be explained, for example, V341 Aql and DG Hya in \citet{kocacs05} and V341 Aql, UY Boo, DG Hya, RZ Cam, and BK And in \citet{wu06}. In the following, we briefly discuss the possible causes of the 11 outliers marked in Figure \ref{fig_fourier}. A detailed investigation of them is beyond the scope of this work.

KIC 11802860: this RRab star in the {\it Kepler} field is not a Blazhko variable, and its PTF/iPTF light curve did not show any obvious peculiarities. We do not believe the spectroscopic measurements are inaccurate because they were obtained with CFHT and this RRab star is quite bright ($\sim13$ mag). We note that the $\phi_{31}$ Fourier parameter in Johnson-Cousin $R$-band derived from other dense ground-based observations is $5.639$ for this RRab star \citep{jeon14}, which is in good agreement with the value listed in Table \ref{tab3} ($5.644\pm0.007$). Hence, there are no obvious reasons to explain why this RRab star appears to be an outlier.

EPIC 60018743 and EPIC 60018755: one of these RRab stars is a Blazhko variable (EPIC 60018743), and we did not remove its modulated component as in other Blazhko variables in the {\it Kepler} field (see Section 6.2). This might explain why this RRab star is an outlier. We do not think the photometric metallicities $[Fe/H]_{Kp}$ are inaccurate because those measurements are based on almost continuous {\it Kepler} light curves. Nevertheless, the quality of the PTF/iPTF light curves for both of them are similar to other RRab stars in the same sample, as demonstrated in left and right panels of Figure \ref{fig_k2e2_lc}. Their Fourier parameters also agree with other RRab stars at a similar period. The relatively small number of data points per light curve, $\sim40$ for both of them, might incorrectly estimate the $\phi_{31}$ Fourier parameters. In the near future, the accumulation of a large number of data points from ZTF could assist in resolving the outlier status of these two RRab stars. 

S12\_RR2, S12\_RR3 and S12\_RR4: as shown in Figure \ref{fig_s12lc}, PTF/iPTF light curves for these three RRab stars, as well as others in the same sample, are noisy because they are faint RRab stars and hence their photometric measurements are less accurate. This could partially explain why these three RRab stars are outliers. However, their Fourier parameters also agree with other RRab stars, except S12\_RR3 in $\phi_{21}$ plot (upper-left panel in Figure \ref{fig_fourier}). \citet{sesar12} noted that the spectroscopic metallicities for S12\_RR3 were differed by $0.5$~dex from two measurements with different instruments. In contrast, the other two RRab stars (S12\_RR1 and S12\_RR5) with three spectroscopic observations show a very good agreement of measured metallicity (within $0.2$~dex). The low-resolution spectroscopic observation could also partially contribute to the outlier status of these three RRab stars, especially for S12\_RR2 and S12\_RR4 at which they only have one measurement taken from P200/DBSP. Our observations with P200/DBSP, presented in Section 6.1, demonstrated that sometimes the P200/DBSP spectra could lead to an inaccurate measurement of metallicity. Finally, we pointed out that \citet{kocacs05} declared a spectroscopic metallicity is inaccurate if the number of measurements is less than three.

S13\_RR6, S13\_RR21, S13\_RR23 and S13\_RR35: Fourier parameters for these four RRab stars are located within the parameter space defined by all other RRab stars, as shown in Figure \ref{fig_fourier}. Their PTF/iPTF light curves (see Figure \ref{fig_s13lc}) also did not exhibit any abnormality, except for S13\_RR6, which is noisier. We therefore believe their $\phi_{31}$ values are reasonably estimated. Nevertheless, with additional accumulated data points per light curves from the ZTF, the accuracy of $\phi_{31}$ values, and hence their photometric metallicities, could be improved. As discussed previously, the possibility of inaccurate spectroscopic metallicity based on single observations from P200/DBSP could not be ruled out either. 

S13\_RR17: this RRab star has the largest errors on $R_{21}$ and $R_{31}$ Four parameters, and the second largest errors on $\phi_{21}$ and $\phi_{31}$ Four parameters. Its phased PTF/iPTF light curve also has a gap around the phase of $\sim 0.4$. Furthermore, the minimum light for this RRab star is near $R_{PTF}\sim19$~mag, hence the photometric measurements are less accurate around phases of minimum light. Combining these reasons, we believe the problems on the light curve lead to the inaccurate measurement of photometric metallicity, which is hence displayed as an outlier in Figure \ref{fig_feh_compare_sesar}. This most metal-poor RRab star worth the collection of additional light curve data in the era of ZTF.

\begin{figure}
  %\epsscale{0.6}
  \plotone{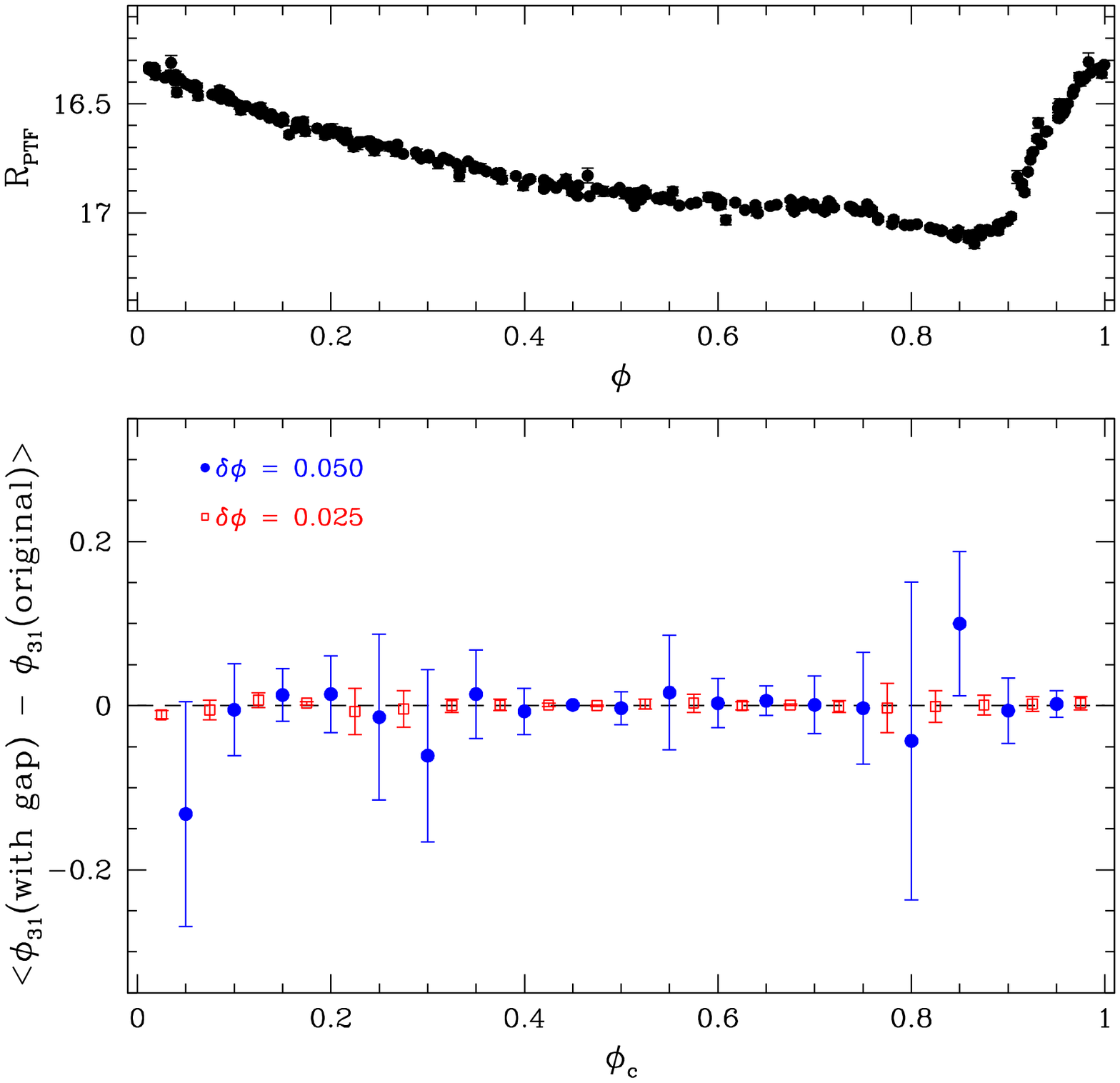}
  \caption{Results on testing the effect of phase gaps in observed light curves on the determination of the $\phi_{31}$ Fourier parameter, and hence the $[Fe/H]_{PTF}$ metallicity. The upper panel displays an example of a well-sampled light curve for a typical non-Blazhko RRab star, with maximum light located at phase zero ($\phi=0$). The lower panel shows the weighted mean variations of $\phi_{31}$ (i.e. the difference of $\phi_{31}$ in the original light curves and after including a phase gap in the light curves) as a function of $\phi_c$, the location of a phase gap that is artificially included in the phased light curves. The phase gap is created by removing those data points with $\phi$ falling within $\phi_c\pm\delta \phi$ of the well-sampled light curves. Open squares and filled circles shown in the lower panel are for the cases of $\delta \phi=0.025$ (for a phase gap with a width of 0.05) and $\delta \phi=0.050$ (for a phase gap with a width of 0.1), respectively. Error bars represent the standard deviations on the weighted means, based on the combined light curves for non-Blazhko RRab stars in the {\it Kepler} field that have large numbers of data points ($N>120$, to ensure a good coverage of the light curve) in their light curves. The same Fourier order $n$ was adopted in both of the original light curves and the light curves with artificial phase gap included. The dashed line indicates $y=0$ and not the fit to the data points.} 
  \label{fig_phasegap}
\end{figure}

Finally, we examined the influence of a gap in phased light curves when determining the $\phi_{31}$ Fourier parameters, which translate to $[Fe/H]_{PTF}$ via Equation (4). We took several of the well-sampled light curves from the non-Blazhko RRab stars in the {\it Kepler} field and artificially removed some data points to mimic a phase gap in the light curves. Figure \ref{fig_phasegap} presents the test results for the phase gap, with a width of 0.05 and 0.1, located at different parts of the phased light curve. In the case of a phase gap that has a width of 0.05, Figure \ref{fig_phasegap} reveals that such a phase gap will not alter the determination of the $\phi_{31}$ Fourier parameters, regardless of the location of the phase gap. When the width of the phase gap is increased to 0.1, the $\phi_{31}$ Fourier parameters could be affected by the presence of the phase gap near the maximum or minimum light, as is indicated by the the two most deviated (filled circle) points in Figure \ref{fig_phasegap}. This result reiterates the finding in \citet{wu06}, who suggested that the photometric metallicity can still be estimated from light curves with insufficient phase coverage when there are data points around the maximum or minimum light. In other phases around the ascending and descending branch of the light curve, the phase gap with width of 0.1 does not greatly affect the determination of the $\phi_{31}$ Fourier parameters. Certainly, the $\phi_{31}$ Fourier parameters would be less accurate when the width of the phase gap increases, and auxiliary techniques such as adding an interpolated data point in the gap (as employed in this work) or applying a polynomial fit \citep{jurcsik96} are needed to recover the $\phi_{31}$ Fourier parameters. We anticipate that the problem of the phase gap presented in light curves will be diminished within the ZTF project because a much larger number of data points per light curve will be collected in the era of ZTF.

\vspace{0.5cm}

{\it Facility:} \facility{PO:1.2m}

\acknowledgments

We thank the referee for valuable input that improved the manuscript. We acknowledge Wee Siang Edmund Yuen, a 2015 summer student at the National Central University, for carrying out the preliminary analysis of the PTF data for RR Lyrae in the {\it Kepler} field. CCN is thankful for the funding from the Ministry of Science and Technology (Taiwan) under the grants 104-2112-M-008-012-MY3 and 104-2119-M-008-024. This research has made use of the NASA/IPAC Infrared Science Archive, which is operated by the Jet Propulsion Laboratory, California Institute of Technology, under contract with the National Aeronautics and Space Administration.

% ===============================================
%               REFERENCE
% ===============================================


\begin{thebibliography}{}

\bibitem[Alcock et al.(2003)]{alcock03} Alcock, C., Alves, D.~R., Becker, A., et al.\ 2003, \apj, 598, 597 

\bibitem[Bellm(2014)]{bellm14} Bellm, E.\ 2014, The Third Hot-wiring the Transient Universe Workshop (HTU-III), Proceedings of the Third Hot-wiring the Transient Universe Workshop, Edited by P. R. Wozniak, M. J. Graham, A. A. Mahabal and R. Seaman, 27

\bibitem[Bellm et al.(2016)]{bellm16} Bellm, E.~C., Kaplan, D.~L., Breton, R.~P., et al.\ 2016, \apj, 816, 74 

\bibitem[Bellm \& Sesar(2016)]{bellm16s} Bellm, E.~C., \& Sesar, B.\ 2016, {\tt pyraf-dbsp: Reduction pipeline for the Palomar Double Beam Spectrograph}, Astrophysics Source Code Library, ascl:1602.002 

\bibitem[Benk{\H o} et al.(2011)]{benko11} Benk{\H o}, J.~M., Szab{\'o}, R., \& Papar{\'o}, M.\ 2011, \mnras, 417, 974 

\bibitem[Bertin \& Arnouts(1996)]{bertin96} Bertin, E., \& Arnouts, S.\ 1996, \aaps, 117, 393 

\bibitem[Braga et al.(2015)]{braga15} Braga, V.~F., Dall'Ora, M., Bono, G., et al.\ 2015, \apj, 799, 165 

\bibitem[Caputo et al.(2000)]{caputo00} Caputo, F., Castellani, V., Marconi, M., \& Ripepi, V.\ 2000, \mnras, 316, 819

\bibitem[Chang et al.(2014)]{chang14} Chang, C.-K., Ip, W.-H., Lin, H.-W., et al.\ 2014, \apj, 788, 17 

\bibitem[De Lee(2008)]{delee08} De Lee, N.\ 2008, Ph.D.~Thesis, Michigan State University

\bibitem[Deb \& Singh(2009)]{deb09} Deb, S., \& Singh, H.~P.\ 2009, \aap, 507, 1729 

\bibitem[Deb \& Singh(2010)]{deb10} Deb, S., \& Singh, H.~P.\ 2010, \mnras, 402, 691 

\bibitem[Demarque et al.(2000)]{demarque00} Demarque, P., Zinn, R., Lee, Y.-W., \& Yi, S.\ 2000, \aj, 119, 1398 

\bibitem[Gratton et al.(2004)]{gratton04} Gratton, R.~G., Bragaglia, A., Clementini, G., et al.\ 2004, \aap, 421, 937 

\bibitem[Guggenberger et al.(2012)]{guggenberger12} Guggenberger, E., Kolenberg, K., Nemec, J.~M., et al.\ 2012, \mnras, 424, 649 

\bibitem[Honeycutt(1992)]{honeycutt1992} Honeycutt, R.~K.\ 1992, \pasp, 104, 435 

\bibitem[Jeon et al.(2014)]{jeon14} Jeon, Y.-B., Ngeow, C.-C., \& Nemec, J.~M.\ 2014, Precision Asteroseismology, edited by Guzik, J. A., Chaplin, W. J., Handler, G. \& Pigulski, A., Proceedings of the International Astronomical Union (Cambridge University Press), IAU Symposium, 301, 427 

\bibitem[Jurcsik(2003)]{jurcsik03} Jurcsik, J.\ 2003, \aap, 403, 587 

\bibitem[Jurcsik \& Kov{\'a}cs(1996)]{jurcsik96} Jurcsik, J., \& Kov{\'a}cs, G.\ 1996, \aap, 312, 111 

\bibitem[Kunder \& Chaboyer(2008)]{kunder08} Kunder, A., \& Chaboyer, B.\ 2008, \aj, 136, 2441 

\bibitem[Kov{\'a}cs(1995)]{kovacs95a} Kov{\'a}cs, G.\ 1995, \aap, 295, 693 

\bibitem[Kov{\'a}cs(2005)]{kocacs05} Kov{\'a}cs, G.\ 2005, \aap, 438, 227 

\bibitem[Kov{\'a}cs \& Zsoldos(1995)]{kovacs95} Kov{\'a}cs, G., \& Zsoldos, E.\ 1995, \aap, 293, L57 

\bibitem[Laher et al.(2014)]{laher14} Laher, R.~R., Surace, J., Grillmair, C.~J., et al.\ 2014, \pasp, 126, 674 

\bibitem[Law et al.(2009)]{law09} Law, N.~M., Kulkarni, S.~R., Dekany, R.~G., et al.\ 2009, \pasp, 121, 1395 

\bibitem[Marconi et al.(2015)]{marconi15} Marconi, M., Coppola, G., Bono, G., et al.\ 2015, \apj, 808, 50 

\bibitem[Martinez-Vazquez et al.(2016)]{mv16} Martinez-Vazquez, C.~E., Monelli, M., Bono, G., et al.\ 2016, RRL2015: High-Precision Studies of RR Lyrae Stars, Communications from the Konkoly Observatory, Vol. 105, edited by L. Szabados, R. Szabó and K. Kinemuchi, 53

\bibitem[McNamara(1999)]{mcnamara99} McNamara, D.~H.\ 1999, \pasp, 111, 489 

\bibitem[Moln{\'a}r et al.(2015)]{molnar15} Moln{\'a}r, L., Szab{\'o}, R., Moskalik, P.~A., et al.\ 2015, \mnras, 452, 4283 

\bibitem[Nemec et al.(2011)]{nemec11} Nemec, J.~M., Smolec, R., Benk{\H o}, J.~M., et al.\ 2011, \mnras, 417, 1022 

\bibitem[Nemec et al.(2013)]{nemec13} Nemec, J.~M., Cohen, J.~G., Ripepi, V., et al.\ 2013, \apj, 773, 181 

\bibitem[Ngeow(2015)]{ngeow15} Ngeow, C.~C.\ 2015, The 2015 International Conference on Space Science and Communication (IconSpace2015), Proc. 2015 IEEE Int. Conf. on Space Science and Communication (IconSpace2015), edited by Mandeep Singh, Mohd Fais Mansor, Wayan Suparta, Siti Aminah Bahari and Ahmad Ridzuan Mohammed Shariff, 302

\bibitem[Ofek et al.(2012a)]{ofek12a} Ofek, E.~O., Laher, R., Law, N., et al.\ 2012a, \pasp, 124, 62 

\bibitem[Ofek et al.(2012b)]{ofek12b} Ofek, E.~O., Laher, R., Surace, J., et al.\ 2012b, \pasp, 124, 854 

\bibitem[Oke \& Gunn(1982)]{oke82} Oke, J.~B., \& Gunn, J.~E.\ 1982, \pasp, 94, 586 

\bibitem[Oke et al.(1995)]{oke95} Oke, J.~B., Cohen, J.~G., Carr, M., et al.\ 1995, \pasp, 107, 375 

\bibitem[Oluseyi et al.(2012)]{oluseyi12} Oluseyi, H.~M., Becker, A.~C., Culliton, C., et al.\ 2012, \aj, 144, 9 

\bibitem[Rahmer et al.(2008)]{rahmer08} Rahmer, G., Smith, R., Velur, V., et al.\ 2008, \procspie, 7014, 70144Y 
%Proceedings of the SPIE

\bibitem[Rau et al.(2009)]{rau09} Rau, A., Kulkarni, S.~R., Law, N.~M., et al.\ 2009, \pasp, 121, 1334 

\bibitem[Sandage(2004)]{sandage04} Sandage, A.\ 2004, \aj, 128, 858 

\bibitem[Sandage \& Tammann(2006)]{sandage06} Sandage, A., \& Tammann, G.~A.\ 2006, \araa, 44, 93 

\bibitem[Sesar et al.(2010)]{sesar10} Sesar, B., Ivezi{\'c}, {\v Z}., Grammer, S.~H., et al.\ 2010, \apj, 708, 717 

\bibitem[Sesar et al.(2012)]{sesar12} Sesar, B., Cohen, J.~G., Levitan, D., et al.\ 2012, \apj, 755, 134 

\bibitem[Sesar et al.(2013)]{sesar13} Sesar, B., Grillmair, C.~J., Cohen, J.~G., et al.\ 2013, \apj, 776, 26 

\bibitem[Sesar et al.(2014)]{sesar14} Sesar, B., Banholzer, S.~R., Cohen, J.~G., et al.\ 2014, \apj, 793, 135

\bibitem[Simon \& Lee(1981)]{simon81} Simon, N.~R., \& Lee, A.~S.\ 1981, \apj, 248, 291 

\bibitem[Simon(1988)]{simon88} Simon, N.~R.\ 1988, \apj, 328, 747

\bibitem[Skowron et al.(2016)]{skowron16} Skowron, D.~M., Soszy{\'n}ski, I., Udalski, A., et al.\ 2016, Acta Astronomica, 66, 269

\bibitem[Smith et al.(2014)]{simth14} Smith, R.~M., Dekany, R.~G., Bebek, C., et al.\ 2014, \procspie, 9147, 914779 
%Proceedings of the SPIE

\bibitem[Smolec(2005)]{smolec05} Smolec, R.\ 2005, \actaa, 55, 59 

\bibitem[Surace et al.(2015)]{surace15} Surace, J., Laher, R., Masci, F., Grillmair, C., \& Helou, G.\ 2015, Astronomical Data Analysis Software an Systems XXIV (ADASS XXIV). Edited by A. R. Taylor and E. Rosolowsky. San Francisco: Astronomical Society of the Pacific, 495, 197 

\bibitem[van Eyken et al.(2011)]{vanE11} van Eyken, J.~C., Ciardi, D.~R., Rebull, L.~M., et al.\ 2011, \aj, 142, 60 

\bibitem[Watkins et al.(2009)]{watkins09} Watkins, L.~L., Evans, N.~W., Belokurov, V., et al.\ 2009, \mnras, 398, 1757 

\bibitem[Wu et al.(2006)]{wu06} Wu, C., Qiu, Y.~L., Deng, J.~S., Hu, J.~Y., \& Zhao, Y.~H.\ 2006, \aap, 453, 895 


\end{thebibliography}
\end{document}